\documentclass[prd,eqsecnum,floats,aps,showpacs,twocolumn,floatfix,superscriptaddress]{revtex4-1}
\usepackage{amsmath}
\usepackage{graphicx}
\usepackage{cancel}
\usepackage{multirow}

\newcommand{\vrr}[0]{\vec{r}}

\newcommand{\ignore}[1]{}
\def \im {{\rm i}}

\newcommand{\q}{\vec{q}}
\newcommand{\ta}{\vec{\tau}}

\newcommand{\el}{\mathcal{L}}

\newcommand{\intl}{\int\frac{d^4l}{(2\pi)^4}}

\newcommand{\intlq}{\int\frac{d^4l}{(2\pi)^4}}
\newcommand{\intlt}{\int\frac{d^3l}{(2\pi)^3}}

\newcommand{\ou}{\overline{u}}
\newcommand{\oee}{\overline{E}}
\newcommand{\om}{\overline{M}}

\newcommand{\vp}{\vec{p}}
\newcommand{\vpp}{\vec{p}\,'}
\newcommand{\kn}{\cancel{k}_N}
\newcommand{\rn}{\cancel{r}_N}
\newcommand{\vl}{\vec{l}}

\newcommand{\vq}{\vec{q}}

\newcommand{\vsu}{\vec{\sigma}_1}
\newcommand{\vsd}{\vec{\sigma}_2}
\newcommand{\vtu}{\vec{\tau}_1}
\newcommand{\vtd}{\vec{\tau}_2}

\newcommand{\intx}{\int_0^1dx}
\newcommand{\inty}{\int_{0}^1dy}
\newcommand{\intz}{\int_0^1dz}
\newcommand{\lm}{l_{\mu}}
\newcommand{\lnn}{l_{\nu}}
\newcommand{\lr}{l_{\rho}}

\newcommand{\cq}{C_q}
\newcommand{\cqp}{C_q'}

\begin{document}
\title{One-loop contributions in the EFT for the 
$\Lambda N \to NN$ transition}

\author{A. P\'erez-Obiol}
\email[]{axel@ecm.ub.edu}
\affiliation{Departament d'Estructura i Constituents de la Mat\`{e}ria,\\
Institut de Ci\`encies del Cosmos (ICC), \\
Universitat de Barcelona, Mart\'i Franqu\`es 1, E--08028, Spain}

\author{D. R. Entem}
\email[]{entem@usal.es} 
\affiliation{Grupo de F\'isica Nuclear and IUFFyM,\\
Universidad de Salamanca, E-37008 Salamanca, Spain}

\author{B. Juli\'a-D\'{\i}az}
\email[]{bruno@ecm.ub.edu} 
\affiliation{Departament d'Estructura i Constituents de la Mat\`{e}ria,\\
Institut de Ci\`encies del Cosmos (ICC), \\
Universitat de Barcelona, Mart\'i Franqu\`es 1, E--08028, Spain}

\author{A. Parre\~no}
\email[]{assum@ecm.ub.edu}
\affiliation{Departament d'Estructura i Constituents de la Mat\`{e}ria,\\
Institut de Ci\`encies del Cosmos (ICC), \\
Universitat de Barcelona, Mart\'i Franqu\`es 1, E--08028, Spain}

\date{\today}

\begin{abstract}
We consider the $\Lambda N\to NN$ weak transition, responsible for 
a large fraction of the non-mesonic weak decay of hypernuclei. 
We follow on the previously derived effective field theory and 
compute the next-to-leading one-loop corrections. Explicit 
expressions for all diagrams are provided, which result in 
contributions to all relevant partial waves.  
\end{abstract}

\maketitle

\section{Introduction}

One of the major challenges in nuclear physics is to understand the
interactions among hadrons from first principles. For more than twenty
years, many research groups have directed their efforts to develop
Effective Field Theories (EFT), working with the idea of separating
the nuclear force in 
long-range  and short-range components. The underlying premise was
that low-energy processes, as the ones encountered in nuclear physics,
should not be affected by the specific details of the high-energy physics.

The typical energies associated to nuclear phenomena suggest that the
appropriate degrees of freedom are nucleons and pions (or the ground
state baryon and pseudo scalar octets for processes involving
strangeness), interacting derivatively as it is dictated by the effective chiral
Lagrangian. The nuclear interaction is characterized by the presence
of very different scales, going from the values of the masses of the
light pseudo-scalar bosons to the ones of the ground-state octet
baryons. The EFT formalism makes use of this separation of scales to
construct an expansion of the Lagrangian in terms of a parameter built
up from ratios of these scales. For example, in the study of the
low-energy nucleon-nucleon interaction, a clear 
 separation of scales is seen between the external momentum of the
 interacting nucleons, a soft scale which typically takes values up to
 the pion mass, and a hard scale corresponding to the nucleon mass. While the long-range part
of this interaction is governed by the light scale through the
pion-exchange mechanism, short-range forces are accounted for by
zero-range contact operators, organized according to an increasing number of derivatives. These
contact terms, which respect chiral symmetry, have values which are
not constrained by the chiral Lagrangian, and therefore, their
relative strength (encapsulated in the size of the low-energy coefficients, LECs) has to be
obtained from a fit to nuclear observables. The large amount of
experimental data for the interaction among pions and nucleons has
made possible to perform successful EFT calculations of the strong nucleon-nucleon
interaction up to fourth order in the momentum expansion (${\cal
  O}(p^4)$), at next-to-next-to-next-to-leading order (N$^3$LO) in the
heavy-baryon formalism~\cite{Epelbaum:2012vx,entem}.
In the weak sector, the study of nucleon-nucleon Parity Violation (PV)
with an Effective Field Theory at leading order has been undertaken in
Ref.~\cite{nucPV05}, where the authors discuss existing and possible
few-body measurements that can help in constraining the relevant (five) low-energy
constants at order $p$ in the momentum expansion and the ones
associated with dynamical pions.

In the strange sector, the experimental situation is less favorable
due to the short life-time of hyperons, unstable against the weak
interaction. This fact complicates the extraction of information
regarding the strong interaction among baryons in free space away from the nucleonic
sector. Nevertheless, SU(3) extensions of the EFT for nucleons and
pions have been developed at leading order (LO)~\cite{SW96, KDT01,
  H02, BBPS05} and next-to-leading (NLO) order~\cite{PHM06}. In the
present work we consider the weak 
four-body $\Lambda N \to NN$ interaction, which is accessible
experimentally by looking at the decay of $\Lambda-$hypernuclei, bound
systems composed by nucleons and one $\Lambda$ hyperon.
%{\it i.e.} looking at systems where nonperturbative effects happen to
%be important (note that, in the calculation of these decays, the weak
%interaction is included in a perturbative way).
These aggregates decay weakly through mesonic ($\Lambda \to N \pi$) and
non-mesonic ($\Lambda N \to NN$) modes, the former being suppressed
for
mass numbers of the order or larger than 5, due to the Pauli blocking
effect
acting on the outgoing nucleon.  In contrast to the weak NN PV
interaction, which is masked by the much stronger Parity Conserving
(PC) strong NN signal, the weak $|\Delta S|=1 \, \Lambda N$
interaction has the advantage of presenting a change of flavor as a
signature, favoring its
detection in the presence of the strong interaction.

The first studies
of the weak $\Lambda N$ interaction using a lowest order effective
theory were presented in Refs.~\cite{Jun, PBH05,PPJ11} . These
works included the exchange of the lighter pseudoscalar mesons while
parametrizing the short-range part of the interaction with contact
terms at order ${\cal O}(q^0)$, where $q$ denotes the momentum
exchanged between the interacting baryons. While the results of
Ref.~\cite{PPJ11} show that it is possible to
reproduce the hypernuclear decay data with the lowest order effective
Lagrangian, the stability of the momentum expansion has to be checked
by including the next order in the EFT.
If an effective field theory
can be built for the weak $\Lambda N \to NN$ transition, the values
for the LECs of the theory, which encode the high-energy components of
the interaction, should vary within a reasonable and natural range
when one includes higher orders in the calculation.
Compared to the LO
calculation, which involves two LECs, the unknown baryon-baryon-kaon
vertices and the pseudoscalar cut-off parameter in the form-factor,
the NLO calculation introduces additional unknowns. Namely, the
parameters associated to the new contact terms (three when one
neglects the small value of the
momentum of the initial particles, a nucleon and a hyperon bound in
the hypernucleus, in front of the momentum of the two outgoing
nucleons) and the couplings appearing in the two-pion exchange
diagrams. Therefore, in order to constrain the EFT at NLO, one needs
to collect enough data, either 
through the accurate measure of hypernuclear decay observables, or
through the measure of the inverse reaction in free space, $n p \to
\Lambda p$. Unfortunately, the small values of the cross-sections for
the weak 
strangeness production mechanism, of the order of  $10^{-12}$
mb~\cite{Haidenbauer1995,Parreno1998,Inoue2001}, has prevented, for
the time being, its consideration as part of the experimental data
set, despite the effort invested in 
extracting different polarization observables for this
process~\cite{Kishimoto2000,Ajimura2001}.
%, they have failed in providing with a clear signal for any of the
%observables. Until the experimental community overcomes this
%difficulty and/or
%nuclear physisicsts come up with new alternative measurements, the
%decay of hypernuclei will be the only quantitative way to obtain
%information on the weak $|\Delta S|=1$ four-fermion interaction.
At present, quantitative experimental information on the $|\Delta
S|=1$ weak interaction in the baryonic sector comes from the measure
of
the total and partial decay rates of hypernuclei, and an asymmetry in
the number of protons detected parallel and antiparallel to the
polarization axis, which comes from the interference between the PC
and PV weak amplitudes. Since observables from one hypernucleus to
another
can be related through hypernuclear structure coefficients, one has to
be careful
in selecting the data that can be used in the EFT calculation. For
example,
while one may indeed
expect measurements from different p-shell hypernuclei, say, A=12 and
16, to provide with the same constraint, the situation is different
when including data from s-shell hypernuclei like A=5. For the latter,
the initial $\Lambda N$ pair can only be in a relative s-state, while
for the former, relative p-states are allowed as well.
%We therefore conclude that only 6 data points (3 observables for a
%given s-shell hypernucleus and 3 for a given p-shell hypernucleus)
%would be in principle independent and could be used in our EFT
%calculation.

In this paper we present the analytic expressions to be included at
next-to-leading order
in the effective theory for the weak $\Lambda N$ interaction. These
expressions have been derived by considering four-fermion contact
terms with a derivative operator insertion together with the two-pion
exchange mechanism.

The paper is organized as follows. In Section II we introduce
the Lagrangians and the power counting scheme we use to calculate the
relevant Feynman diagrams. In Sections \ref{ss:loc} and \ref{ss:nloc}
we present the LO and NLO potentials for the $\Lambda N\rightarrow NN$
transition,
and a comparison between both contributions is performed in Section
\ref{sec:bc}. We conclude and summarize in Section
\ref{sec:conclusions}.

\section{Interaction Lagrangians and counting scheme}
\label{sec2}

The non-mesonic weak decay of the $\Lambda$ involves both the 
strong and electroweak interactions. The $\Lambda$ decay is 
mediated by the presence of a nucleon which in the simplest 
meson-exchange picture, exchanges a meson, e.g. $\pi$, $K$, 
with the $\Lambda$. Thus, computing the transition requires 
the knowledge of the strong and weak Lagrangians
involving all the hadrons entering in the process. In this 
section we describe the strong and weak Lagrangians entering 
at leading order (LO) and next-to-leading order (NLO) in the 
$\Lambda N\to NN$ interaction. 

%\subsection{Weak interaction}

\begin{figure}[t]
\includegraphics[scale=0.25]{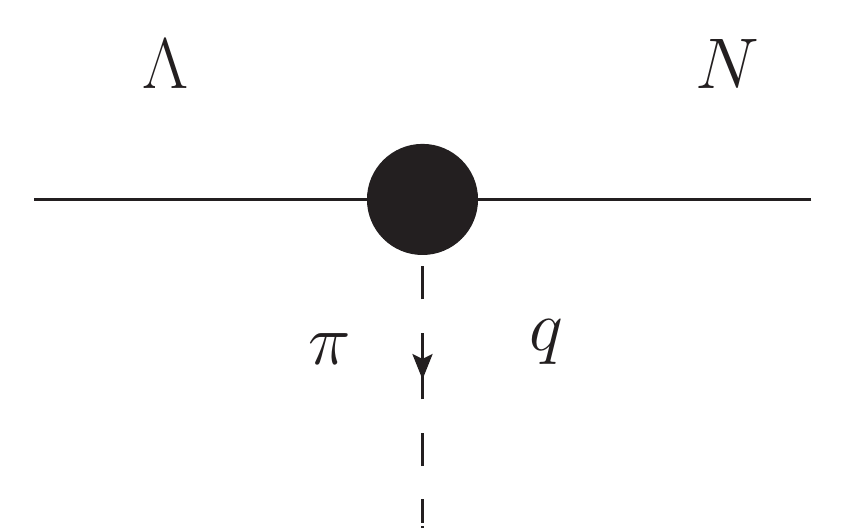}
\includegraphics[scale=0.25]{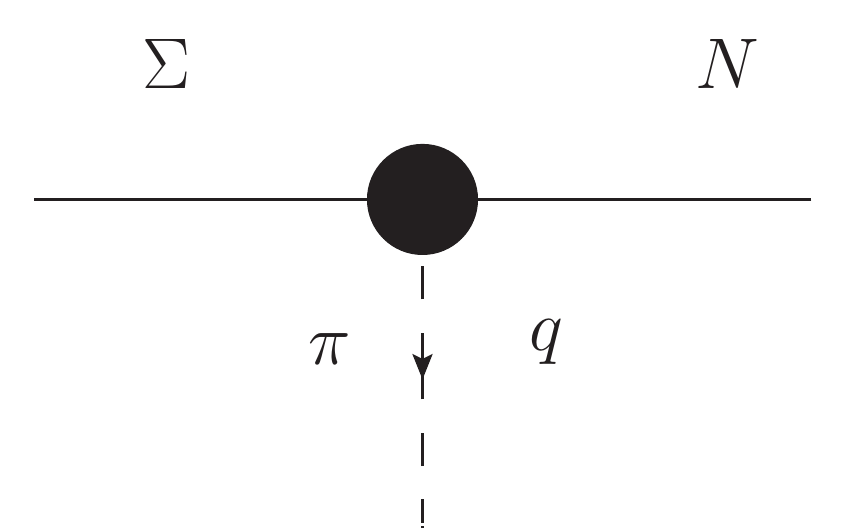}
\includegraphics[scale=0.25]{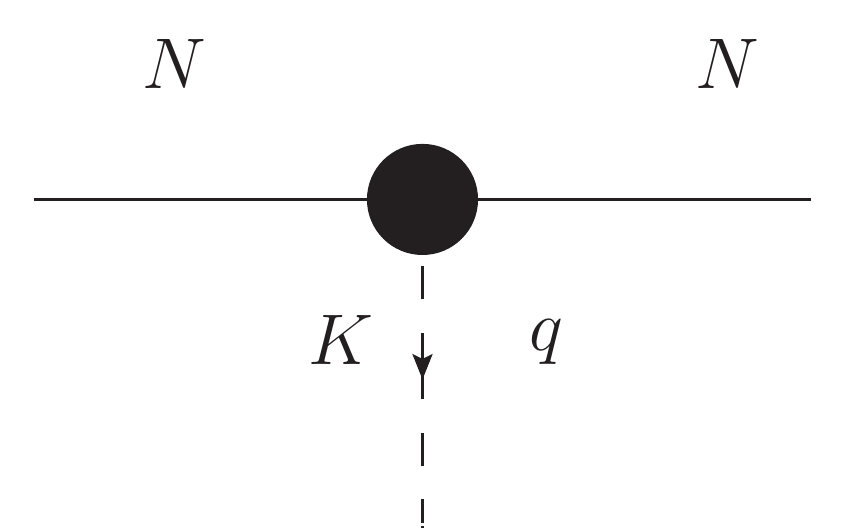}
\caption{Weak vertices for the $\Lambda N\pi$, 
$\Sigma N\pi$ and $NNK$ stemming from the Lagrangians in Eq.~(\ref{eq:weakl}). 
The weak vertex is represented by a solid black circle.
\label{vf2}}      
\end{figure}

The weak interaction between the $\Sigma$, $\Lambda$ and $N$ baryons
and the pseudoscalar $\pi$ and $K$ mesons is described by the
phenomenological Lagrangians:
\ignore{There are five weak interaction Lagrangians involving $N$, $\pi$, 
$\Sigma$ and $\Lambda$ that enter at LO and NLO in the 
$\Lambda N\to NN$ transition. The first two are related to the 
free space decay of the $\Lambda$ and $\Sigma$ particles, and the
third is derived from the first using SU(3) symmetry. In order 
to have a quantitative description of the $\Lambda$ and $\Sigma$ 
weak decay, we consider the following phenomenological Lagrangians, }
\begin{align}
\label{eq:weakl}
\el_{\Lambda N\pi}^w=&-iG_Fm_\pi^2\overline{\Psi}_N(A+B\gamma^5)
\ta\cdot\vec{\pi}\Psi_\Lambda
\\\nonumber
\el_{\Sigma N\pi}^w=&
-iG_Fm_\pi^2\overline{\Psi}_N(\vec{A}_{\Sigma_i}+\vec{B}_{\Sigma_i}\gamma^5)
\cdot\vec{\pi}\Psi_{\Sigma_i}\,,
\\\nonumber
{\cal L}^{w}_{NN K} =&
-iG_Fm_\pi^2 \, \left[ \, \overline{\psi}_{N} \left( ^0_1 \right)
\,\,( C_{K}^{PV} + C_{K}^{PC} \gamma_5) \,\,(\phi^{K})^\dagger
\psi_{N} \right.
\\ \nonumber
 & \left. + \, \overline{\psi}_{N} \psi_{N}
\,\,( D_{K}^{PV}
+ D_{K}^{PC}
\gamma_5 ) \,\,(\phi^{K})^\dagger \,\,
\left( ^0_1 \right) \right] \ ,
\end{align}
where $G_Fm_\pi^2=2.21\times10^{-7}$ is the weak Fermi coupling
constant, $\gamma$ are the Dirac matrices and 
$\tau$ the Pauli matrices.
The index $i$ appearing in the $\Sigma$ field refers to the different
isospurion states for the $\Sigma$ hyperon:
\begin{equation}
\Psi_{\Sigma\frac12}=
\left(\begin{array}{c}-\sqrt{\frac23}\Sigma_+\\\frac{1}{\sqrt3}\Sigma_0\end{array}\right)\,,
~~
\Psi_{\Sigma\frac32}=
\left(\begin{array}{c}0\\-\frac{1}{\sqrt3}\Sigma_+\\\sqrt{\frac23}\Sigma_0\\\Sigma_-\,
\end{array}\right)\,.
\end{equation}
The PV and PC structures, $\vec{A}_{\Sigma_i}$ and
$\vec{B}_{\Sigma_i}$ contain the corresponding weak coupling constants
together with the isospin operators $\tau^a$ for $\frac12\to\frac12$
transitions and $T^a$ for $\frac12\to\frac32$ transitions.
The weak couplings $A=1.05$, $B=-7.15$, $A_{\Sigma\frac12}=-0.59$,
$A_{\Sigma\frac32}=2.00$,  
$B_{\Sigma\frac12}=-15.68$, and $B_{\Sigma\frac32}=-0.26$ \cite{DF96}
are fixed to reproduce the experimental data of the corresponding
hyperon decays, while the ones involving kaons,
$C_K^{PC}=-18.9$, $D_K^{PC}=6.63$, $C_{K}^{PV}=0.76$ and
$D_K^{PV}=2.09$, are derived using SU(3) symmetry.

\ignore{$\vec{A}_{\Sigma\frac12}=A_{\Sigma\frac12}\tau^a$, 
$\vec{A}_{\Sigma\frac32}=A_{\Sigma\frac32}T^a$,
$\vec{B}_{\Sigma\frac12}=B_{\Sigma\frac12}\tau^a$ and
$\vec{B}_{\Sigma\frac32}=B_{\Sigma\frac32}T^a$,
$C_K^{PC}=-18.9$, $D_K^{PC}=6.63$, $C_{K}^{PV}=0.76$ and $D_K^{PV}=2.09$.}
\begin{figure}[t]
\includegraphics[scale=0.4]{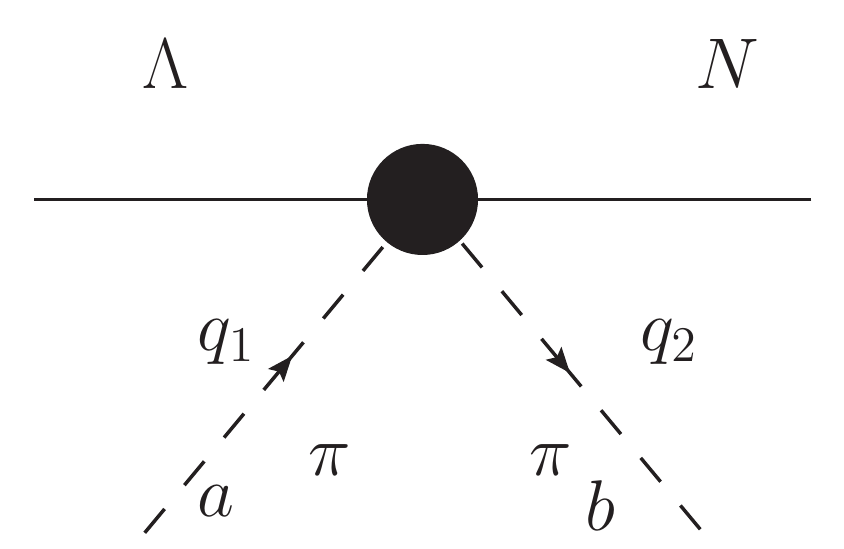}
\includegraphics[scale=0.4]{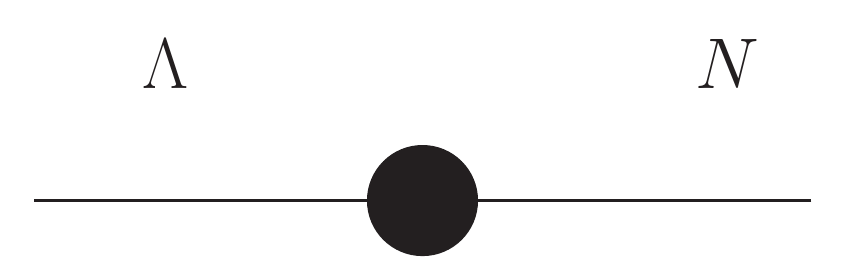}
\caption{Weak vertices corresponding to the $\Lambda N\pi\pi$, and 
$\Lambda N$ interactions. The solid black circle represents the weak 
vertex. The corresponding Lagrangians are given in Eq.~(\ref{eq:weakl2}).
\label{vf3}}      
\end{figure}
\ignore{The matrices $T^a$ connect isospin $\frac12$ and $\frac32$ states.
The couplings  For thep
$\Lambda$ and $\Sigma$ mesonic decays it is found, respectively,
$A=1.05$, $B=-7.15$ and $A_{\Sigma\frac12}=-0.59$, $A_{\Sigma\frac32}=2.00$,
$B_{\Sigma\frac12}=-15.68$, $B_{\Sigma\frac32}=-0.26$ \cite{DF96}. }

The other two weak vertices entering at the considered order
(Fig.~\ref{vf3}) are 
obtained from the weak SU(3) chiral Lagrangian,
\begin{align}
\el^w_{\Lambda N\pi\pi}=&
G_Fm_\pi^2\frac{h_{2\pi}}{f_\pi^2}(\vec{\pi}\cdot\vec{\pi})
\overline{\Psi}\Psi_\Lambda\,,  \label{eq:weakl2}\\
\el^w_{\Lambda N}     =&  iG_Fm_\pi^2 h_{\Lambda N}
\overline{\Psi}\Psi_\Lambda \nonumber\,,
\end{align}
with 
$
h_{2\pi}=(D+3F)/(8\sqrt6 G_Fm_\pi^2)=-10.13\text{ MeV}
$
and
$
h_{\Lambda N}=-(D+3F)/(\sqrt6 G_F m_\pi^2)=81.02\text{ MeV} \,.
$
$D$ and $F$ are the couplings parametrizing the weak chiral 
SU(3) Lagrangian, and can be fitted 
through the pole model to the experimentally known  hyperon decays. In
that case, one finds that when s-wave amplitudes are correctly 
reproduced, p-wave amplitude predictions disagree with the 
experiment~\cite{donoghue}. 

\begin{figure}[tb]
\begin{tabular}{cc}
\includegraphics[scale=0.3]{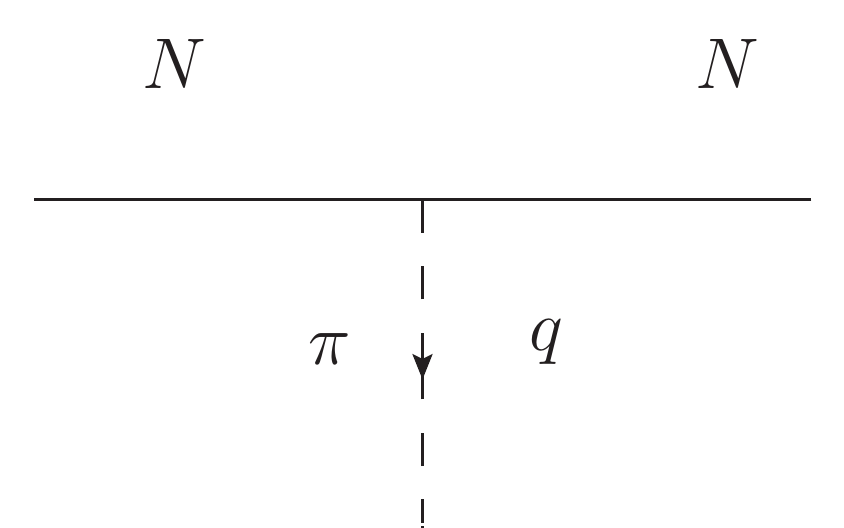}&
\includegraphics[scale=0.3]{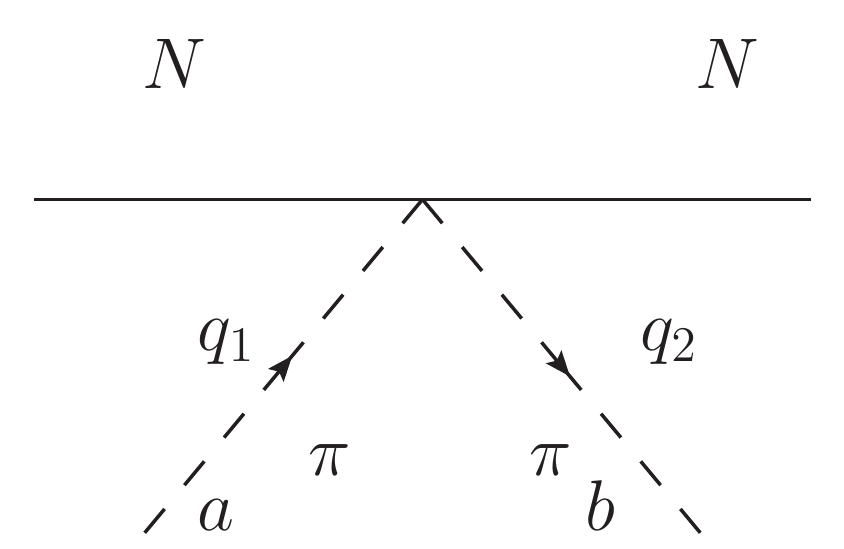}
\\\\\includegraphics[scale=0.3]{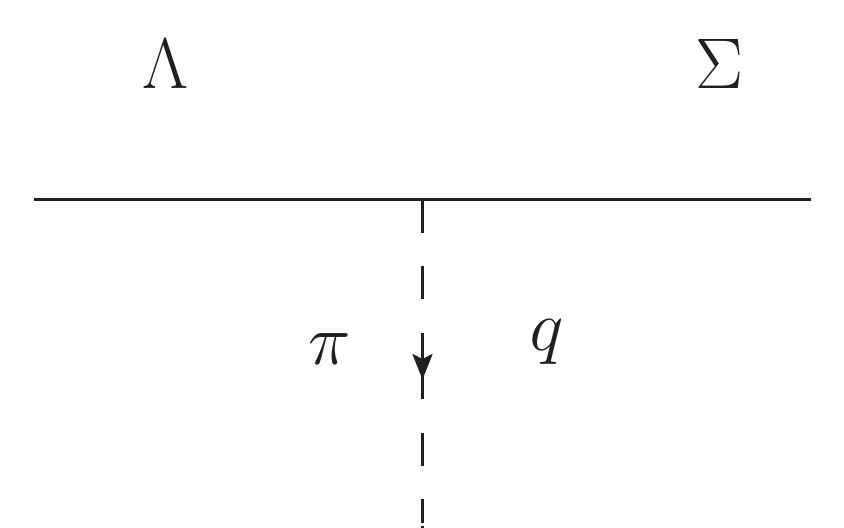}&
\includegraphics[scale=0.3]{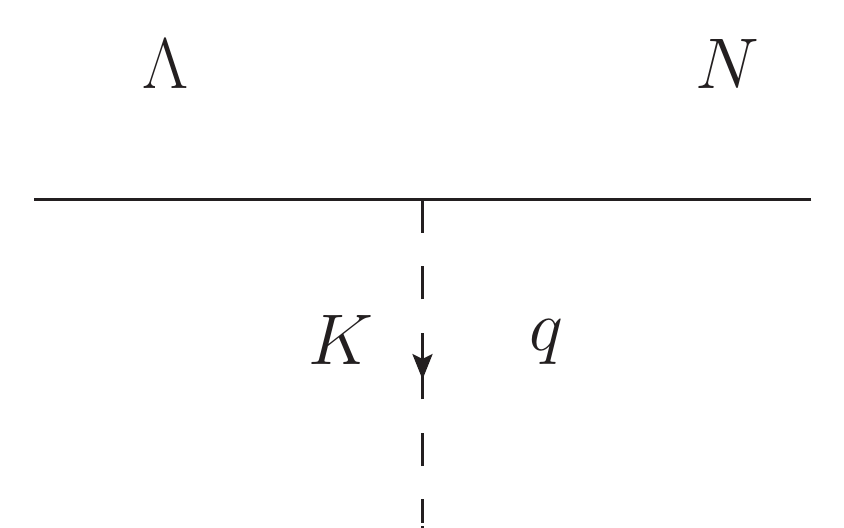}
\end{tabular}
\caption{Strong vertices for the $NN\pi$, $NN\pi\pi$, 
$\Lambda \Sigma\pi$ and $\Lambda N K$ which arise from the 
Lagrangians in Eq.~(\ref{eq:str}).
\label{vf1}}      
\end{figure}
The strong vertices for the interaction between our baryonic and
mesonic degrees of freedom are obtained from the strong SU(3) chiral Lagrangian~\cite{donoghue}, 
\begin{align}
\el^s_{NN\pi}=&-\frac{g_A}{2f_\pi}\overline{\Psi}\gamma^\mu\gamma_5
\ta\Psi\cdot\partial_\mu\vec{\pi}\,,\nonumber\\
\el^s_{NN\pi\pi} =&-\frac{1}{4f_\pi^2}\overline{\Psi}\gamma^\mu
\ta\cdot(\vec{\pi}\times\partial_\mu\vec{\pi})\Psi\,,\nonumber\\
\el^s_{\Lambda\Sigma\pi}=&-\frac{D_s}{\sqrt{3}}\,\overline{\Psi}_\Lambda\gamma^\mu\gamma_5
\Psi_\Sigma\cdot\partial_\mu\vec{\pi} \,,\label{eq:str}
\\\nonumber
\el^{s}_{\Lambda N K} =&
 \, \frac{D_s+3F_s}{2\sqrt3f_\pi} \, \overline{\Psi}_{N}
\gamma^\mu\gamma_5 \,\partial_\mu\phi_{K}
 \Psi_\Lambda  \,,
\end{align}
where we have taken the convention which gives us
$\Psi_\Sigma\cdot\vec{\pi}=\Psi_{\Sigma_+}\pi_-+\Psi_{\Sigma_-}\pi_++\Psi_{\Sigma_0}\pi_0$, and we consider, 
$g_A=1.290$, $f_\pi=92.4$ MeV, $D_s=0.822$, and $F_s=0.468$.
These strong coupling constants are
taken from $NN$ interaction  models such as the
J\"ulich~\cite{JB} or Nijmegen~\cite{nij99} potentials.  
The four interaction vertices corresponding to these Lagrangians 
are depicted in Fig.~\ref{vf1}.

Once the interaction Lagrangians involving the relevant 
degrees of freedom have been presented, we need to define the power
counting scheme which allows us to organize the different
contributions to the full amplitude.   
 
\subsection{Power counting scheme}
\label{ss:cs}

The amplitude for the $\Lambda N\to NN$ transition is built as 
the sum of a medium and long-range one meson exchanges (i.e. $\pi$ 
and $K$), the contribution from the two-pion exchanges, and the 
contribution of the contact interactions up to ${\cal O} (q^2/M^2)$ 
as described below. The order at which the different terms enter 
in the perturbative expansion of the amplitudes is given by the 
so-called Weinberg power counting scheme~\cite{W9091}. 

In our calculations we will employ the heavy baryon formalism~\cite{jm}. 
This technique introduces a perturbative expansion in the baryon 
masses appearing in the Lagrangians, so that this new large scale 
does not disrupt the well-defined Weinberg power counting. 
It is worth noting that, in the heavy baryon formalism, terms 
of the type $\overline{\Psi}_B\gamma^5\Psi_B$ 
are subleading in front of terms like $\overline{\Psi}_B\Psi_B$, since 
they show up at one order higher in the heavy baryon expansion. 
In our calculation, we choose to keep both terms in our 
Lagrangians of Eqs.~(\ref{eq:weakl}) because the experimental values 
for the couplings $B_\Lambda$ and $B_\Sigma$ are  much larger than 
$A_\Lambda$ and $A_\Sigma$. For example, $A_\Lambda=1.05$ and 
$B_\Lambda=-7.15$~\cite{donoghue}.

Our calculation is characterized by the presence of different octet
baryons in the relevant Feynman diagrams, contributing in both, the
spinors and propagators.
\ignore{A relevant feature of the interaction we are considering is that 
different octet baryons appear in the corresponding Feynman diagrams: 
$N$, $\Lambda$ and $\Sigma$. 
Specifically, these masses appear in the spinors and propagators.}
The spinors for the incoming $\Lambda$ and $N$ with masses $M_\Lambda$
and $M_N$, energies $E_p^{\Lambda}$ and $E_p^N$,  and momenta $\vp$ and $-\vp$ are
\begin{align}\nonumber
u_1(E_p^\Lambda,\vp\,)=
\sqrt{\frac{E_p^\Lambda+M_\Lambda}{2M_\Lambda}}
\left(\begin{array}{c}
1\\\nonumber
\frac{\vsu\cdot\vp}{E_p^\Lambda+M_\Lambda}
\end{array}\right)\,,
\\\\\nonumber
u_2(E_p^N,-\vp\,)=
\sqrt{\frac{E_p^N+M_N}{2M_N}}
\left(\begin{array}{c}
1\\
-\frac{\vsd\cdot\vp}{E_p^N+M_N}
\end{array}\right)\,,
\end{align}
and for the outgoing nucleons with momenta $\vpp$ and $-\vpp$, and
energy $E'\equiv\frac12\left(E_p^\Lambda+E_p^N\right)$,
\begin{align}\nonumber
\bar{u}_1(E',\vpp)=
\sqrt{\frac{E'+M_N}{2M_N}}
\left(\begin{array}{cc}
1&
-\frac{\vsu\cdot\vpp}{E'+M_N}
\end{array}\right)\,,
\\\\\nonumber
\bar{u}_2(E',-\vpp)=
\sqrt{\frac{E'+M_N}{2M_N}}
\left(\begin{array}{cc}
1&
ñ\frac{\vsd\cdot\vpp}{E'+M_N}
\end{array}\right)\,.
\end{align}
The relativistic propagator of a baryon with mass $M_B$ and momentum
$p$ reads 
\begin{equation}
\frac{i}{\cancel{p}-M_B+i\epsilon}
=\frac{i(\cancel{p}+M_B)}{p^2-M_B^2+i\epsilon} \,.
\end{equation}
Making the heavy baryon expansion with these spinors and propagators
introduces mass differences ($M_\Lambda-M_N$, $M_\Sigma-M_\Lambda$) in
the baryonic propagators. A reasonable approach would be to consider
these mass differences  of order ${\cal O}\left(\vq^{\,2}/\Lambda^2\right)$ 
($M_B=\om+{\cal O}\left(\vq^{\,2}/\Lambda^2\right)$), and thus they would 
not enter in the loop diagrams. We have chosen to leave the physical
masses in both the initial and final spinors and also in the 
intermediate propagators; i.e. we consider the mass differences as 
another scale in the heavy baryon expansion. The corresponding SU(3) 
symmetric limit is also given at the end of section~\ref{ss:tped}, and
can be easily obtained from our expressions by  
setting the mass differences, which we explicitly retain, to zero.

The procedure we follow to compute the different Feynman diagrams 
entering the transition amplitude is the following: first we 
write down the relativistic expressions for each diagram, and 
then afterwards, we perform the heavy baryon expansion.

In the next sections we will describe the LO and NLO 
contributions to the process $\Lambda N\to NN$, following the scheme 
presented here. The explicit expressions and details of the calculations
are given in the Appendices.

\section{Leading order Contributions}
\label{ss:loc}

\begin{figure}[th]
\centering
\begin{tabular}{ccc}
\includegraphics[scale=0.2]{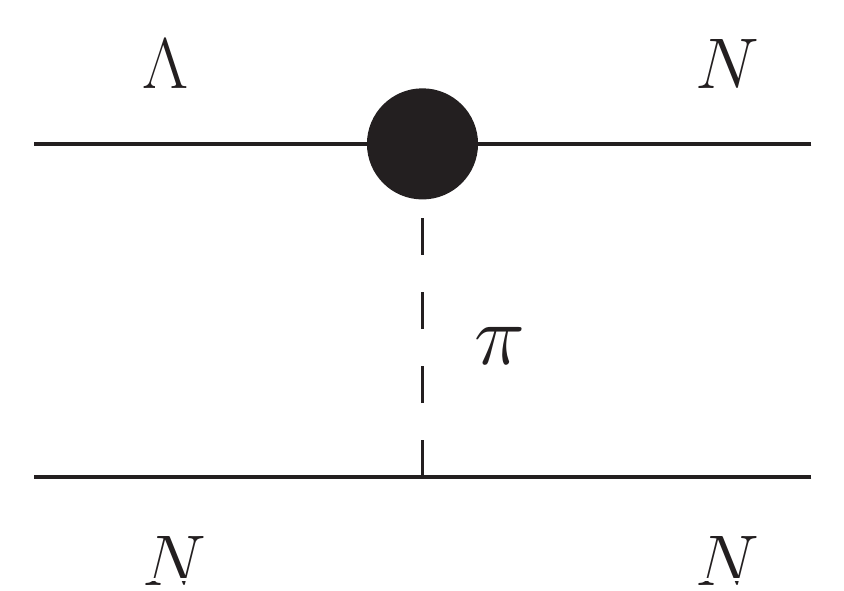}
&
\includegraphics[scale=0.2]{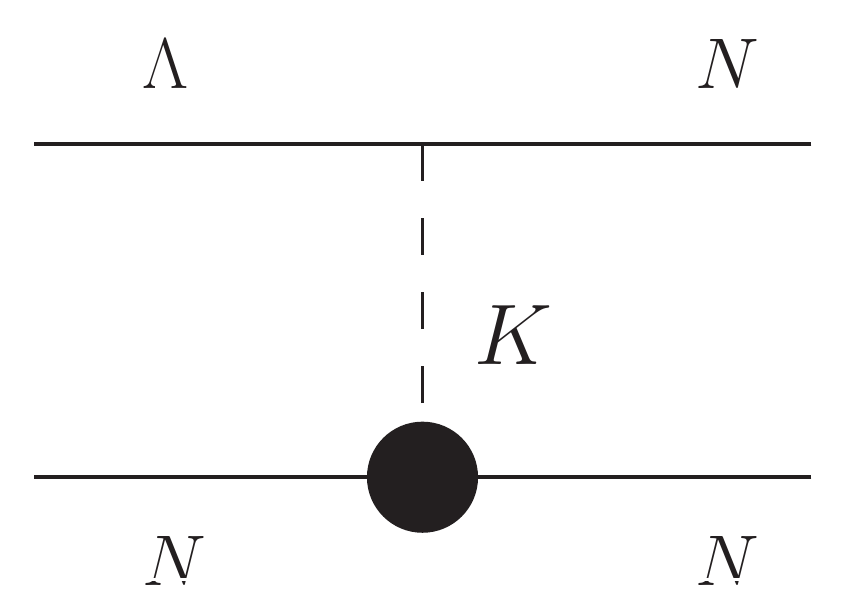}
&
\includegraphics[scale=0.2]{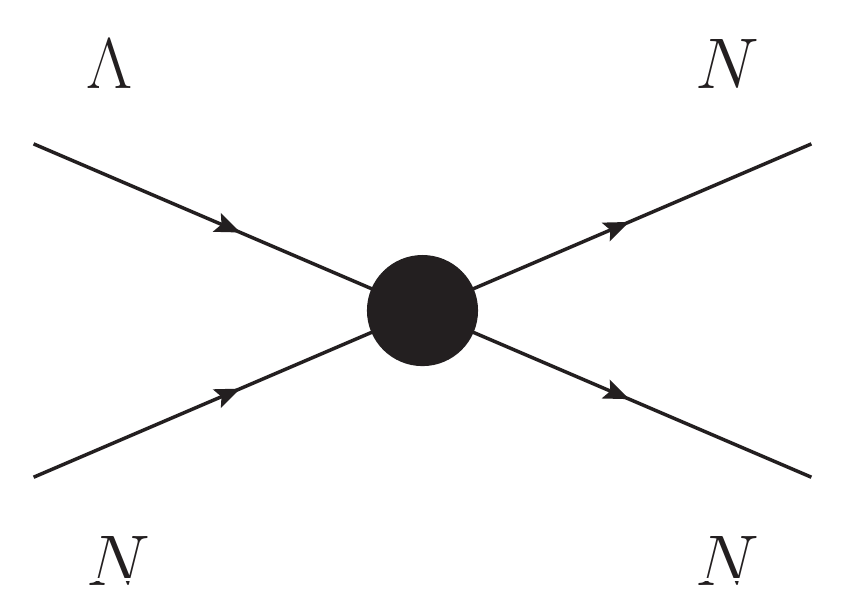}
\end{tabular}
\caption{One-pion and one-kaon exchange
contributions to the transition.\label{fig:loc}}
\end{figure}

For completeness, we rewrite here the LO EFT already presented 
in Ref.~\cite{PPJ11}, and then build the NLO contributions in the next 
section. 

At tree level, the transition potential $\Lambda N\to NN$ involves 
the LO contact terms, and $\pi$ and $K$ exchanges, 
as depicted in Fig.~\ref{fig:loc}. First, the contact interaction 
can be written as the most general Lorentz invariant potential with no
derivatives. 
The four-fermion (4P) interaction in momentum space at leading order 
(in units of $G_F$) is
\begin{eqnarray}
V_{4P} ({\vec q} \, ) &=&
C_0^0 + C_0^1 \; {\vec \sigma}_1
{\vec \sigma}_2  \,,\label{eq:vlo}
\end{eqnarray}
where $C_0^0$ and $C_0^1$ are low energy constants which need to 
be fitted by direct comparison to experimental data. In 
Ref.~\cite{PPJ11} we presented several sets of values which were 
to a large extent compatible with the scarce data on hypernuclear 
decay. 

The potentials for the one pion and one kaon exchanges, as functions 
of transferred momentum $\vq\equiv\vpp-\vp$, read,
respectively~\cite{PRB97}
\begin{align}
{V_{\pi}} ({\vec q}\,) =&
\nonumber
-\frac{G_F m_\pi^2g_{NN\pi}}{2 M_N}
\left(
A_\pi - \frac{B_\pi}{2 \om}{\vec \sigma}_1 \, {\vec q} \,
\right)
\frac{{\vec \sigma}_2 \, {\vec q}\,}{-q_0^2+{\vec q}^{\; 2}+m_\pi^2} \,
\\&\times \vtu\cdot\vtd
{\rm ,}
\label{eq:pion}\\
{V_{K}} ({\vec q}\,) =&
-\frac{G_F m_\pi^2g_{\Lambda N K}}{2\om}
\left(
\hat{A} - \frac{\hat{B}}{2 M_N}{\vec \sigma}_1 \, {\vec q} \,
\right)
\nonumber\\&\times\frac{{\vec \sigma}_2 \, {\vec q}\,} {-q_0^2+{\vec q}^{\; 2}+m_K^2} \,
{\rm ,}
\label{eq:kaon}
\end{align}
where $m_\pi=138$ MeV and $m_K=495$ MeV, $q_0\equiv
\frac12(M_\Lambda-M_N)$, 
$g_{NN\pi}\equiv \frac{g_A M_N}{f_\pi}$, $g_{\Lambda N 
  K}\equiv-\frac{D_s+3F_s}{2\sqrt3f_\pi}$,
$\om\equiv\frac12(M_N+M_\Lambda)$, and
\begin{align*}
{\hat A} &=\left( \frac{ C^{PV}_{K}}{2} +
D^{PV}_{K} + \frac{
C^{PV}_{K}}{2} {\vec \tau}_1 {\vec \tau}_2
\,\right),
\\
{\hat B}&= \left( \frac{
C^{PC}_{K}}{2} +
D^{PC}_{K} + \frac{
C^{PC}_{K}}{2}
{\vec \tau}_1 \, {\vec \tau}_2 \right) \,.
\end{align*}
\section{Next-to-leading order contributions}
\label{ss:nloc}
\begin{table}[b]
\centering
\begin{tabular}{ccc}
Order & Parity & Structures
\\\hline
0 & PC & $1$, $\vsu\cdot\vsd$
\\
\multirow{2}{*}{1} & \multirow{2}{*}{PV} & 
$\vsu\cdot\vq$, $\vsu\cdot\vp$, $\vsd\cdot\vq$, 
\\& & $\vsd\cdot\vp$, $(\vsu\times\vsd)\cdot\vq$,
$(\vsu\times\vsd)\cdot\vp$, 
\\\noalign{\smallskip}
\multirow{2}{*}{2} & \multirow{2}{*}{PC} & 
$\vq^2$, $\vp^2$, $(\vsu\cdot\vsd)\vq^2$, 
$(\vsu\cdot\vsd)\vp^2$, 
$(\vsu\cdot\vq)(\vsd\cdot\vq)$,\\&&
$(\vsu\cdot\vp)(\vsd\cdot\vp)$, $(\vsu+\vsd)\cdot(\vq\times\vp)$
\\\noalign{\smallskip}
\multirow{2}{*}{2} & \multirow{2}{*}{PV} & 
$\vq\cdot\vp$, $(\vsu\cdot\vsd)\vq\cdot\vp$,
$(\vsu\cdot\vq)(\vsd\cdot\vp)$,
\\&&$(\vsu\cdot\vp)(\vsd\cdot\vq)$, $(\vsu-\vsd)\cdot(\vq\times\vp)$
\end{tabular}
\caption{All possible PC and PV NLO operational structures 
connecting the initial and final spin and angular momentum 
states. There are a total of 18. 
 \label{tab:contacts}}      
\end{table}

The NLO contribution to the weak decay process, $\Lambda N\to NN$, 
includes contact interactions with one and two derivative operators,
caramel diagrams and two-pion-exchange diagrams. 

\subsection{NLO contact potential}
\label{sec:nlocontact}
In principle the NLO contact potential should include, in the center 
of mass, structures involving both the initial ($\vp\,$) and final 
($\vpp$) momenta, or independent linear combinations, e.g. 
$\vq\equiv\vpp-\vp$ and $\vp$. Table~\ref{tab:contacts} lists all 
these possible structures. At NLO there are 18 LECs ---6 PV ones at order
${\cal O}\left(q/M\right)$, 7 PC ones at order ${\cal
  O}\left(q^2/M^2\right)$ and 5 PV ones at order ${\cal
  O}\left(q^2/M^2\right)$---, which must be 
fitted to experiment. This is not feasible with current experimental 
data on hypernuclear decay. A reasonable way to reduce the number 
of LECs and render the fitting procedure more tractable is to note 
that the pionless weak decay mechanism we are interested in 
takes place inside a bound hypernucleus. Thus, one can consider 
that in the $\Lambda N\rightarrow NN$ transition potential the 
initial baryons have a fairly small momentum. Moreover, the final 
nucleons gain an extra momentum from the surplus mass of the 
$\Lambda$ ($M_\Lambda-M_N=116$ MeV), which in most cases allow to 
consider, $\vpp\gg\vp$. In this case, one may approximate 
$\vq\simeq\vpp$ and $\vp=0$. Within this approximation, the NLO part 
of the contact potential reads (in units of $G_F$):
\begin{align}
V_{4P} ({\vec q} \, ) &=
 C_1^0 \; \displaystyle\frac{{\vec \sigma}_1{\vec q}}{2 M_N}
\label{eq:vnlo}
+ \,C_1^1 \; \displaystyle\frac{{\vec \sigma}_2{\vec q}}{2 M_N}
+ {\im} \, C_1^2 \; \displaystyle
\frac{({\vec \sigma}_1 \times {\vec\sigma}_2)\;{\vec q}}{2 M_N}
\\
&+ C_2^0 \; \displaystyle\frac{{\vec \sigma}_1{\vec q}
  \;{\vec\sigma}_2{\vec q}}{4 M_N^2}+
C_2^1 \; \displaystyle\frac{{\vec \sigma}_1
{\vec \sigma}_2 \; {\vec q}^{\; 2}}
{4 M_N^2}+
C_2^2 \; \displaystyle\frac{{\vec q}^{\,2}}{4 M_N^2} \,.
\nonumber
\end{align}
\begin{figure}[t]
\centering

\begin{tabular}{cccc}
\includegraphics[scale=0.2]{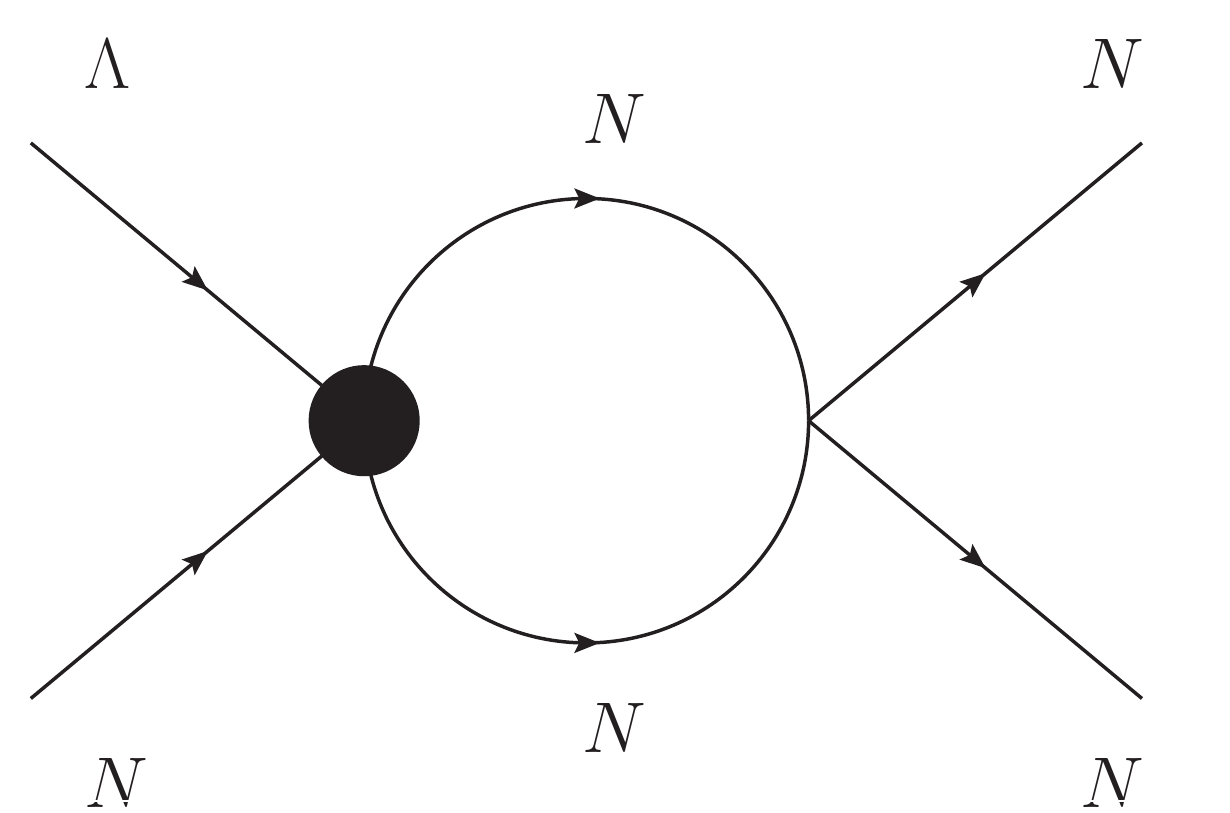}
&
\includegraphics[scale=0.2]{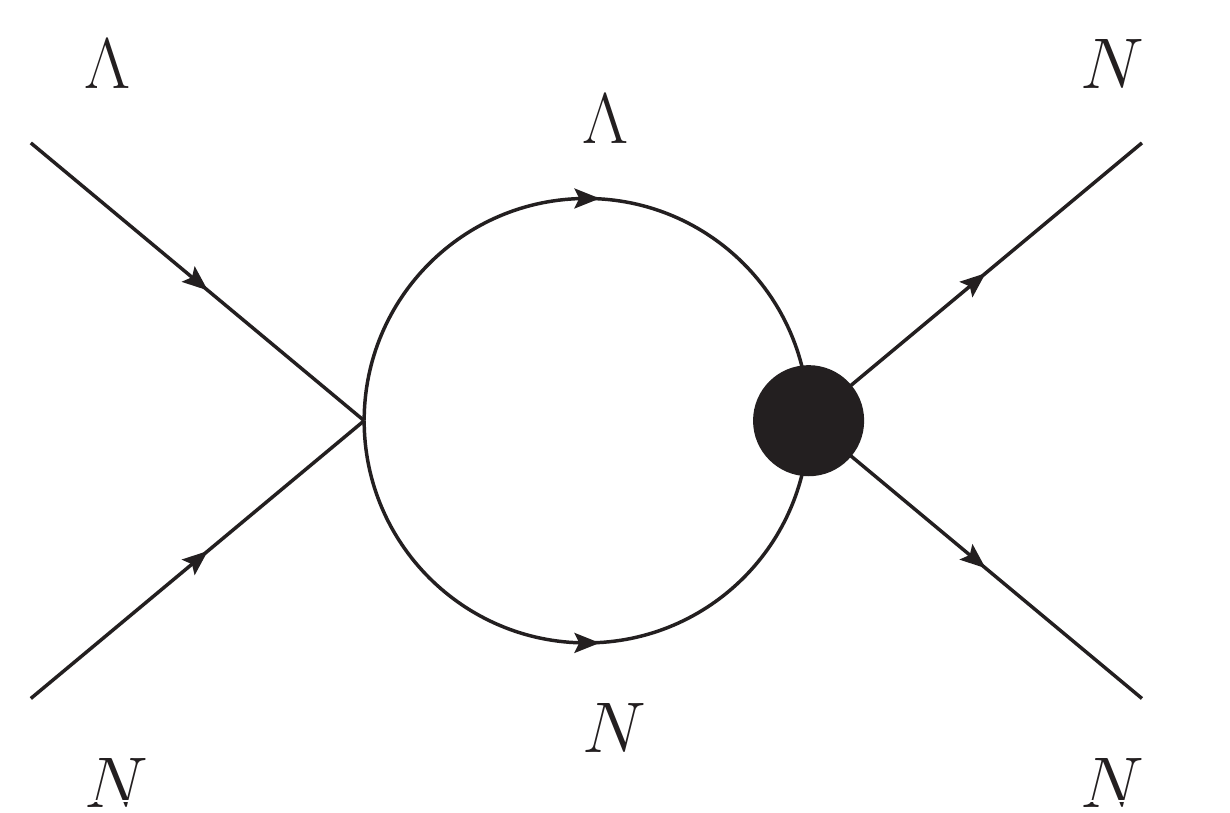}
&
\includegraphics[scale=0.2]{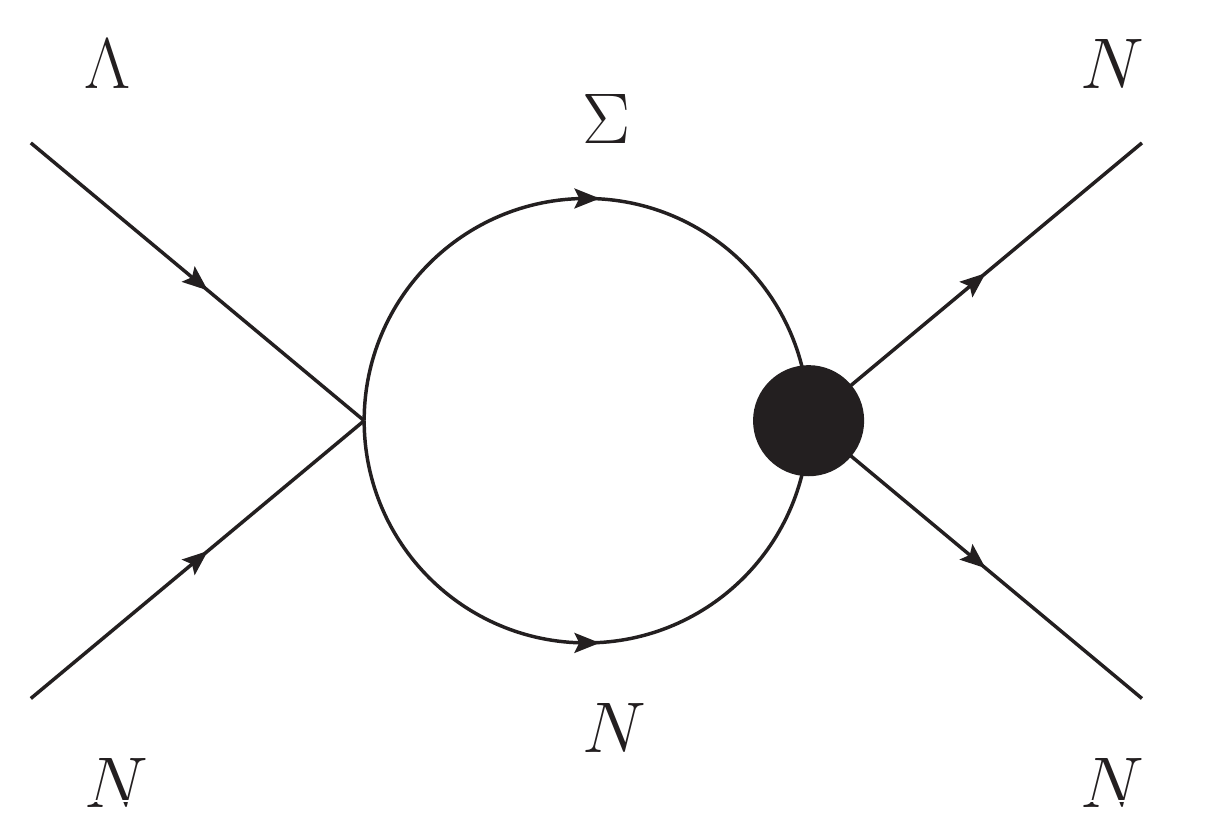}
\\(a)&(b)&(c)
\end{tabular}
\caption{
Caramel diagrams contributing to the process at NLO.
The solid circle represents the weak vertex.
\label{fig:caramels}}
\end{figure}

Using strong and weak LO contact interactions and two baryonic
propagators one can also build three diagrams that enter at NLO.
These caramel-like diagrams are shown in Fig.~\ref{fig:caramels}.
They only differ in the position of the strong and weak vertices and in
the mass of upper-leg baryonic propagator. 
In order to write a general expression for the three caramel diagrams
we label the mass of the upper-leg propagating baryon $M_\alpha$
($M_a=M_N$, $M_b=M_\Lambda$ and $M_c=M_\Sigma$) and the corresponding
strong and weak contact vertices 
$C_{S(s)}^\alpha+C_{T(s)}^\alpha\vsu\cdot\vsd$ 
and $C_{S(w)}^\alpha+C_{T(w)}^\alpha\vsu\cdot\vsd$, where
$\alpha=a,b,c$ corresponds to the labels of
Fig.~\ref{fig:caramels}. It is also convenient to define
$M_\alpha=M_N+\Delta_\alpha$. In the heavy baryon formalism these
diagrams only contribute with an imaginary part of the form
\begin{align}
V_\alpha&=i\frac{G_Fm_\pi^2}{16\pi M_N}
(C_{S(s)}^{\alpha}+C_{T(s)}^{\alpha}\vsu\cdot\vsd)\,
\\&\times\nonumber
(C_{S(w)}^{\alpha}+C_{T(w)}^{\alpha}\vsu\cdot\vsd)\,
\\&\times\nonumber
\sqrt{(\Delta_b-\Delta_\alpha)(\frac12(\Delta_b+\Delta_\alpha)+M_N)+\vp^2}\,.
\end{align}
Few more details are given in App.~\ref{sec:caramels}.

\begin{figure}[t]
\vspace{10pt}
\centering
\includegraphics[scale=0.2]{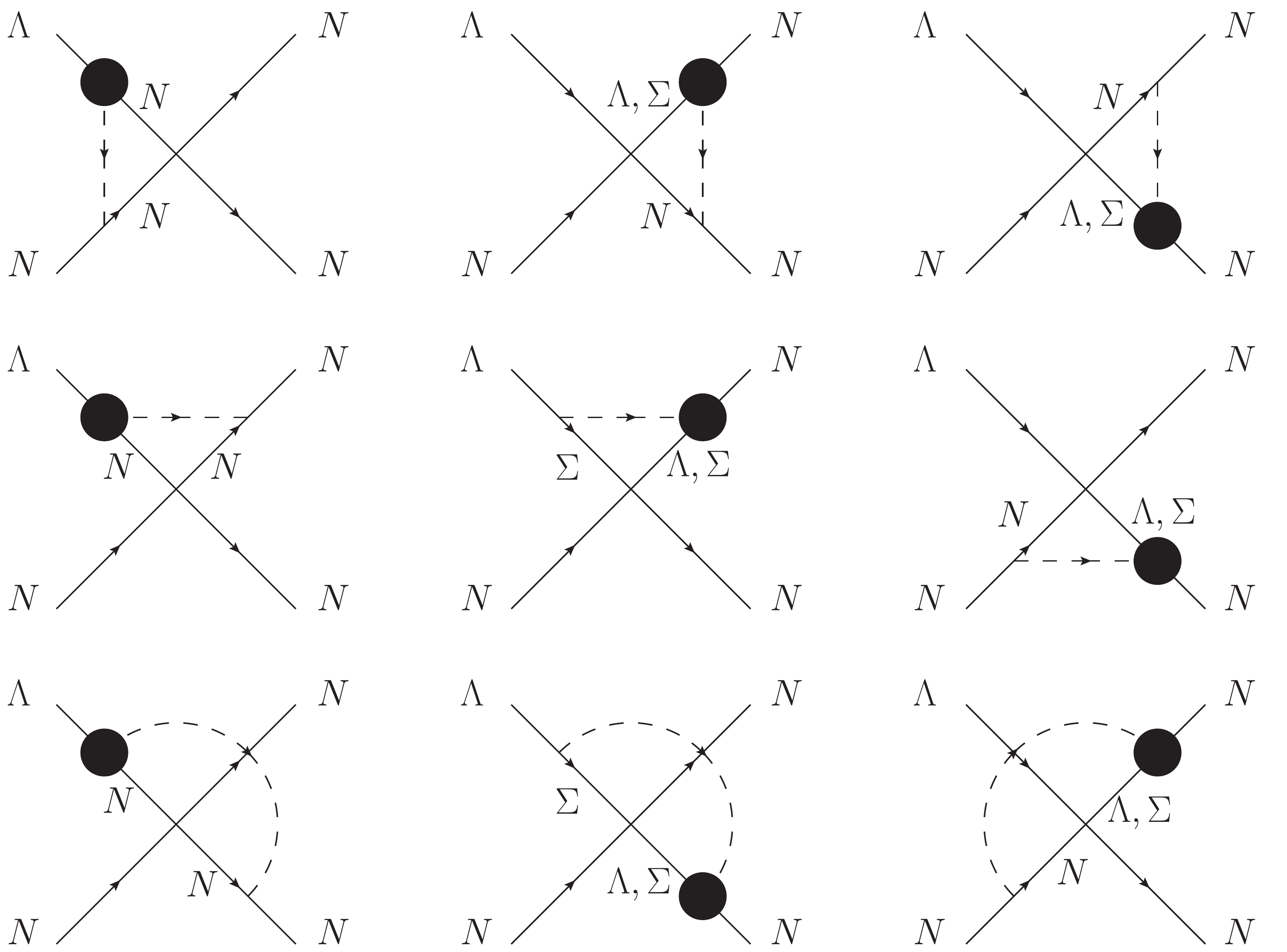}
\\
\includegraphics[scale=0.2]{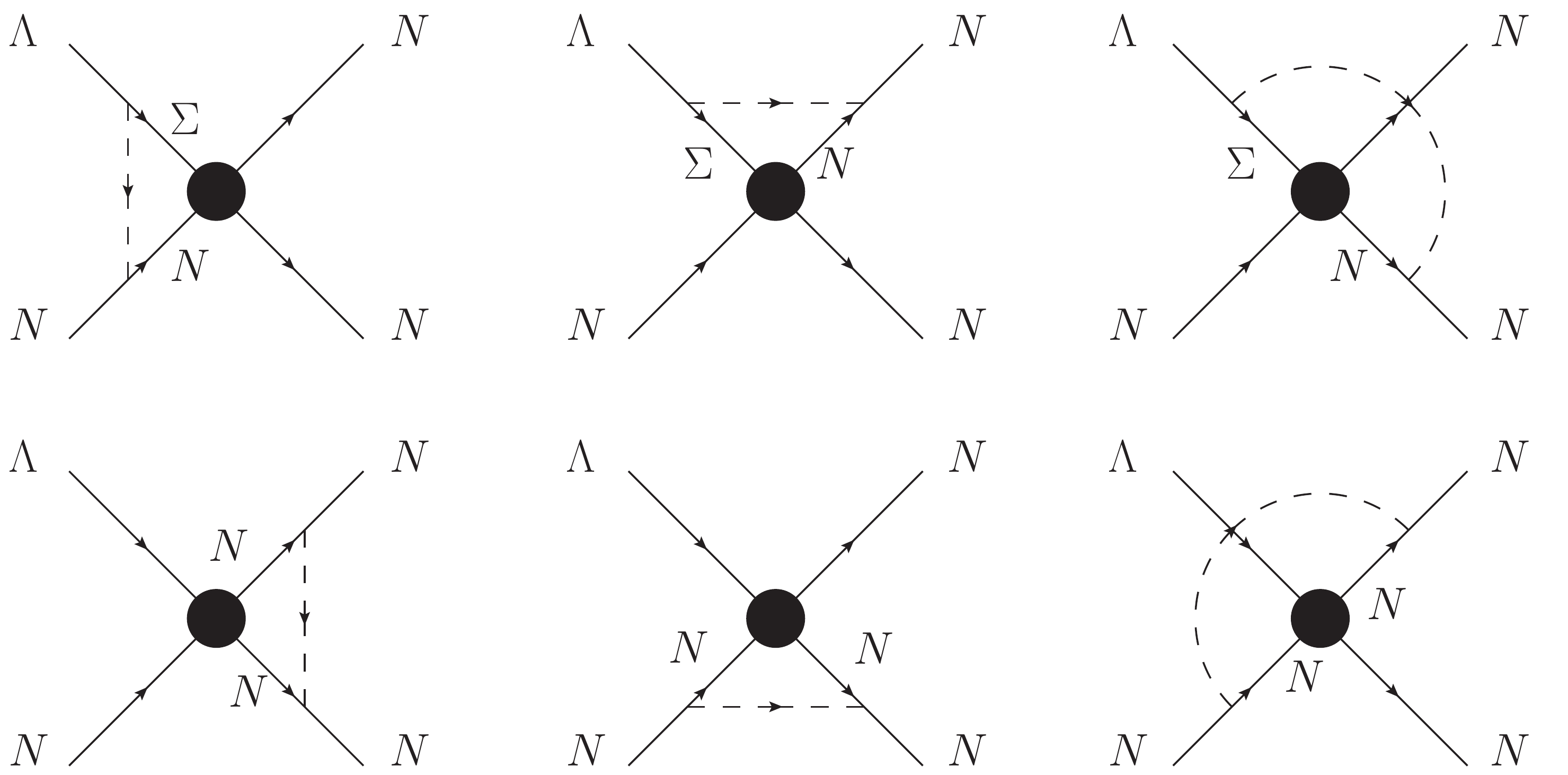}
\caption{Corrections to the LO contact interactions. The 
contributions of all these diagrams can be accounted for 
by an adequate shift of the coefficients of the LO contact terms.}
\label{fig:contact.corrections}
\end{figure}

One pion corrections to the LO contact interactions, shown 
in Fig.~\ref{fig:contact.corrections}, also enter at NLO. The 
net contribution of these diagrams is to shift the coefficients 
of the LO contact terms with functions dependent on $m_\pi$, $M_\Lambda-M_N$ 
and $M_\Sigma-M_N$. 

%\end{tabular}
%\end{table}

\subsection{Two-pion-exchange diagrams}
\label{ss:tped}

\begin{figure}[th]
\centering
\begin{tabular}{c}
\includegraphics[scale=0.3]{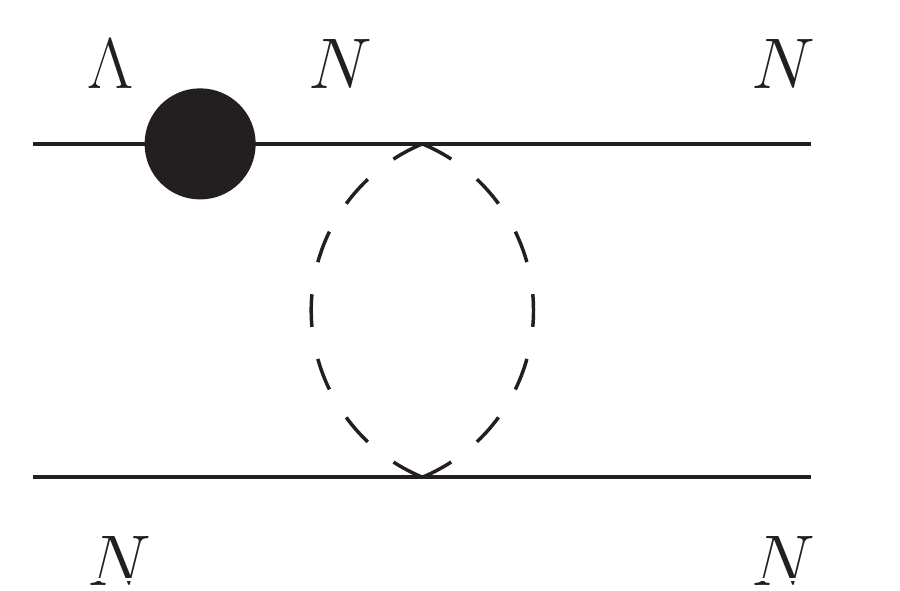}
\\(a)
\end{tabular}
\caption{The ball diagram contributing to the process at NLO. 
The solid circle represents the weak vertex.\label{fig:ball}}
\end{figure}

\begin{figure}[t]
\centering
\begin{tabular}{cccc}
\includegraphics[scale=0.2]{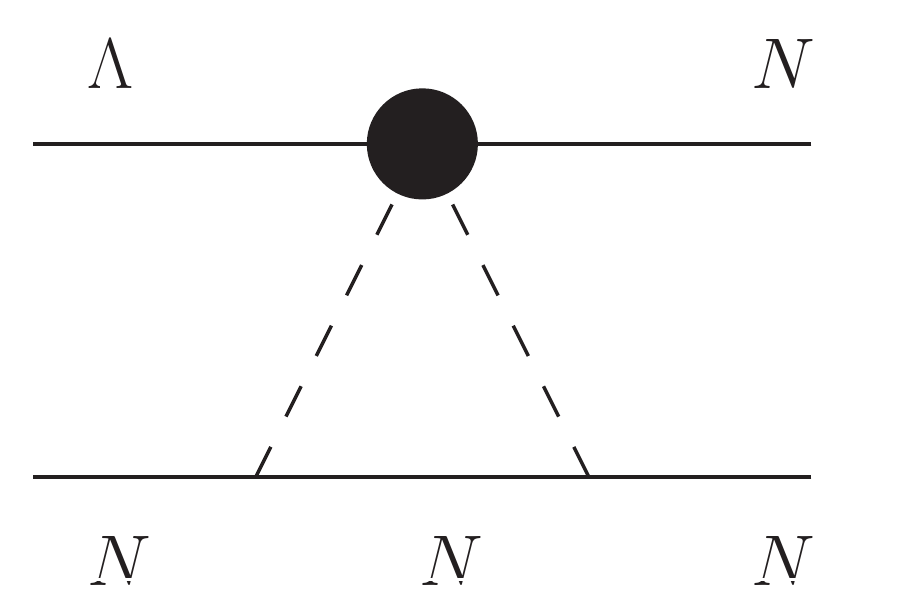}
&
\includegraphics[scale=0.2]{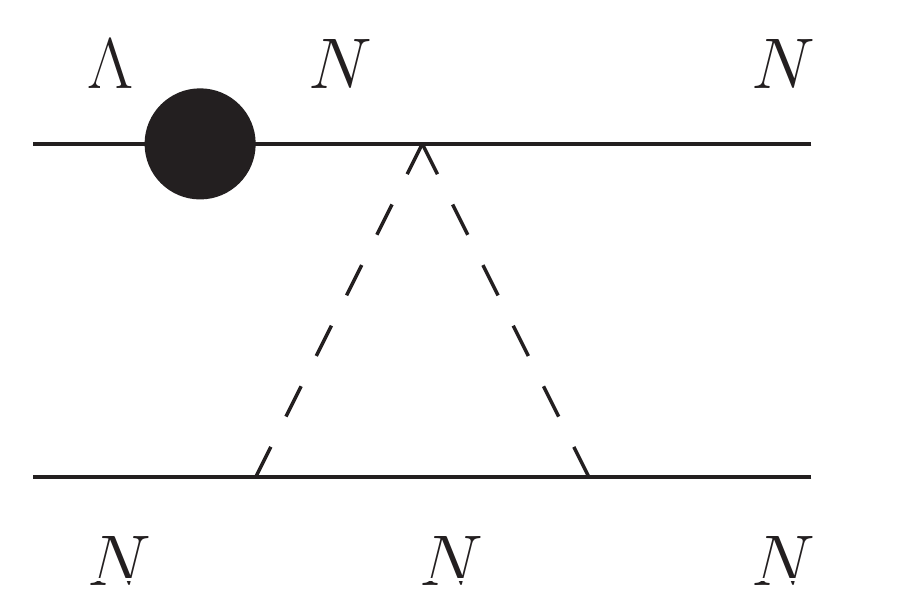}
&
\includegraphics[scale=0.2]{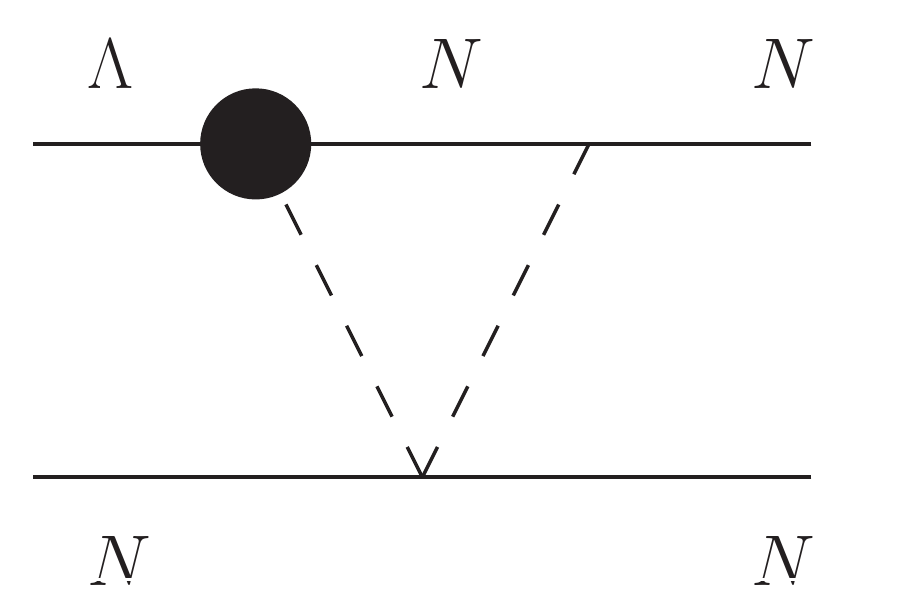}
&
\includegraphics[scale=0.2]{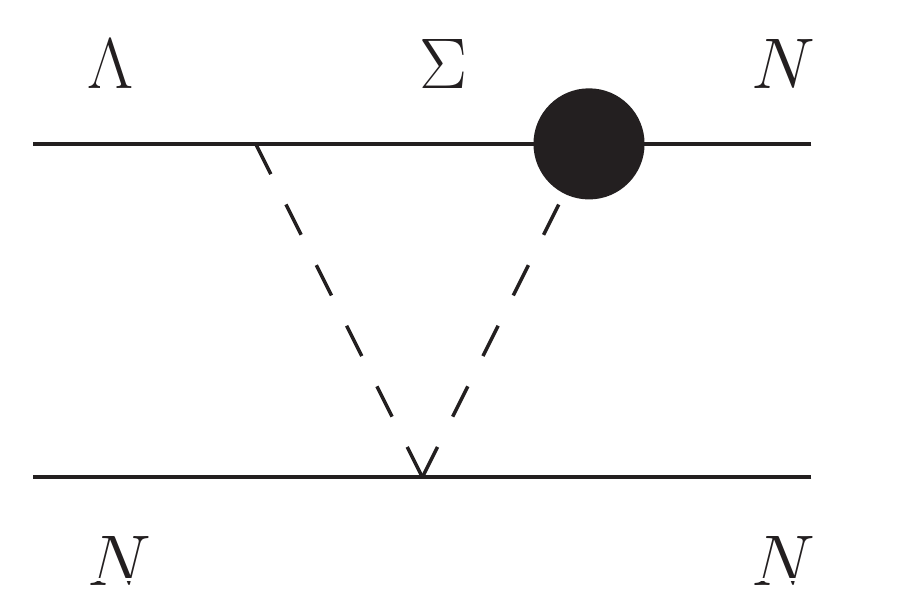}
\\(b)&(c)&(d)&(e)
\end{tabular}
\caption{Triangle diagrams which contribute to the 
process at NLO. The solid circle represents the weak interaction 
vertex.\label{fig:triangle}}
\end{figure}

\begin{figure}[th]
\centering
\begin{tabular}{cccc}
\includegraphics[scale=0.2]{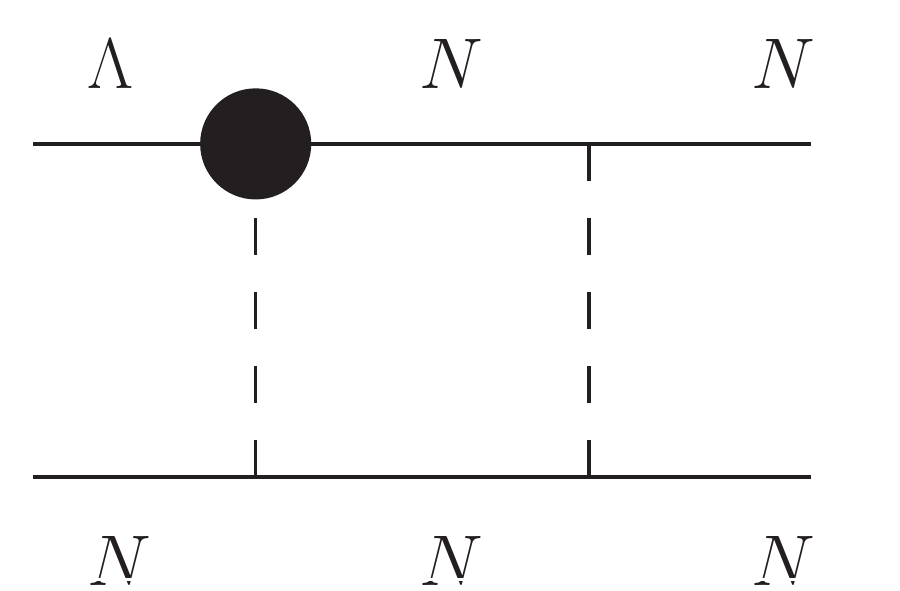}
&
\includegraphics[scale=0.2]{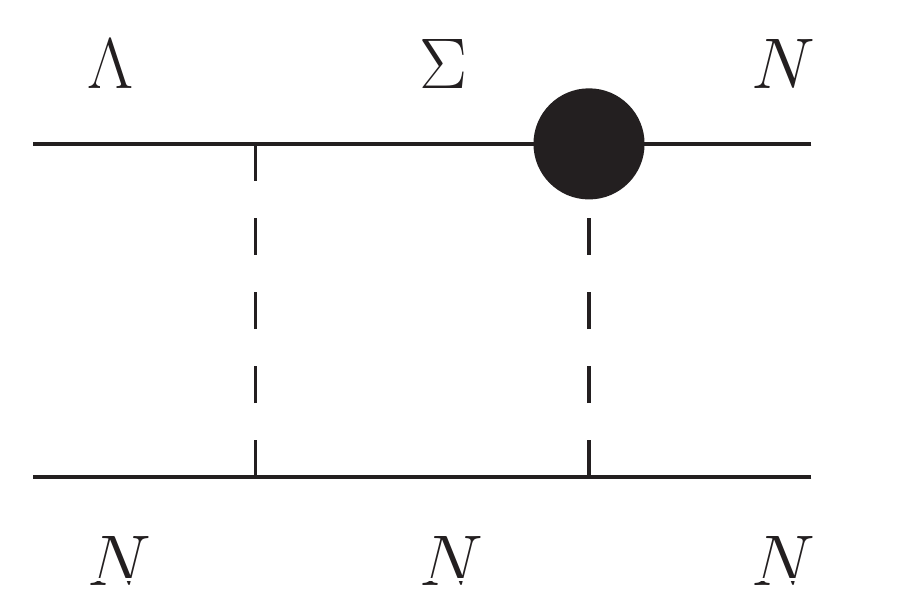}
&
\includegraphics[scale=0.2]{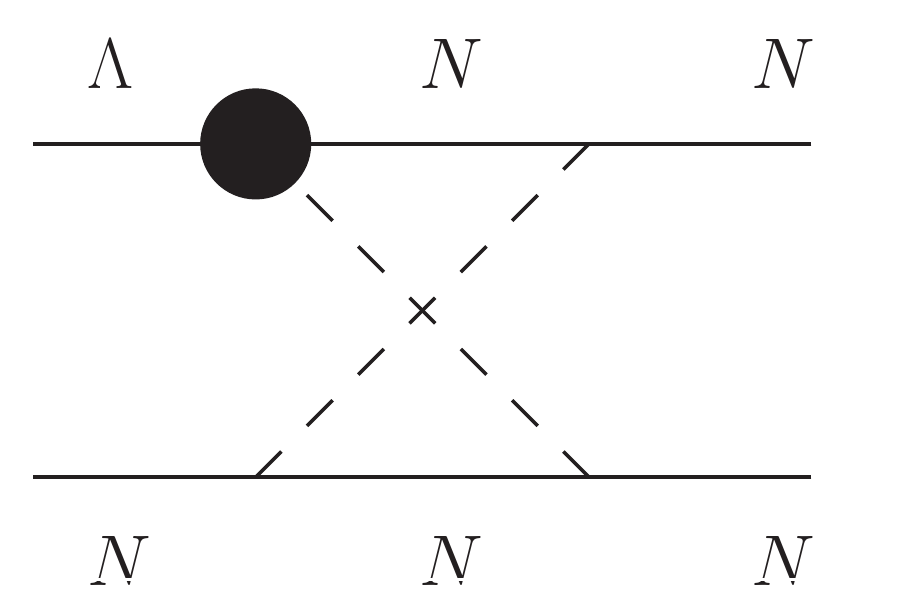}
&
\includegraphics[scale=0.2]{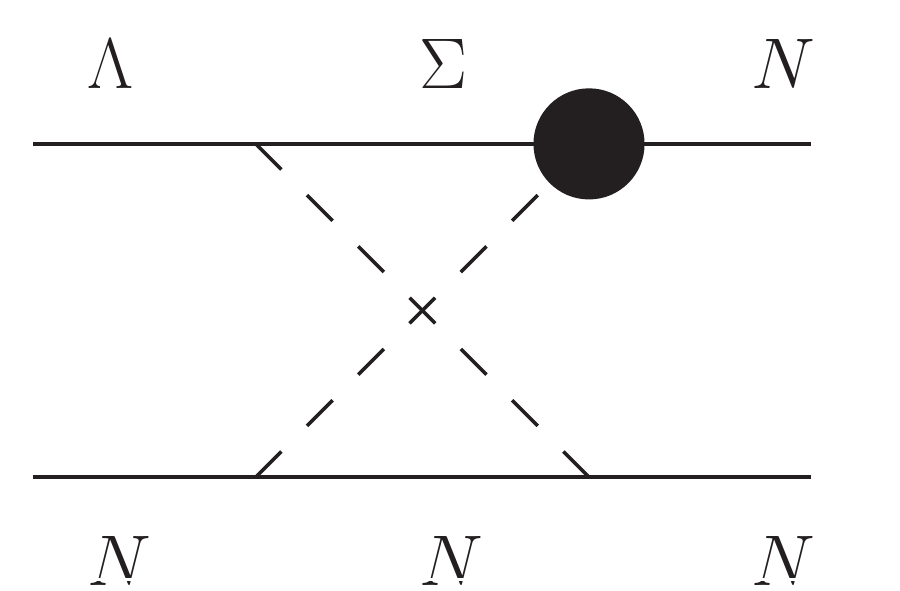}
\\(f)&(g)&(h)&(i)
\end{tabular}
\caption{Box diagrams which contribute to the process at 
NLO. The solid circle stands for the weak interaction 
vertex.\label{fig:box}}
\end{figure}

The two-pion-exchange contributions are organized according 
to the different topologies --- balls, triangles, and boxes---, 
such that most of the integration techniques are shared by 
each class of diagrams. There are two types of ball diagrams, 
of which only one gives a non-zero contribution, depicted in
Fig.~\ref{fig:ball}. In addition, there are four triangle  
diagrams, shown in Fig.~\ref{fig:triangle}, and two box and 
crossed box diagrams, shown in Fig.~\ref{fig:box}. The 
topologies contain, respectively, zero, one, and two baryonic 
propagators, which may correspond to $N$ or $\Sigma$ baryons. All the
diagrams contain two relativistic propagators from the 2$-\pi$
exchange.


The technical details of the evaluation of the Feynman diagrams 
for the ball, triangle and box diagrams are given in the
App.~\ref{sec:balls},~\ref{sec:triangles}, and \ref{sec:boxs}
respectively.
The main technique used is to introduce a number of master 
integrals, which appear in different diagrams, and which reduce 
 the mathematical complexity of the problem
(see App.~\ref{sec:mi}). Once they are defined, we derive a number of 
relations between the master integrals, which can in most cases
be easily checked. Full details are provided to ensure the 
future use of these expressions.

Using the labels defined in Figs.~\ref{fig:ball},~\ref{fig:triangle} 
and~\ref{fig:box} we organize the contributions of all the 
$2-\pi$ exchange diagrams in Eq.~(\ref{eq:tots}). The 
corresponding coefficients in terms of the coupling 
constants, baryon and meson masses, and momenta can be 
read off from the full expressions given in the 
Appendices~\ref{sec:balls},~\ref{sec:triangles} and~\ref{sec:boxs}.
\begin{align}
\label{eq:tots}
V_a=&c_{a1}\,\vtu\cdot\vtd
\\\nonumber
V_b=&c_{b1}
\\\nonumber
V_c=&c_{c1}\,\vtu\cdot\vtd
\\\nonumber
V_d=&
\left[c_{d1}+c_{d2}\,\vsu\cdot\vq
+c_{d3}\,(\vq\cdot\vp)
+c_{d4}\,\vsu\cdot(\vq\times\vp)\right]
(\vtu\cdot\vtd)
\\\nonumber
V_e=&(c_{e1}+c_{e2}\vsu\cdot\vq)(\vtu\cdot\vtd)
\end{align}
\begin{align}
\label{eq:tots2}
V_f=&
\Big[
 c_{f1}
+c_{f2}\vsu\cdot\vsd
+c_{f3}\vsu\cdot\vq
+c_{f4}(\vsu\times\vsd)\cdot\vq
\nonumber\\
+&c_{f5}(\vsu\cdot\vq)(\vsd\cdot\vq)
+c_{f6}(\vsu\cdot\vq)(\vsd\cdot\vp)
\\+&c_{f7}\vsu\cdot(\vp\times\vq)\nonumber
+c_{f8}\vsd\cdot(\vp\times\vq)\Big]
(c_{f1}'+c_{f2}'\,\vtu\cdot\vtd)
\\\nonumber
V_g=&
\Big[
c_{g1}
+c_{g2}\vsu\cdot\vsd
+c_{g3}(\vsu\cdot\vq)(\vsd\cdot\vq)
\Big]
\\\times&\nonumber
(c_{g1}'+c_{g2}'\,\vtu\cdot\vtd)
\\+&\nonumber
\Big[
c_{g4}\vsu\cdot\vq
+c_{g5}(\vsu\times\vsd)\cdot\vq
\Big]
(c_{g1}''+c_{g2}''\,\vtu\cdot\vtd)
\\\nonumber
V_h=&
\Big[
c_{h1}
+c_{h2}\vsu\cdot\vsd
+c_{h3}\vsu\cdot\vq
+c_{h4}(\vsu\times\vsd)\cdot\vq
\\+&c_{h5}(\vsu\cdot\vq)(\vsd\cdot\vq)\nonumber
+c_{h6}(\vsu\cdot\vq)(\vsd\cdot\vp)
\\+&c_{h7}\vsu\cdot(\vp\times\vq)\nonumber
+c_{h8}\vsd\cdot(\vp\times\vq)\Big]
(c_{h1}'+c_{h2}'\,\vtu\cdot\vtd)
\\\nonumber
V_i=&
\Big[
c_{i1}
+c_{i2}\vsu\cdot\vsd
+c_{i3}(\vsu\cdot\vq)(\vsd\cdot\vq)
\Big]
\\\times&\nonumber
(c_{i1}'+c_{i2}'\,\vtu\cdot\vtd)
\\+&\nonumber
\Big[
c_{i4}\vsu\cdot\vq
+c_{i5}(\vsu\times\vsd)\cdot\vq
\Big]
(c_{i1}''+c_{i2}''\,\vtu\cdot\vtd)\,.
\end{align}

Considering the SU(3) limit where all the baryon masses are
considered to take the same value ($q_0=q_0'=0$) the expressions above
become much more simple. Defining
\begin{align*}
At(q)\equiv&\frac{1}{2q}\arctan\left(\frac{q}{2m_\pi}\right)
\\
L(q)\equiv&\frac{\sqrt{4m_\pi^2+q^2}}{q}\ln\left(\frac{\sqrt{4m_\pi^2+q^2}+q}{2m_\pi}\right),
\\q\equiv &\sqrt{\vq^{\,2}},
\end{align*}
and extracting the baryonic poles and the  polynomial terms,
one obtains, 
\begin{align}
V_a=&\label{eq:va}
-\frac{h_{\Lambda N}}{192\pi^2f_\pi^4(M_\Lambda-M_N)}
(4m_\pi^2+q^2)L(q)(\vtu\cdot\vtd)
\\
V_b=&
\frac{3g_A^2h_{2\pi}}{32\pi f_\pi^4}(2m_\pi^2+q^2)At(q)
\\\label{eq:vc}
V_c=&
-\frac{g_A^2h_{\Lambda N}}{384\pi^2
  f_\pi^4(M_\Lambda-M_N)}(8m_\pi^2+5q^2)L(q)(\vtu\cdot\vtd)
\\
V_d=&\nonumber
\frac{g_A}{64\pi^2 f_\pi^3 M_N}
L(q)(\vtu\cdot\vtd)
\left(
-2Bm_\pi^2-B\vq^2+B(\vq\cdot\vp)
\right.\\&\left.+6A M_N (\vsu\cdot\vq)-3iB \vsu\cdot(\vq\times\vp)
\right)
\\
V_e=&
\frac{\sqrt{3}D_s}{384\pi^2 f_\pi^3 M_N}
L(q)
\left(
B_{\Sigma1}(4m_\pi^2+3\vq^2)-4A_{\Sigma1} M_N(\vsu \cdot\vq)
\right)\,,
\end{align}

\begin{align}
V_f=&\nonumber
\frac{g_A^3}{512\pi^2 f_\pi^3 M_N(4m_\pi^2+\vq^2)}L(q)
(-3+2\vtu\cdot\vtd)
\\&\nonumber
\times\left[
\frac{1}{6}B(448m_\pi^4+4m_\pi^2(-24\vq\cdot\vp+47\vq^2)+25\vq^4
\right.\\&\left.\nonumber
-36\vq^2(\vq\cdot\vp))
+4iB(4m_\pi^2+\vq^2)\vsd\cdot(\vq\times\vp)
\right.\\&\left.\nonumber
-4A M_N(8m_\pi^2+3\vq^2)\vsu\cdot\vq
\right.\\&\left.\nonumber
+2iB(8m_\pi^2+3\vq^2)\vsu\cdot(\vq\times\vp)
\right.\\&\left.\nonumber
+4B(4m_\pi^2+\vq^2)(\vsu\cdot\vq)(\vsd\cdot\vp)
\right.\\&\left.\nonumber
-4B(4m_\pi^2+\vq^2)(\vsu\cdot\vq)(\vsd\cdot\vq)
\right.\\&\left.\nonumber
-4B(4m_\pi^2+\vq^2)(\vq\cdot\vp-\vq^2)(\vsu\cdot\vsd)
\right.\\&\left.
-8iAM_N(4m_\pi^2+\vq^2)(\vsu\times\vsd)\cdot\vq
\right]
\\\nonumber\\
V_g=&\nonumber
\frac{D_s g_A^2}{256\sqrt{3}\pi^2 f_\pi^3 M_N(4m_\pi^2+\vq^2)}L(q)
\\&\nonumber
\times\left[
-\frac{1}{6}B_{\Sigma2}(448m_\pi^4+188m_\pi^2\vq^2+25\vq^4)
\right.\\&\nonumber\left.
+4A_{\Sigma2}M_N(8m_\pi^2+3\vq^2)(\vsu\cdot\vq)
\right.\\&\nonumber\left.
+4B_{\Sigma2}(4m_\pi^2+\vq^2)(\vsu\cdot\vq)(\vsd\cdot\vq)
\right.\\&\left.\nonumber
-4B_{\Sigma2}(4m_\pi^2+\vq^2)\vq^2(\vsu\cdot\vsd)
\right.\\&\left.
-8iA_{\Sigma2}M_N(4m_\pi^2+\vq^2)(\vsu\times\vsd)\cdot\vq
\right]\,,
\end{align}

\begin{align}
V_h=&\nonumber
\frac{g_A^3}{512\pi^2 f_\pi^3 M_N(4m_\pi^2+\vq^2)}L(q)
(3+2\vtu\cdot\vtd)
\\&\nonumber
\times\left[
\frac{1}{6}B(448m_\pi^4+4m_\pi^2(-24\vq\cdot\vp+47\vq^2)+25\vq^4
\right.\\&\nonumber\left.
-36\vq^2(\vq\cdot\vp))
-4iB(4m_\pi^2+\vq^2)\vsd\cdot(\vq\times\vp)
\right.\\&\nonumber\left.
-4A M_N(8m_\pi^2+3\vq^2)\vsu\cdot\vq
\right.\\&\nonumber\left.
-2iB(8m_\pi^2+3\vq^2)\vsu\cdot(\vq\times\vp)
\right.\\&\nonumber\left.
+4B(4m_\pi^2+\vq^2)(\vsu\cdot\vq)(\vsd\cdot\vp)
\right.\\&\nonumber\left.
-4B(4m_\pi^2+\vq^2)(\vsu\cdot\vq)(\vsd\cdot\vq)
\right.\\&\nonumber\left.
-4B(4m_\pi^2+\vq^2)(\vq\cdot\vp-\vq^2)(\vsu\cdot\vsd)
\right.\\&\left.
+8iAM_N(4m_\pi^2+\vq^2)(\vsu\times\vsd)\cdot\vq
\right]
\\\nonumber\\
V_i=&\nonumber
\frac{D_s g_A^2}{256\sqrt{3}\pi^2 f_\pi^3 M_N(4m_\pi^2+\vq^2)}L(q)
\\&\nonumber
\times\left[
\frac{1}{6}B_{\Sigma3}(448m_\pi^4+188m_\pi^2\vq^2+25\vq^4)
\right.\\&\nonumber\left.
+A_{\Sigma3}M_N(8m_\pi^2+3\vq^2)(\vsu\cdot\vq)
\right.\\&\nonumber\left.
+4B_{\Sigma3}(4m_\pi^2+\vq^2)(\vsu\cdot\vq)(\vsd\cdot\vq)
\right.\\&\nonumber\left.
-4B_{\Sigma3}(4m_\pi^2+\vq^2)\vq^2(\vsu\cdot\vsd)
\right.\\&\left.
+4iA_{\Sigma3}M_N(4m_\pi^2+\vq^2)(\vsu\times\vsd)\cdot\vq
\right]\,.
\end{align}

The isospin part for the potentials that contain $\Sigma$ propagators
($V_e$, $V_g$, $V_i$) is taken into account by making the replacements:
\begin{align*}
A_{\Sigma1}\to&\frac{2}{3}\left(\sqrt3 A_{\Sigma\frac12}+
A_{\Sigma\frac32}\right)
\vec{\tau_1}\cdot\vec{\tau_2}
\\
B_{\Sigma1}\to&
\frac{2}{3}\left(\sqrt3 B_{\Sigma\frac12}+
B_{\Sigma\frac32}\right)
\vec{\tau_1}\cdot\vec{\tau_2}\,,
\end{align*}

\begin{align*}
A_{\Sigma2}\to&
-\sqrt3A_{\Sigma\frac12}+2A_{\Sigma\frac32}
+\frac23(\sqrt3A_{\Sigma\frac12}+A_{\Sigma\frac32})\vtu\cdot\vtd
\\
B_{\Sigma2}\to&
-\sqrt3B_{\Sigma\frac12}+2B_{\Sigma\frac32}
+\frac23(\sqrt3B_{\Sigma\frac12}+B_{\Sigma\frac32})\vtu\cdot\vtd\,.
\end{align*}

\begin{align*}
A_{\Sigma3}\to&
-\sqrt3A_{\Sigma\frac12}+2A_{\Sigma\frac32}
-\frac23(\sqrt3A_{\Sigma\frac12}+2A_{\Sigma\frac32})\vtu\cdot\vtd
\\
B_{\Sigma3}\to&
-\sqrt3B_{\Sigma\frac12}+2B_{\Sigma\frac32}
-\frac23(\sqrt3B_{\Sigma\frac12}+2B_{\Sigma\frac32})\vtu\cdot\vtd\,.
\end{align*}

Note that Eqs.~(\ref{eq:va}) and (\ref{eq:vc}) only have physical meaning
away from the SU(3) limit.
\section{Brief comparison of LO and NLO contributions}
\label{sec:bc}

\begin{figure}[t]
\centering
\begin{tabular}{l}
\includegraphics[scale=0.8]{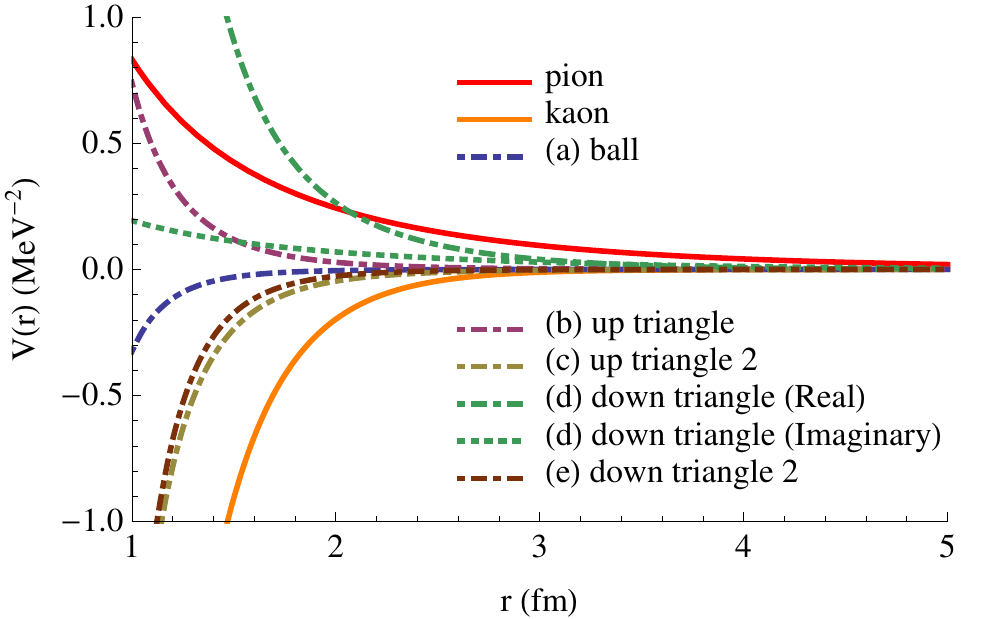}\cr
\includegraphics[scale=0.8]{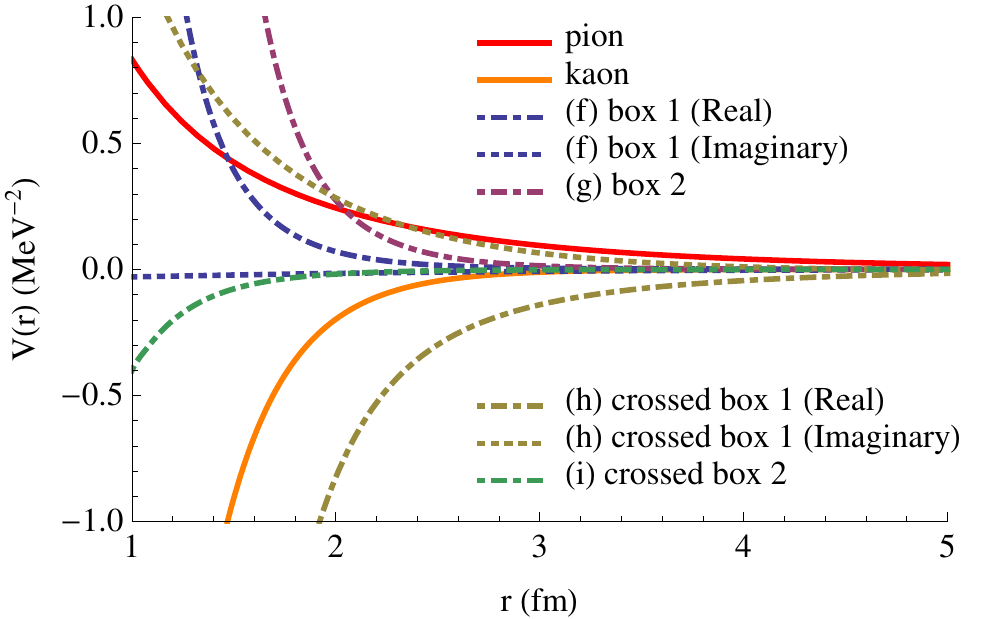}
\end{tabular}
\caption{(UP) Medium-Long range part of the potentials for 
the one-pion-exchange, one-kaon-exchange, ball diagram and 
triangle diagrams. (DOWN) Medium-Long range part of the potentials 
for the one-pion-exchange, one-kaon-exchange, box and crossed box diagrams.
\label{fig:pot}}
\end{figure}

\begin{figure}[t]
\centering
\begin{tabular}{l}
\includegraphics[scale=0.8]{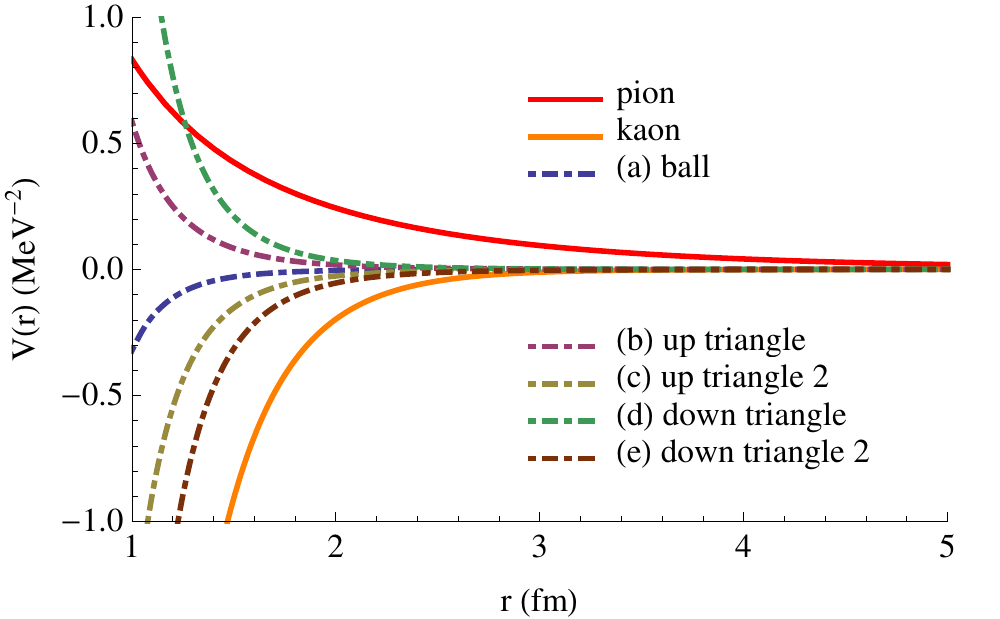}\cr
\includegraphics[scale=0.8]{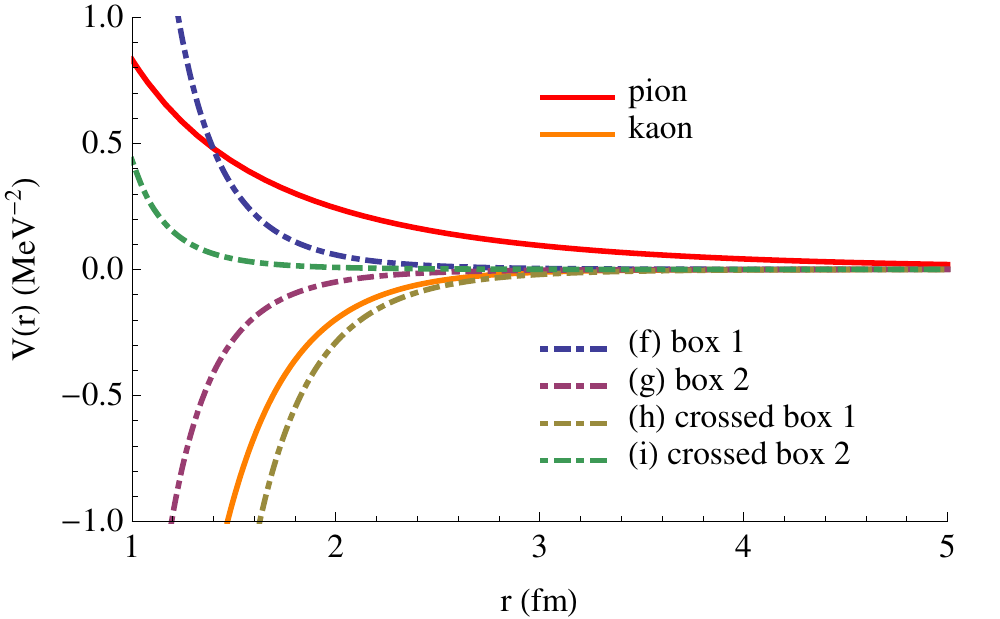}
\end{tabular}
\caption{(UP) Medium-Long range part of the potentials in the SU(3)
  limit 
for the one-pion-exchange, one-kaon-exchange, ball diagram 
and triangle diagrams. (DOWN) Medium-Long range part of the 
potentials in the SU(3) limit for the one-pion-exchange,
one-kaon-exchange, box and crossed box diagrams. 
\label{fig:pot0}}
\end{figure}

In Eqs.~(\ref{eq:tots}) and~(\ref{eq:tots2}) we provide the 
explicit momentum and spin structures arising from the different 
Feynman diagrams. Some features can be easily read off from the 
different terms. First, the ball (a) and first two triangle 
diagrams (b,c) only contribute to the parity conserving part of 
the transition potential. Most other diagrams have a non-trivial 
contribution, involving all allowed momenta and spin structures. 

To provide a sample of the contribution of the different 
diagrams to the full amplitude, we consider one particular 
transition, $^3 S_1\rightarrow ^3S_1$. In particular, we 
compare the $\pi$ and $K$ exchanges with the ball, triangle 
and box diagrams for the  $\Lambda n\rightarrow nn$ interaction.
Since the transition is parity conserving, none of the parity
violating structures of Table~\ref{tab:contacts} contribute. 
For structures of the type $(\vsu\cdot\vq)(\vsd\cdot\vq)$ 
we have that
\begin{align}
(\vsu\cdot\vq)(\vsd\cdot\vq)=\frac{\vq^{\,2}}{3}(\vsu\cdot\vsd)+\frac{\vq^2}{3}\hat{S}_{12}(\hat{q}),
\end{align}
where the tensor operator $\hat{S}_{12}(\hat{q})$ changes two units of 
angular momentum and does not contribute to this transition. 
The potential, therefore, depends only on the modulus of the momentum 
 (or $\vq^{\,2}$). To obtain the potential in position space 
we Fourier-transform the expressions for the 
one-meson-exchange contributions, Eqs.~\ref{eq:pion} and~\ref{eq:kaon},
and the loop expressions in the appendices~\ref{sec:balls},
\ref{sec:triangles} and \ref{sec:boxs}. More explicitly,
\begin{align*}
\tilde{V}(r)= {\cal F}\left[V(\vq^{\,2}) F(\vq^{\,2})\right]&\equiv
\int_{-\infty}^{\infty}\frac{d^3q}{(2\pi)^3}
e^{i\vq\cdot\vrr}V(\vq^2)F(\vq^{\,2})
\ignore{\\&=
\frac{1}{2\pi^2r}\int_0^\infty dq\sin\left(qr\right)q V(q^2),}
\end{align*}
with $q\equiv|\vq|$ and $r\equiv|\vrr|$ and where we have included a
form factor in order to regularize the potential. Following the
formalism developed in Ref.~\cite{PRB97} we use a monopole form factor
for the meson exchange contribution at each vertex, while the $2-\pi$
terms use a Gaussian form of the type $F(\vq^2)\equiv e^{-\vq^{\,4}/\Lambda^4}$.
\ignore{
\begin{equation}
F(\vq^2)\equiv e^{-\frac{\vq^4}{\Lambda^4}}\,,
\end{equation}
have been included.}

\ignore{ We use the cut-off 
value of $\Lambda=4000$ MeV taken from the J\"ulich 
hyperon-nucelon interaction~\cite{HH89}.}

The expressions for each loop have been calculated
using dimensional regularization and are shown in the
appendices $B$, $C$ and $D$. They are written in terms of the
couplings appearing in Sec.~\ref{sec2} and of the master 
integrals appearing in App.~\ref{sec:mi}. $\eta$ is the 
regularization parameter that appears when integrating in 
$D\equiv4-\eta$ dimensions. The modified minimal subtraction 
scheme ($\overline{MS}$) has been used---we have expanded 
in powers of $\eta$ the expressions for the different loop 
contributions and then subtracted the term
$R\equiv-\frac{2}{\eta}+\gamma-1-\ln\left(4\pi\right)$---. 

In Fig.~\ref{fig:pot}, we show the 
respective contributions to the potential in position space. 
The contribution from the different $2-\pi$ exchange potentials 
are seen to be sizable at all distances. In particular, the 
box (f, g, h) and triangle (d) diagrams give larger contributions
than the pion in the medium and long-range. The ball diagram (a) and
the triangles (c), (e), (h) and (i) are attractive while all the others are repulsive.
Notice that diagrams (d), (f) and (h) contribute with an imaginary
part. This is characteristic of diagrams with a $\Lambda N\pi$ vertex,
which may be on shell since $M_\Lambda>M_N+m_\pi$. This imaginary part
is taking into account the amplitude for the possible $\Lambda
N\rightarrow NN\pi$ transition. We stress that the imaginary part
of the box diagram (f) that comes from the baryonic pole has been
extracted, so no iterated part is considered in Fig.~\ref{fig:pot}.

Fig.~\ref{fig:pot0}, shows the same potentials but taking
$q_0=q_0'=0$. All diagrams seem to have a smaller contribution when
the baryon mass differences are neglected. The attractive and
repulsive character of the different potentials does not change except
for the second box diagram and the second crossed box diagram, which turn
to be attractive and repulsive, respectively, when taking the SU(3)
limit.

\section{Conclusions}
\label{sec:conclusions}

The weak decay of hypernuclei is dominated for large
enough number of nucleons by the non-mesonic weak decay
modes. In these modes, the bound $\Lambda$ particle decays
in the presence of nucleons by means of a process which
involves weak and strong interaction vertices describing
the production and absorption of mesons. The relevant,
experimentally known, partial and total decay rates of hypernuclei, are
successfully described by meson-exchange models and also
by a lowest-order effective field theory description
of the weak $\Lambda N\to NN$ process, when appropriate nuclear
wave functions are used for the initial and final nuclear systems.
Nevertheless, the stability of the EFT approach which has to be tested
by looking at higher orders in the theory, could not be analyzed yet,
mainly because of the very scarce world-database for such
observables, a situation which should be improved in the
near future.

In this article we have presented the
one-loop contribution to the previously obtained LO EFT
for the weak $\Delta S=1$ $\Lambda N$ transition.

As expected, the structure of the transition amplitude
is considerably more involved than the corresponding LO amplitude and
contains 
more low-energy coefficients which ought to be fitted to data.
In the present formal work we have solely presented the calculation
of the amplitude terms and have not attempted to make any comparison
to experimental data, therefore, no fit in order to extract the new
unknowns has been performed.
The
different structures which appear in the obtained transition
amplitude, involving spin, isospin and orbital degrees of freedom,
produce sizable contributions to all relevant partial waves.
To illustrate this fact, we have presented the potential
in $r$ space corresponding to the different Feynman diagrams
for the $^3S_1- ^3S_1$ partial wave. Box and cross-box diagrams
are found to produce substantial contributions at distances of the
order of 1 fm, larger than the
ones corresponding to the one-pion-exchange and one-kaon-exchange
mechanisms. 
In view of this result, it would be interesting to see if one-loop
contributions play an equivalent role in other partial wave
transitions, testing possible cancellations or enhancements that would
leave the results for the decay rates either unchanged or modified.
A complete analysis of the higher order terms would require a larger
set of independent hypernuclear decay measurements and a more accurate
measure of some observables, specially those related to the parity
violating asymmetry for s-shell and p-shell hypernuclei.
Moreover, it would be desirable to arrange for alternative experiments
focused to obtain information on the weak $\Delta S=1$ interaction. A
step in this direction was taken 
more than ten years ago by experimental groups at RCNP in Osaka
(Japan) [15,16], by looking 
at the weak strangeness production reaction $np \to \Lambda
p$. Unfortunately, the small value for the cross-section
for this process precluded the compilation of new data. We think that
it is important to foster 
new experimental avenues of approaching the weak interaction among
baryons in the strange 
sector, and even try to recover the Osaka experiment within the
research plan of the new experimental
facilities devoted to the study of strange systems.

To ease the use of the obtained EFT amplitudes,
we have provided with the explicit analytic expressions for all
diagrams which will in future work be implemented in the
calculation of hypernuclear decay observables.

\begin{acknowledgements}
We thank J. Soto, J. Tarr\'us, J. Haidenbauer and A.
Nogga for the helpful comments and discussions.
This work is partly supported by grants 
FPA2010-21750-C02-02 and
FIS2011-24154
from MICINN, 283286 from
European Community-Research Infrastructure Integrating Activity `Study
of Strongly Interacting Matter',
CSD2007-00042 from Spanish Ingenio-Consolider 2010 Program CPAN, and
2009SGR-1289 from Generalitat de Catalunya.
A.P-O. acknowledges support by the APIF Ph.D. program of the
University of Barcelona. B.J.D. is supported by the Ramon y Cajal program.

\end{acknowledgements}

\clearpage

\appendix

\section{Caramel diagrams}
\label{sec:caramels}
\begin{figure}[th]
\includegraphics[scale=0.4]{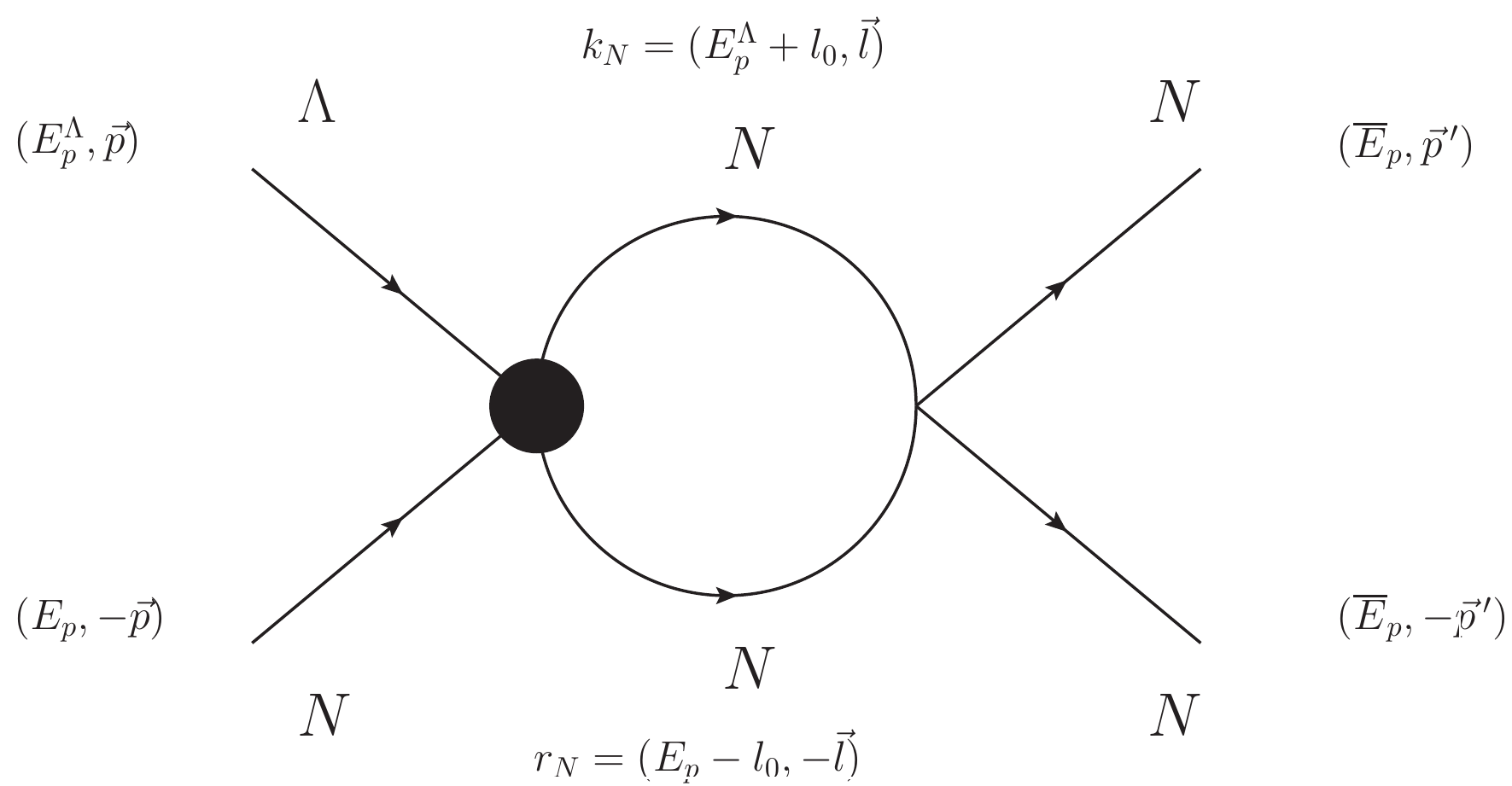}
\caption{First caramel-type Feynman diagram
\label{caramel1}}
\end{figure}
\begin{figure}[th]
\includegraphics[scale=0.4]{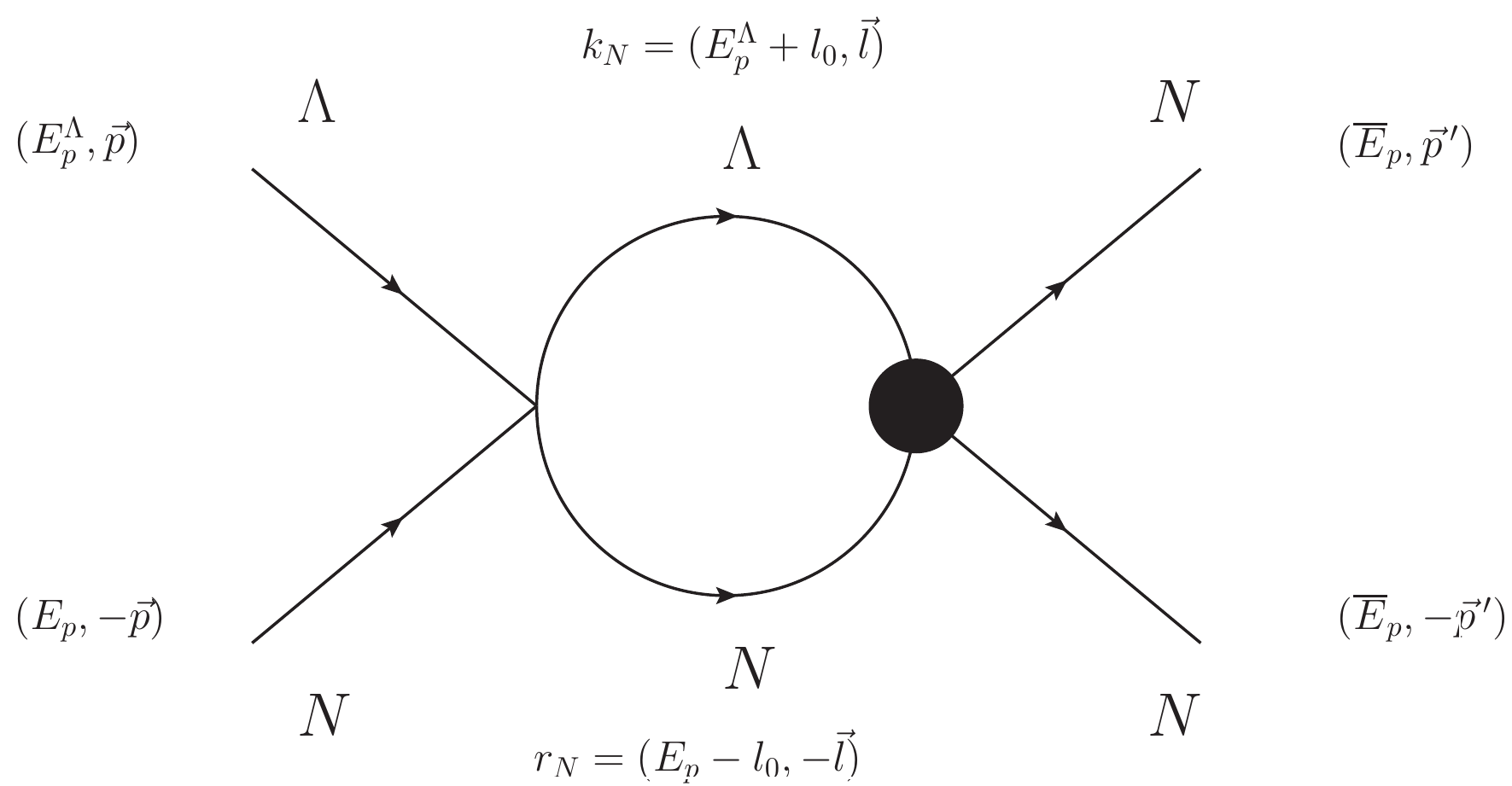}
\caption{Second caramel-type Feynman diagram
\label{caramel2}}
\end{figure}
\begin{figure}[th]
\includegraphics[scale=0.4]{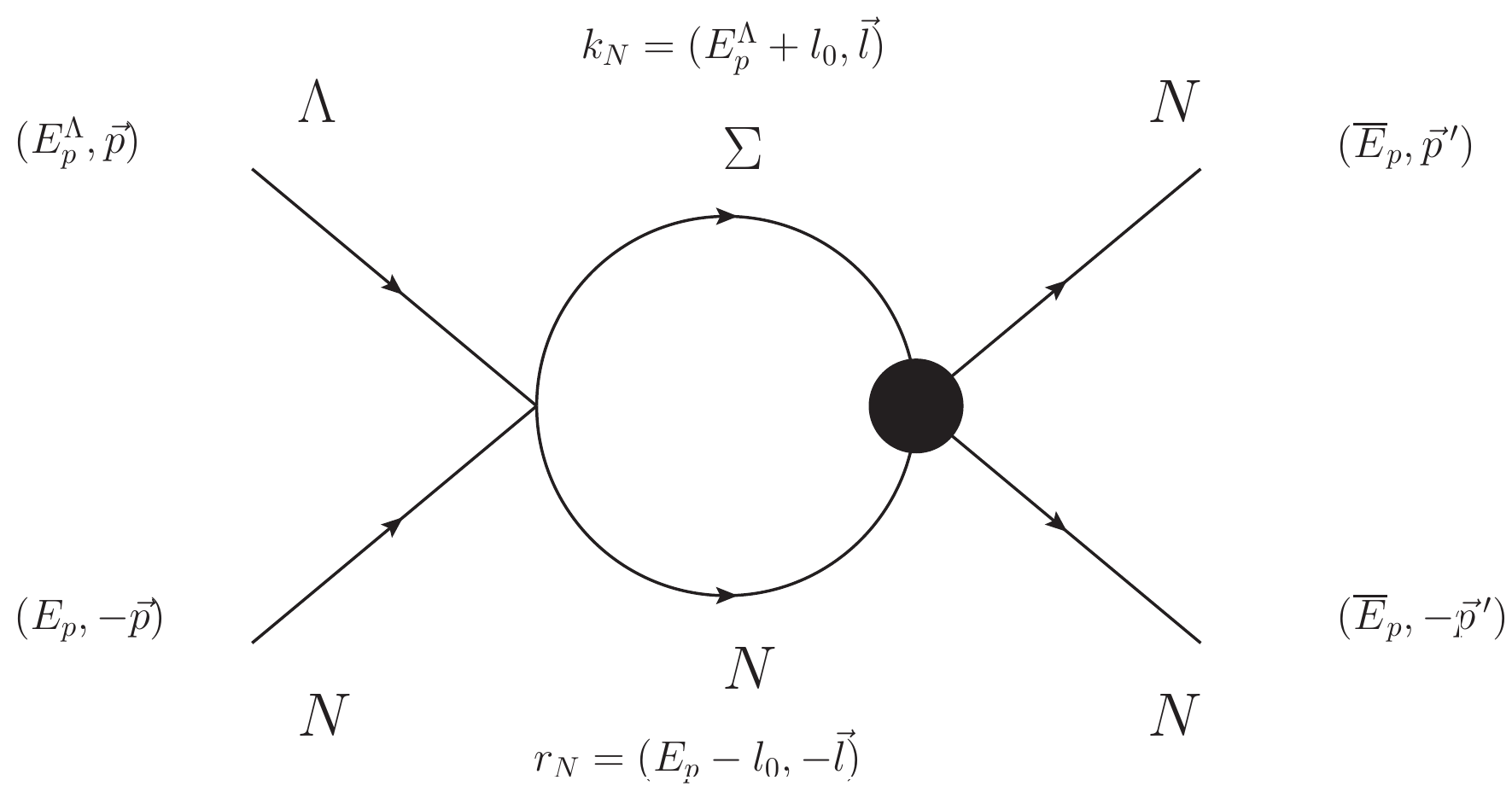}
\caption{Third caramel-type Feynman diagram
\label{caramel3}}
\end{figure}

Using the same notation that is described in section
\ref{sec:nlocontact}
we write a general expression for the three caramel diagrams that
depends on the label $\alpha=a,b,c$, which corresponds, respectively,
to the masses and vertices of Figs.~\ref{caramel1},~\ref{caramel2}, and
\ref{caramel3}. 
The relativistic expression for our caramel diagrams is,
\begin{align*}
V_\alpha&=iG_Fm_\pi^2
(C_{S(s)}^{\alpha}+C_{T(s)}^{\alpha}\vsu\cdot\vsd)\,
(C_{S(w)}^{\alpha}+C_{T(w)}^{\alpha}\vsu\cdot\vsd)\,
\\&\times
\intlq
\frac{1}{(E_p-l_0)^2-\vl^2-M_N^2+i\epsilon}
\\&\times\frac{1}{(E_p^\Lambda+l_0)^2-\vl^2-M_\alpha^2}
\end{align*}
In order to not miss the relativistic pole we must first integrate the
temporal part ($l_0$) before heavy-baryon expand the
expression. Proceeding in this manner one obtains a purely imaginary
part (the real is suppressed in the heavy baryon expansion).
\begin{align*}
V_\alpha&=-\frac{G_Fm_\pi^2}{4M_N}
(C_{S(s)}^{\alpha}+C_{T(s)}^{\alpha}\vsu\cdot\vsd)\,
(C_{S(w)}^{\alpha}+C_{T(w)}^{\alpha}\vsu\cdot\vsd)\,
\\&\times
\intlt\,
\frac{1}{(\Delta_b-\Delta_\alpha)(\frac12(\Delta_b+\Delta_\alpha)+M_N)+\vp^2-\vl^2}
\\&=i\frac{G_Fm_\pi^2}{16\pi M_N}
(C_{S(s)}^{\alpha}+C_{T(s)}^{\alpha}\vsu\cdot\vsd)\,
(C_{S(w)}^{\alpha}+C_{T(w)}^{\alpha}\vsu\cdot\vsd)\,
\\&\times
\sqrt{
(\Delta_b-\Delta_\alpha)(\frac12(\Delta_b+\Delta_\alpha)+M_N)+\vp^2}
\end{align*}

\section{Ball diagrams}
\label{sec:balls}

In our calculation we have two different kind of ball 
diagrams depending on the position of the weak vertex, 
although only one of them actually contributes. Their 
contribution can be written in terms of the $B$ 
integrals defined in Appendix ~\ref{sec:mi}.

Here and in the following sections we first write 
the relativistic amplitude using $V=i \ M$ and then 
the corresponding heavy baryon expression.

\begin{figure}[b]
\includegraphics[scale=0.4]{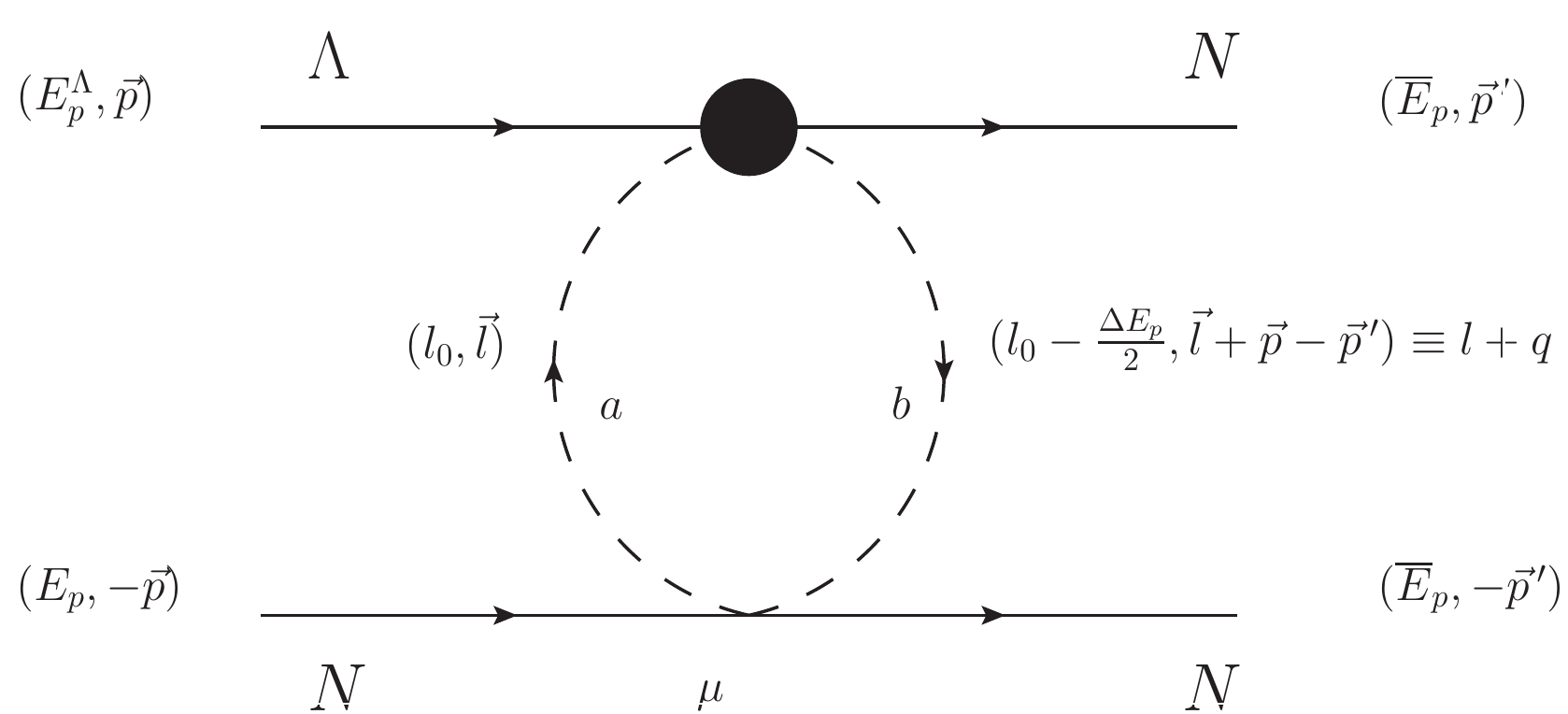}
\caption{Kinematical variables of the 
first kind of ball-diagram.\label{fball1}}
\end{figure}

For the first type of ball diagram, depicted in 
Fig.~\ref{fball1}, we obtain the following contribution,
\begin{align*}
V_{\text{ball 1}}=& \frac{G_Fm_\pi^2 h_{2\pi}}{4f_\pi^4}
\delta_{ab}\ \epsilon^{abc}\tau^c
\nonumber\\
&\times \intlq 
\frac{1}{l^2-m_\pi^2+i\epsilon} 
\frac{1}{(l-q)^2-m_\pi^2+i\epsilon}\nonumber
\\&\times
\ou_1(\oee,\vpp) 
u_1(E_p^\Lambda,\vp)\nonumber\\
&\times
\ou_2(\oee_p,-\vpp)
\gamma_\mu(q^\mu-2l^\mu)
u_2(E_p,-\vp)
\\=&0\,,
\end{align*}
which is shown to vanish due to the isospin factor, 
$\delta_{ab}\epsilon^{abc}\tau^c=0$.

\begin{figure}[b]
\includegraphics[scale=0.4]{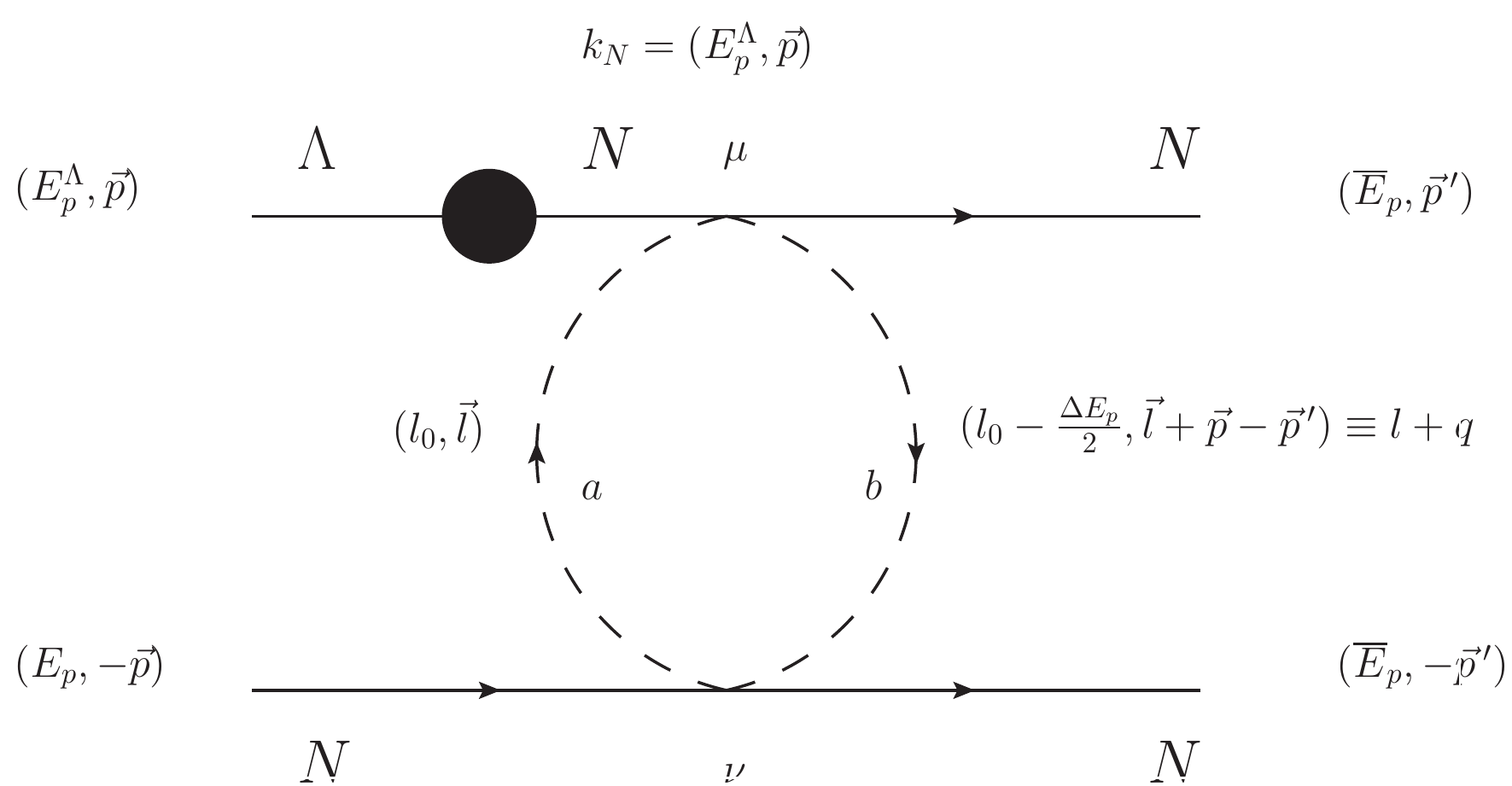}
\caption{Kinematical variables of the second kind of 
ball-diagram.\label{fball2g}}
\end{figure}

The amplitude corresponding to the diagram in 
Fig.~\ref{fball2g} reads, 
\begin{eqnarray}
V_a&=&-i 
\frac{G_Fm_\pi^2h_{\Lambda N}}{8f_\pi^4} (\vtu\cdot\vtd) \nonumber\\
&\times&\intlq 
\frac{1}{l^2-m_\pi^2+i\epsilon}
\,\frac{1}{(l+q)^2-m_\pi^2+i\epsilon} \nonumber\\
&\times&
\frac{(2l^\mu+q^\mu)(q^\nu+2l^\nu)}{k_N^2-M_N^2+i\epsilon}
\nonumber\\
&\times&
\ou_1(\oee,\vpp) \gamma_\mu (\kn+M_N)
u_1(E_p^\Lambda,\vp) \nonumber\\
&\times&
\ou_2(\oee_p,-\vpp) 
\gamma_\nu
u_2(E_p,-\vp)\,.
\end{eqnarray}
Using heavy baryon expansion,
\begin{eqnarray}
V_a
&=&\ \frac{G_Fm_\pi^2h_{\Lambda N}}{8\Delta Mf_\pi^4}
(\vtu\cdot\vtd)
(4 { B}_{20}
+4q_0 { B}_{10}
+q_0^2 { B}) \,,\nonumber\\
\end{eqnarray}
where we have used the master integrals with
$q_0=-\frac{M_\Lambda-M_N}{2}$ and $\vq=\vpp-\vp$.

\section{Triangle diagrams}
\label{sec:triangles}

\begin{figure}[th]
\includegraphics[scale=0.4]{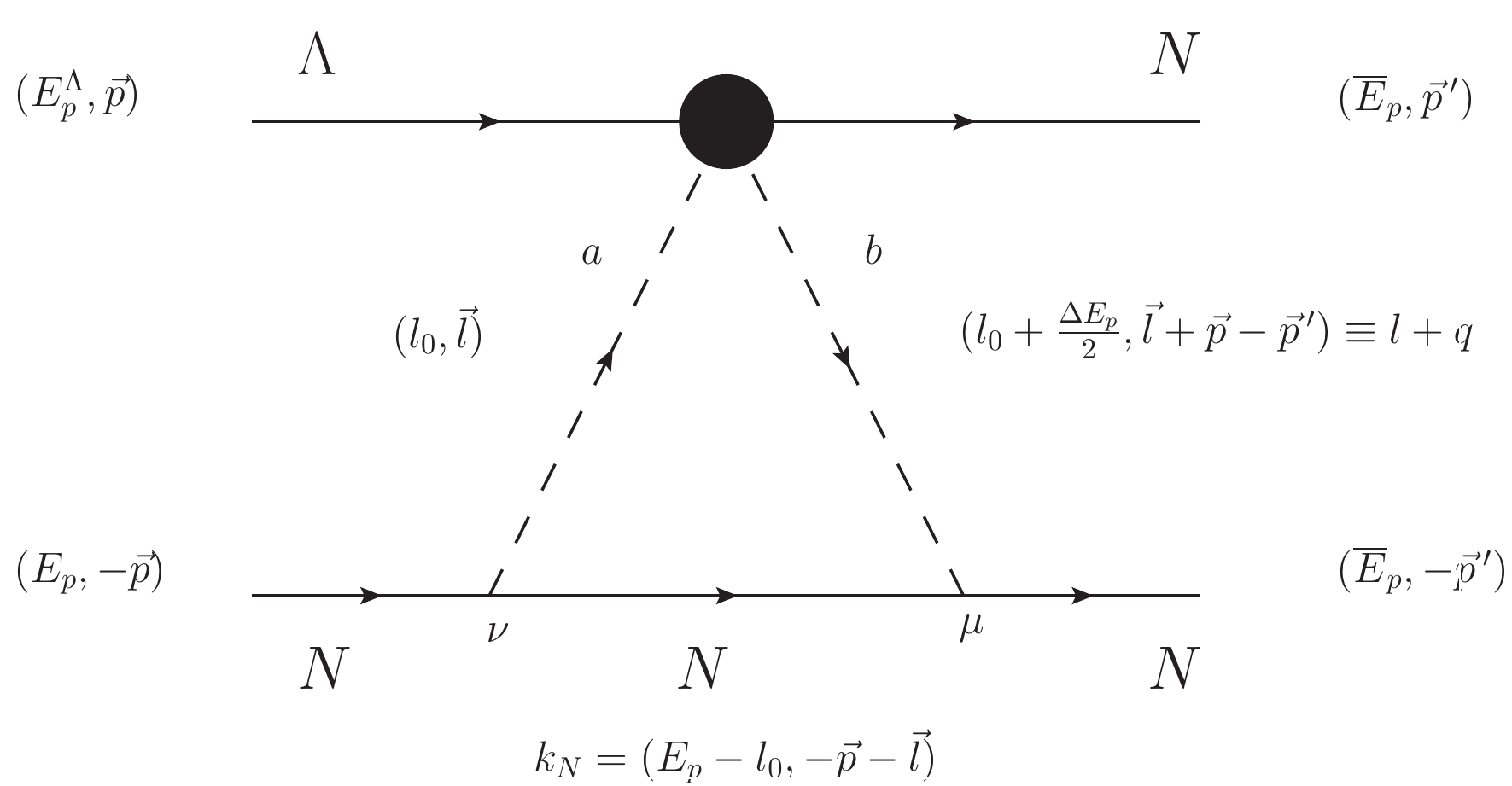}
\caption{Up triangle diagram contributing at NLO.
\label{uptri}}
\end{figure}

Two up triangles and two down triangles contribute to the
interaction. The final expressions are written in terms of 
the integrals $I$ defined in Appendix~\ref{sec:mi}.
The amplitude for the first up triangle, depicted in 
Fig.~\ref{uptri}, is
\begin{align}
V_b=&-i\frac38\frac{G_F m_\pi^2h_{2\pi}g_A^2}{M_N f_\pi^4}
\intlq
\frac{1}{l^2-m_\pi^2+i\epsilon}
\nonumber\\\times&\,
\frac{1}{(l+q)^2-m_\pi^2+i\epsilon}\,
\frac{(l^\mu+q^\mu)l^\nu}{k_N^2-M_N^2+i\epsilon}
\\\times&\nonumber\,
\,\ou_1(\oee_p,\vpp)\ou_1(E_p^\Lambda,\vp)
\\\times&\nonumber\,
\ou_2(\oee_p,-\vpp)\gamma_\mu\gamma_5(\kn+M_N)\gamma_\nu\gamma_5
u_2(E_p,-\vp)\,.
\end{align}
Using heavy baryon expansion,
\begin{align}
V_b
=\frac34\frac{G_F m_\pi^2h_{2\pi}g_A^2}{f_\pi^4}
\left[
(3-\eta)I_{22}+\vq^2I_{23}+\vq^2I_{11}
\right]\,,
\end{align}
where, we have used the master integrals with
$q_0=\frac{M_\Lambda-M_N}{2}$, $q_0'=0$ and $\vq=\vpp-\vp$.

\begin{figure}[th]
\includegraphics[scale=0.4]{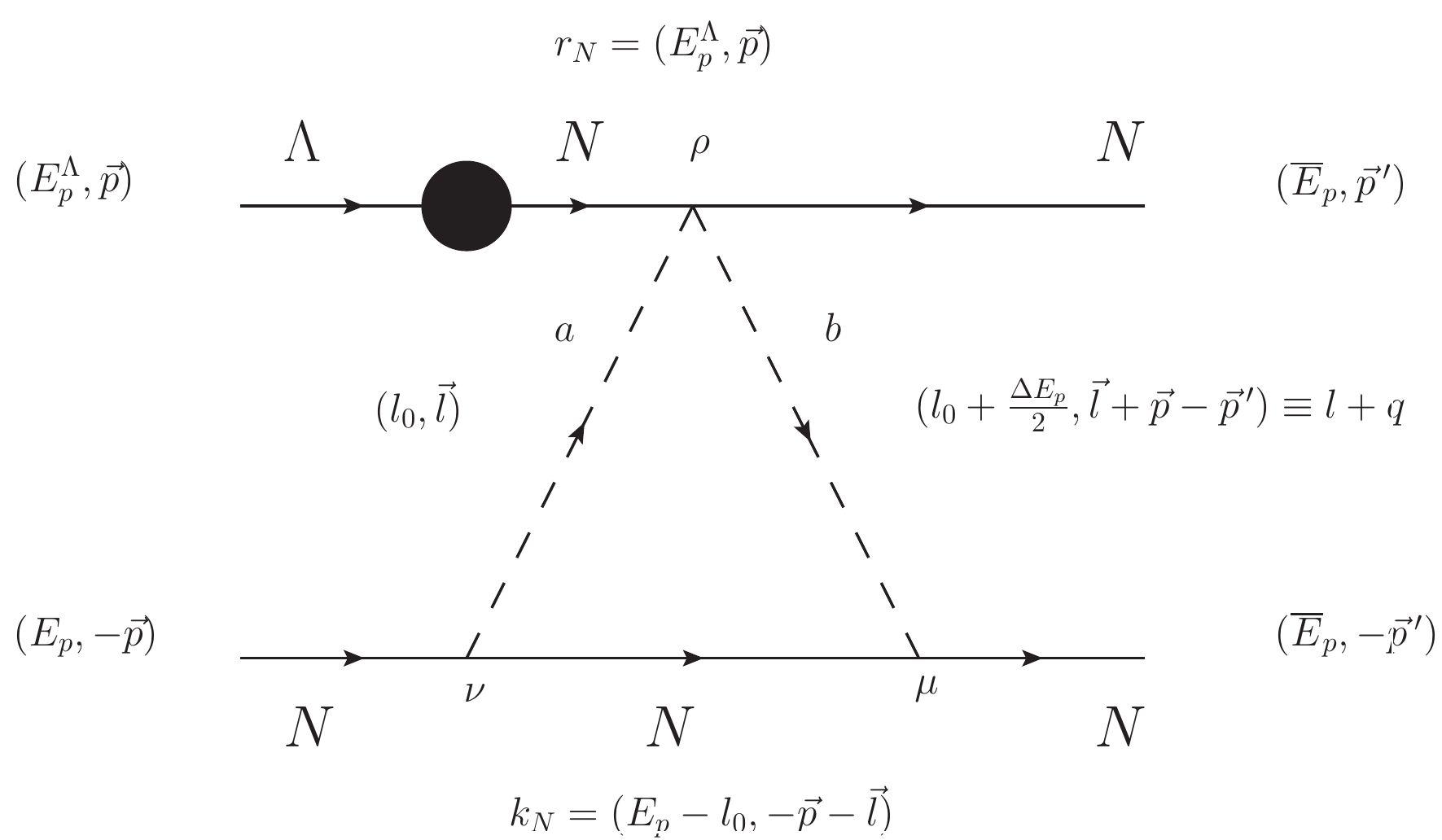}
\caption{Second up triangle contribution at NLO.}
\label{uptri2}
\end{figure}
For the second up triangle, depicted in Fig.~\ref{uptri2}, 
the relativistic amplitude is
\begin{align}
V_c=&
-i\frac{G_Fm_\pi^2h_{\Lambda N} g_A^2}{8f_\pi^4(r_N^2-M_N^2)}
\vtu\cdot\vtd
\intlq
\frac{1}{l^2-m_\pi^2+i\epsilon}
\nonumber\\\times&\,
\frac{1}{(l+q)^2-m_\pi^2+i\epsilon}\,
\frac{(2l^\rho+q^\rho)(l^\mu+q^\mu)l^\nu}{k_N^2-M_N^2+i\epsilon}
\\\times&\nonumber\,
\ou_1(\oee,\vpp)\gamma_\rho(\kn'+M_N)
 u_1(E_p^\Lambda,\vp)
\\\times&\nonumber\,
\ou_2(\oee_p,-\vpp)\gamma_\mu\gamma_5(\kn+M_N)
\gamma_\nu\gamma_5 u_2(E_p,-\vp)\,.
\end{align}
Using heavy baryon expansion,
\begin{align}
V_c=&
\frac{G_Fm_\pi^2h_{\Lambda N} g_A^2}{8\Delta M f_\pi^4}
\vtu\cdot\vtd
\left[
2(3-\eta)I_{32}+2\vq^2I_{33}+2\vq^2I_{21}
\nonumber\right.\\+&\left.
(3-\eta)q_0I_{22}+q_0\vq^2I_{23}+q_0\vq^2I_{11}
\right]\,,
\end{align}
where, we have used the master integrals with
$q_0=\frac{M_\Lambda-M_N}{2}$, $q_0'=0$  and $\vq=\vpp-\vp$.

\begin{figure}[th]
\includegraphics[scale=0.4]{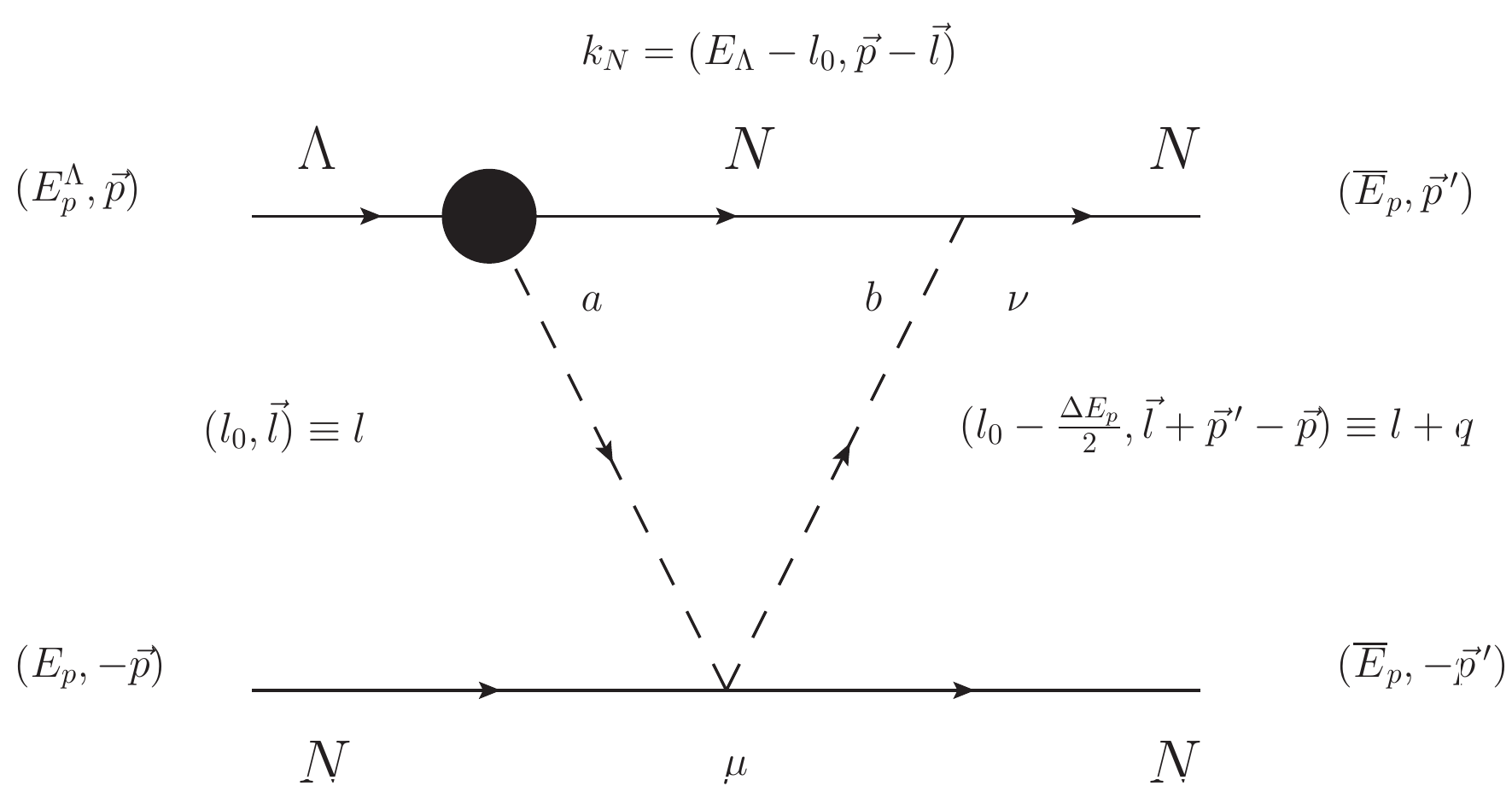}
\caption{``Down''-triangle contribution at NLO.\label{downtri}}
\end{figure}
The amplitude for the first down triangle (Fig.~\ref{downtri}) 
is
\begin{align}
V_d=&
i\frac{G_Fm_\pi^2g_A}{4 f_\pi^3}
(\vec{\tau}_1\cdot\vec{\tau}_2)
\intlq
\frac{1}{l^2-m_\pi^2+i\epsilon}
\\\times&\nonumber\,
\frac{1}{(l+q)^2-m_\pi^2+i\epsilon}
\frac{(l^\nu+q^\nu)(2l^\mu+q^\mu)}{k_N^2-M_N^2+i\epsilon}
\\\times&\nonumber\,
\ou_1(\oee,\vpp)
\gamma_{\nu}\gamma_5
(\kn+M_N)
(A+B\gamma_5)
 u_1(E_p^\Lambda,\vp)
\\\times&\,\nonumber
\ou_2(\oee_p,-\vpp)\gamma_\mu u_2(E_p,-\vp)\,,
\end{align}
with the heavy baryon expansion, it reduces to, 
\begin{align}
V_d=&
-\frac{G_Fm_\pi^2g_A}{8M_N f_\pi^3}
(\vec{\tau}_1\cdot\vec{\tau}_2)
\Big[
B(2I_{30}+7q_0I_{20}+7q_0^2I_{10}
\nonumber\\+&
2q_0^3I
-2(3-\eta)I_{32}-(3-\eta)q_0I_{22})
\\-&\nonumber
B(2I_{21}+q_0I_{11}+2I_{33}+q_0I_{23})\vq^2
\\-&\nonumber
B(2I_{10}+2I_{21}+q_0 I+q_0I_{11})(\vq\cdot\vp)
\\+&\nonumber
2A\,M_N(2I_{21}
+q_0I_{11}-2I_{10}-q_0I)\vsu\cdot\vq
\\+&\nonumber
iB(-2I_{21}-q_0I_{11}+2I_{10}+q_0I)\vsu(\vq\times\vp)
\Big]\,.
\end{align}
We have used the master integrals with
$q_0=-\frac{M_\Lambda-M_N}{2}$, $q_0'=-M_\Lambda+M_N$ and
$\vq=\vp'-\vp$.

\begin{figure}[th]
\includegraphics[scale=0.4]{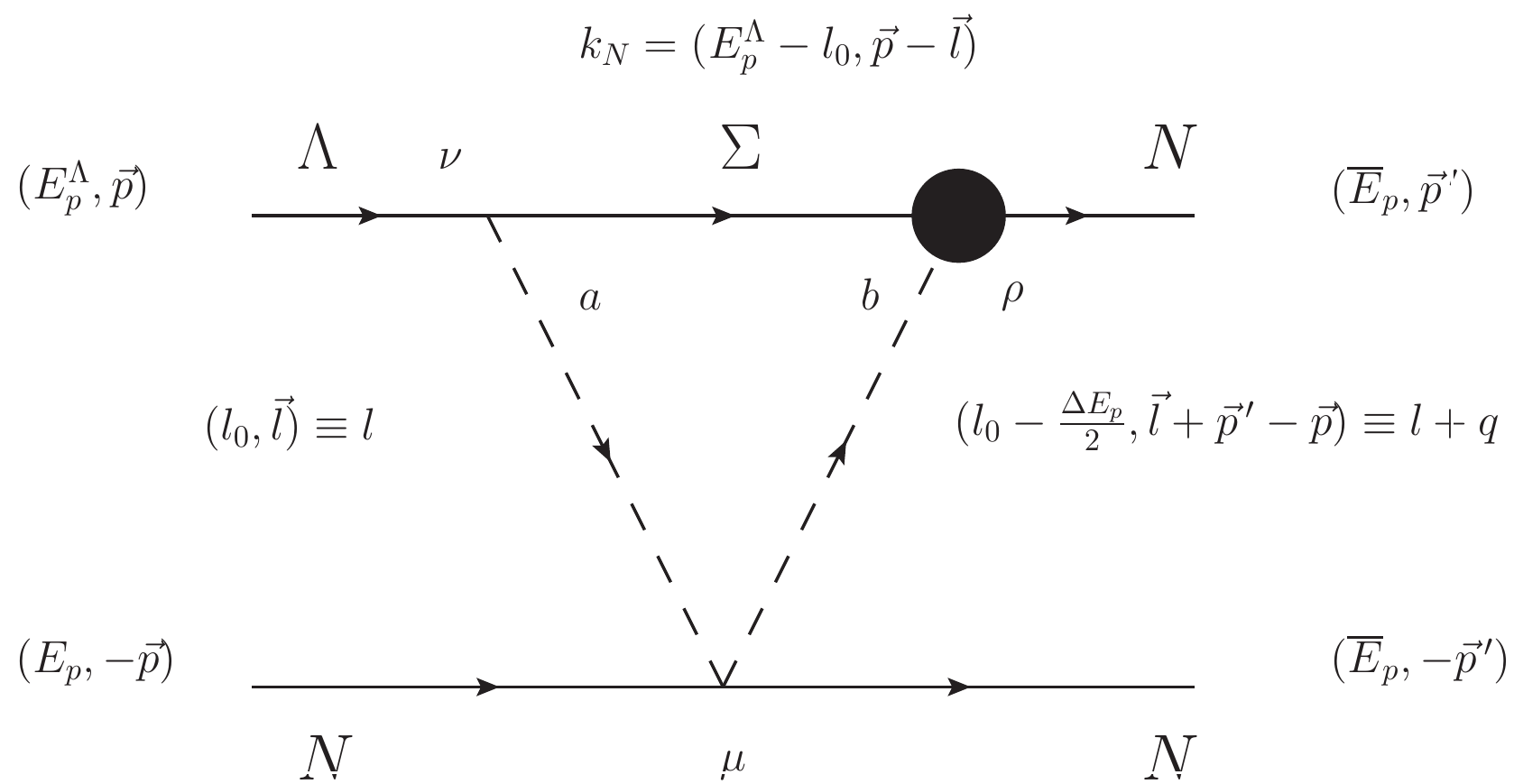}
\caption{Second type of down-triangle involving the 
intermediate exchange of a $\Sigma$.}
\label{downtri2}
\end{figure}

The second type of down-triangle diagram 
involves the intermediate exchange of the 
$\Sigma$ (Fig.~\ref{downtri2}). Its amplitude is
\begin{align*}
V_e=&
\frac{G_Fm_\pi^2D_s}{4\sqrt{3}f_\pi^3}
\intlq
\frac{1}{l^2-m_\pi^2+i\epsilon}
\nonumber\\\times&\nonumber\,
\frac{1}{(l+q)^2-m_\pi^2+i\epsilon}\,
\frac{(2l^\mu+q^\mu)l^\nu}{k_N^2-M_\Sigma^2+i\epsilon}
\\\times&\nonumber\,
\ou_1(\oee,\vpp)(A_{\Sigma}+B_{\Sigma}\gamma_5)(\kn+M_\Sigma)
\gamma_{\nu}\gamma_5 u_1(E_p^\Lambda,\vp)
\\\times&\nonumber\,
\ou_2(\oee_p,-\vpp)\gamma_\mu u_2(E_p,-\vp) \,.
\end{align*}
Using the heavy baryon expansion
\begin{align}
V_e=&
-\frac{G_Fm_\pi^2D_s}{8\sqrt{3}M_N f_\pi^3}
\Big[
B_{\Sigma}\Big(-2I_{30}+(-5q_0-2\Delta M_\Sigma)I_{20}
\\+&\nonumber
2(3-\eta)I_{32}+2\vq^2I_{33}+2\vq^2I_{21}+\vq^2I_{21}
\\+&\nonumber
q_0(-2q_0-\Delta M_\Sigma)I_{10}
+(3-\eta)q_0I_{22}+q_0\vq^2I_{23}+q_0\vq^2I_{11}\Big)
\\-&\nonumber
2A_{\Sigma} M_N(2I_{21}
+q_0I_{11})(\vsu\cdot\vq)
\Big]\,.
\end{align}
The isospin is taken into account by replacing every $A_\Sigma$ and
$B_\Sigma$ by
\begin{align*}
\frac{2}{3}\left(\sqrt3 A_{\Sigma\frac12}+
A_{\Sigma\frac32}\right)
\vec{\tau_1}\cdot\vec{\tau_2},
~~~
\frac{2}{3}\left(\sqrt3 B_{\Sigma\frac12}+
B_{\Sigma\frac32}\right)
\vec{\tau_1}\cdot\vec{\tau_2}\,,
\end{align*}
where, we have used the master integrals with
$q_0=-\frac{M_\Lambda-M_N}{2}$, $q_0'=M_\Sigma-M_\Lambda$ and
$\vq=\vp'-\vp$.

\section{Box diagrams}
\label{sec:boxs}

We have two kind of direct box diagrams and two cross-box 
ones. Direct box diagrams usually present a pinch singularity. 
This is because the poles appearing in the baryonic propagators 
get infinitesimally close to one another. In our integrals 
the denominators appearing in the baryonic propagators also 
contain terms proportional to $M_\Lambda-M_N$ and $M_\Sigma-M_\Lambda$, 
and this avoids the singularity. 

\begin{figure}[th]
\includegraphics[scale=0.4]{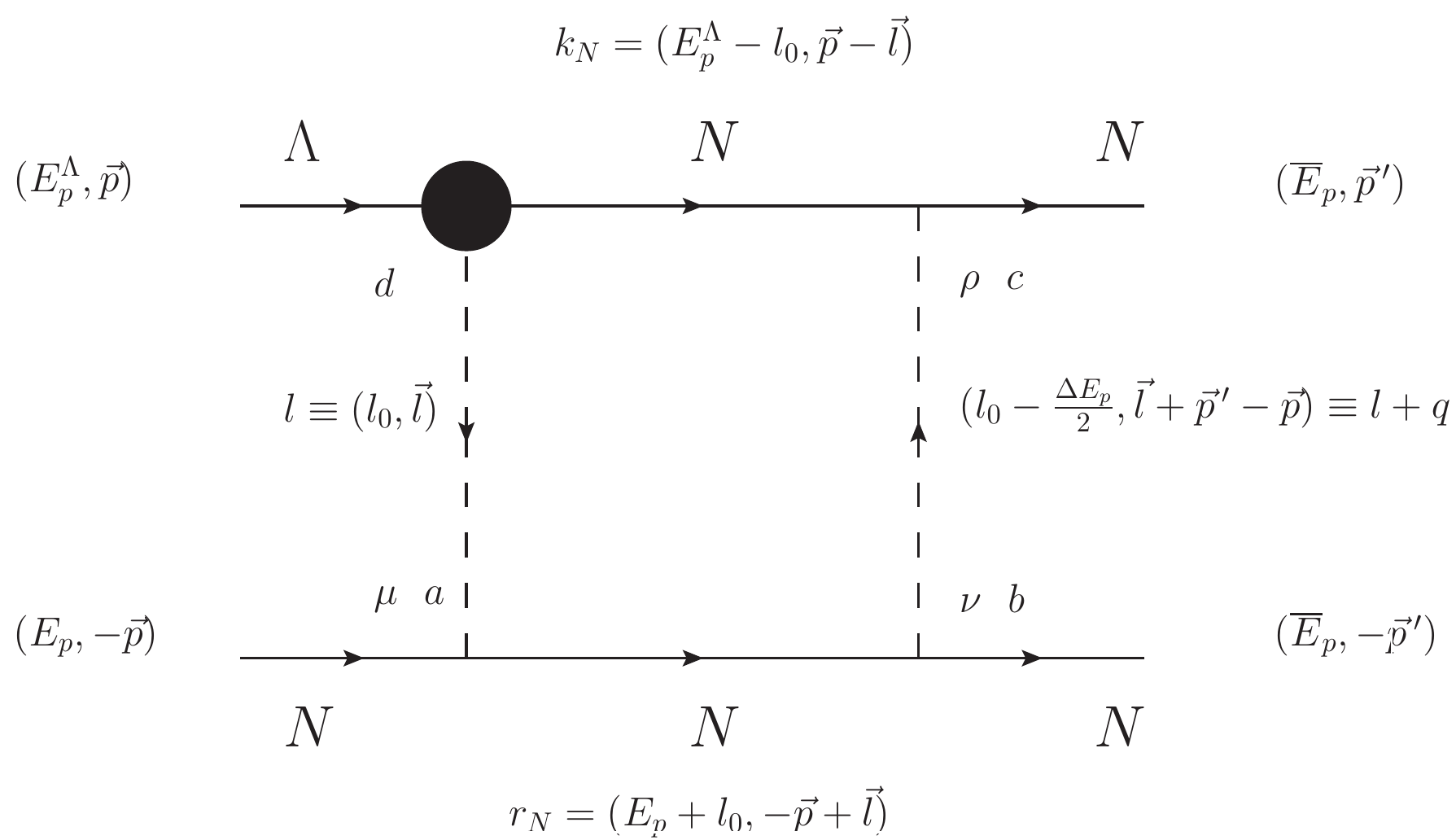}
\caption{Box diagram contributing at NLO.}
\label{box1}
\end{figure}

The integrals entering in the expression of the amplitudes are 
the $J$ and $K$ defined in Appendix~\ref{sec:mi}. The amplitude 
for the first type of box diagram (Fig.~\ref{box1}) is
\begin{align}
V_f=&
i\frac{G_Fm_\pi^2g_A^3}{8f_\pi^3}
(3-2\vtu\cdot\vtd)
\intlq
\frac{1}{l^2-m_\pi^2+i\epsilon}
\nonumber\\\times&\nonumber\,
\frac{1}{(l+q)^2-m_\pi^2+i\epsilon}\,
\frac{1}{k_N^2-M_N^2+i\epsilon}a
\\\times&\nonumber\,
\frac{(l^\rho+q^\rho)(l^\nu+q^\nu)l^\mu}{r_N^2-M_N^2+i\epsilon}
\\\times&\nonumber\,
\ou_1(\oee,\vpp)
\gamma_\rho\gamma_5
(\kn+M_N)
(A+B\gamma_5)
 u_1(E_p^\Lambda,\vp)
\\\times&\nonumber\,
\ou_2(\oee_p,-\vpp)
\gamma_\nu\gamma_5
(\rn+M_N)
\gamma_\mu\gamma_5
 u_2(E_p,-\vp) \,.
\end{align}
Using the heavy baryon expansion,
\begin{align*}
V_f&=-\frac{G_Fm_\pi^2g_A^3}{32M_N f_\pi^3}
(3-2\vtu\cdot\vtd)\Bigg[
-4A M_N \left(4K_{22}
+ K_{11}   \vq^2
\nonumber\right.\\+&\left.\nonumber
2 K_{23} \vq^2+K_{35} \vq^2+(5-\eta)
K_{34}\right)\vsu\cdot \vq
\nonumber\\-&\nonumber
2B K_{22} (\vsu\cdot \vq)(\vsd\cdot \vq)
+
2B K_{22} (\vsu\cdot \vq)(\vsd\cdot \vp)
\nonumber\\-&\nonumber
4i A M_N K_{22} \left(\vsu\times\vsd\right)\cdot\vq
-
2 B \left(\vp\cdot
 \vq-\vq^2\right)
K_{22} \vsu\cdot \vsd
\nonumber\\+&\nonumber
2 i B  \left(K_{11} \vq^2+2
   K_{23} \vq^2+K_{35} \vq^2
\nonumber\right.\\+&\left.\nonumber
(4-\eta) K_{22}+(5-\eta) K_{34}\right)
\vsu\cdot\left(\vp\times \vq\right)
+2 i B K_{22}
   \vsd\cdot\left(\vp\times \vq\right)
\nonumber\\-&\nonumber
2 B \Big(K_{11}\vq^2
   \left(\vp\cdot \vq+2 q_0{}^2\right)+ K_{23}( 2\vp\cdot
   \vq \vq^2+2q_0^2\vq^2+\vq^4)
\nonumber\\+&\nonumber
   K_{35}( \vp\cdot \vq \vq^2+2\vq^4)
   +K_{22}((4-\eta)   \vp\cdot\vq+\vq^2+(6-2\eta)q_0^2)
\nonumber\\+&\nonumber
   (5-\eta)K_{34}(\vp\cdot \vq+2\vq^2)
   +K_{48} \vq^4+K_{21}
   \vq^2 q_0+K_{33} \vq^2 q_0
\nonumber\\-&\nonumber
K_{31} \vq^2
   -K_{43} \vq^2 
   +2(5-\eta) K_{47} \vq^2
   +(3-\eta) K_{32}
   q_0
\nonumber\\-&\nonumber
(3-\eta) K_{42}+(15-8\eta) K_{46}\Big)
\Bigg]\,,
\end{align*}
where we have used the master integrals with
$q_0=-\frac{M_\Lambda-M_N}{2}$, $q_0'=M_N-M_\Lambda$,  and
$\vq=\vp'-\vp$.

\begin{figure}[th]
\includegraphics[scale=0.4]{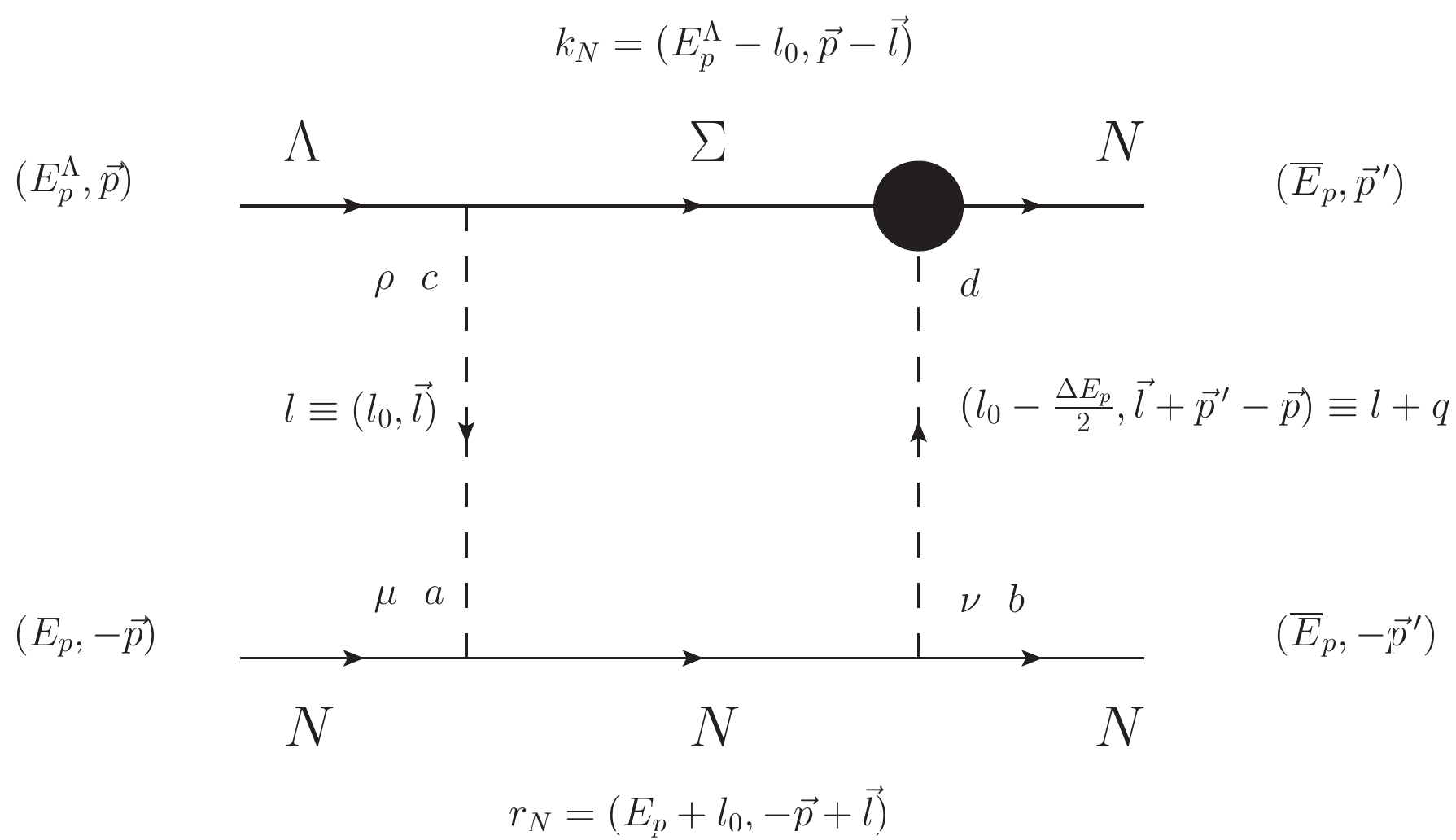}
\caption{Second box-type Feynman diagram.
\label{box2}}
\end{figure}
The second box diagram (Fig.~\ref{box2}), which involves a $\Sigma$
propagator, contributes with 
\begin{align}
V_g=&
-i\frac{G_Fm_\pi^2g_A^2D_s}{4\sqrt{3} f_\pi^3}
\intlq
\frac{1}{l^2-m_\pi^2+i\epsilon}
\\\times&\nonumber\,
\frac{1}{(l+q)^2-m_\pi^2+i\epsilon}\,
\frac{1}{k_N^2-M_\Sigma^2+i\epsilon}
\\\times&\nonumber\,
\frac{l^\rho(l^\nu+q^\nu)l^\mu}{r_N^2-M_N^2+i\epsilon}
\\\times&\nonumber\,
\ou_1(\oee,\vpp)(A_\Sigma+B_{\Sigma}\gamma_5)(\kn+M_N)\gamma_\rho\gamma_5 u_1(E_p^\Lambda,\vp)
\\\times&\nonumber\,
\ou_2(\oee_p,-\vpp)\gamma_\nu\gamma_5(\rn+M_N)\gamma_\mu\gamma_5 u_2(E_p,-\vp)\,.
\end{align}
Using the heavy baryon expansion
\begin{align*}
V_g=&
\frac{G_Fm_\pi^2g_A^2D_s}{16\sqrt{3}M_N f_\pi^3}
\Big[
-2B_{\Sigma} K_{22} \vq^2 \vsu\cdot \vsd
\nonumber\\-&\nonumber\,
 4A_{\Sigma} K_{22} M_N i\left(\vsu\times
  \vsd\right)\vq 
\\-&\nonumber\,
4 A_{\Sigma} M_N\left(\vq^2 K_{23}+5 K_{34}+\vq^2
  K_{35}+K_{22}\right)\vsu\cdot \vq
\\+&\nonumber\,
  2B_{\Sigma} K_{22} (\vsu\cdot \vq)(\vsd\cdot \vq)
+
2 B_{\Sigma} \left(\vq^2 K_{22}+\vq^4
  K_{23}
\right.\\-&\nonumber\left.\,
\vq^2 K_{31}+(3-\eta) (\Delta M-\Delta M_\Sigma) K_{32}
\right.\\+&\nonumber\left.\,
\vq^2 (\Delta M -\Delta M_\Sigma)K_{33}+2(5-\eta) \vq^2 K_{34}
+2 \vq^4 K_{35}
\right.\\-&\nonumber\left.\,
(3-\eta)
  K_{42}-\vq^2 K_{43}+(15-8\eta) K_{46}
\right.\\+&\nonumber\left.\,
2(5-\eta) \vq^2 K_{47}
+
\vq^4K_{48}
+\vq^2 K_{21} \left(\text{$\Delta $M}-\text{$\Delta
      $M}_{\Sigma }\right) \right)
\Big]\,.
\end{align*}
To take into account the isospin we must replace every $A_\Sigma$ and
$B_\Sigma$ by
\begin{align*}
A\to&
-\sqrt3A_{\Sigma\frac12}+2A_{\Sigma\frac32}
+\frac23(\sqrt3A_{\Sigma\frac12}+A_{\Sigma\frac32})\vtu\cdot\vtd
\\
B\to&
-\sqrt3B_{\Sigma\frac12}+2B_{\Sigma\frac32}
+\frac23(\sqrt3B_{\Sigma\frac12}+B_{\Sigma\frac32})\vtu\cdot\vtd\,.
\end{align*}
We have used the master integrals with
$q_0=-\frac{M_\Lambda-M_N}{2}$, $q_0'=M_\Sigma-M_\Lambda$,  and
$\vq=\vp'-\vp$.

\begin{figure}[th]
\includegraphics[scale=0.4]{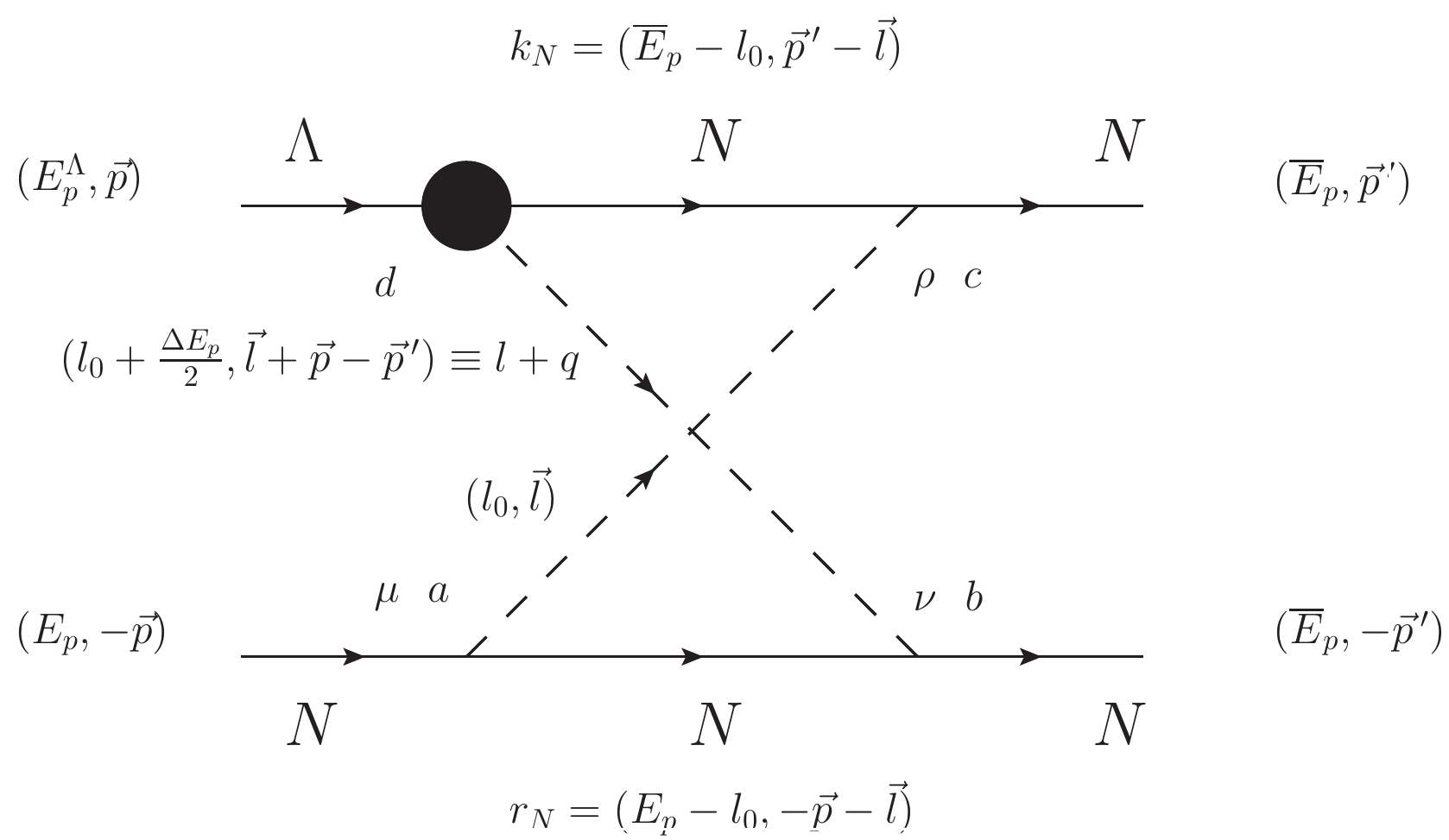}
\caption{Crossed-box diagram contributing at NLO.}
\label{box2g}
\end{figure}
The second crossed box diagram (Fig.~\ref{box2g}) includes a
$\Sigma$-propagator and contributes to the potential with
\begin{align}
V_h=&
i\frac{G_Fm_\pi^2g_A^3}{8f_\pi^3}
(3+2\vtu\cdot\vtd)
\intlq
\frac{1}{(l+q)^2-m_\pi^2+i\epsilon}
\nonumber\\\times&\nonumber\,
\frac{1}{l^2-m_\pi^2+i\epsilon}\,
\frac{1}{r_N^2-M_N^2+i\epsilon}\,
\frac{(l^\rho)(l^\nu+q^\nu)(l^\mu)}{k_N^2-M_N^2+i\epsilon}
\\\times&\nonumber\,
\ou_1(\oee,\vpp)
\gamma_\rho\gamma_5(\kn+M_N)(A+B\gamma_5) u_1(E_p^\Lambda,\vp)
\\\times&\nonumber\,
\ou_2(\oee_p,-\vpp)\gamma_\nu\gamma_5(\rn+M_N)\gamma_\mu\gamma_5
 u_2(E_p,-\vp)\,.
\end{align}
Using heavy baryon expansion and the master integrals of
Sec.~\ref{sec:mi}, and redefining $\vq\equiv\vpp-\vp$,
\begin{align}
V_h=&
-\frac{G_Fm_\pi^2g_A^3}
{32M_N f_\pi^3}
(3+2\vtu\cdot\vtd)\Big[
-2 iB J_{22}\vsd \left(\vp\times \vq\right)
\nonumber\\+&
2B J_{22} \left(-\vp\cdot
  \vq+\vq^2\right)
\vsu\cdot \vsd 
\\+&\nonumber
2 i B\left(J_{22}+\vq^2 J_{23}+(5+\eta) J_{34}+\vq^2
  J_{35}\right)
\vsu\cdot\left(\vp\times \vq\right)
\\+&\nonumber
4i A J_{22} M_N\left(\vsu\times \vsd\right)\vq
\\+&\nonumber
4 AM_N \left(\vq^2 J_{23}+5 J_{34}+\vq^2
  J_{35}+J_{22}\right)\vsu\cdot\vq
\\+&\nonumber
2BJ_{22} (\vsu\cdot \vq)(\vsd\cdot \vp)
-2BJ_{22} (\vsu\cdot \vq)(\vsd\cdot\vq)
\\-&\nonumber
2B\left(\vq^2 q_0 J_{21}+\left(-\vp\cdot \vq+\vq^2\right)
  J_{22}+(-\vp\cdot \vq \vq^2 +\vq^4)
  J_{23}
\right.\\-&\left.\nonumber
\vq^2 J_{31}+(3-\eta) q_0 J_{32}+\vq^2 q_0 J_{33}
\right.\\+&\left.\nonumber
(5-\eta)(-\vp\cdot \vq +2 \vq^2 )J_{34}
+
(-\vp\cdot \vq  \vq^2 +2 \vq^4 )J_{35}
\right.\\-&\left.\nonumber
(3-\eta) J_{42}
-\vq^2 J_{43}+(15-8\eta)J_{46}
+
2(5-\eta) \vq^2 J_{47}
\right.\\+&\left.\nonumber
\vq^4 J_{48}\right)
\Big]\,.
\end{align}
We have used the master integrals with
$q_0=\frac{M_\Lambda-M_N}{2}$, $q_0'=-\frac{M_\Lambda-M_N}{2}$,  and
$\vq=\vp'-\vp$.

\begin{figure}[th]
\includegraphics[scale=0.4]{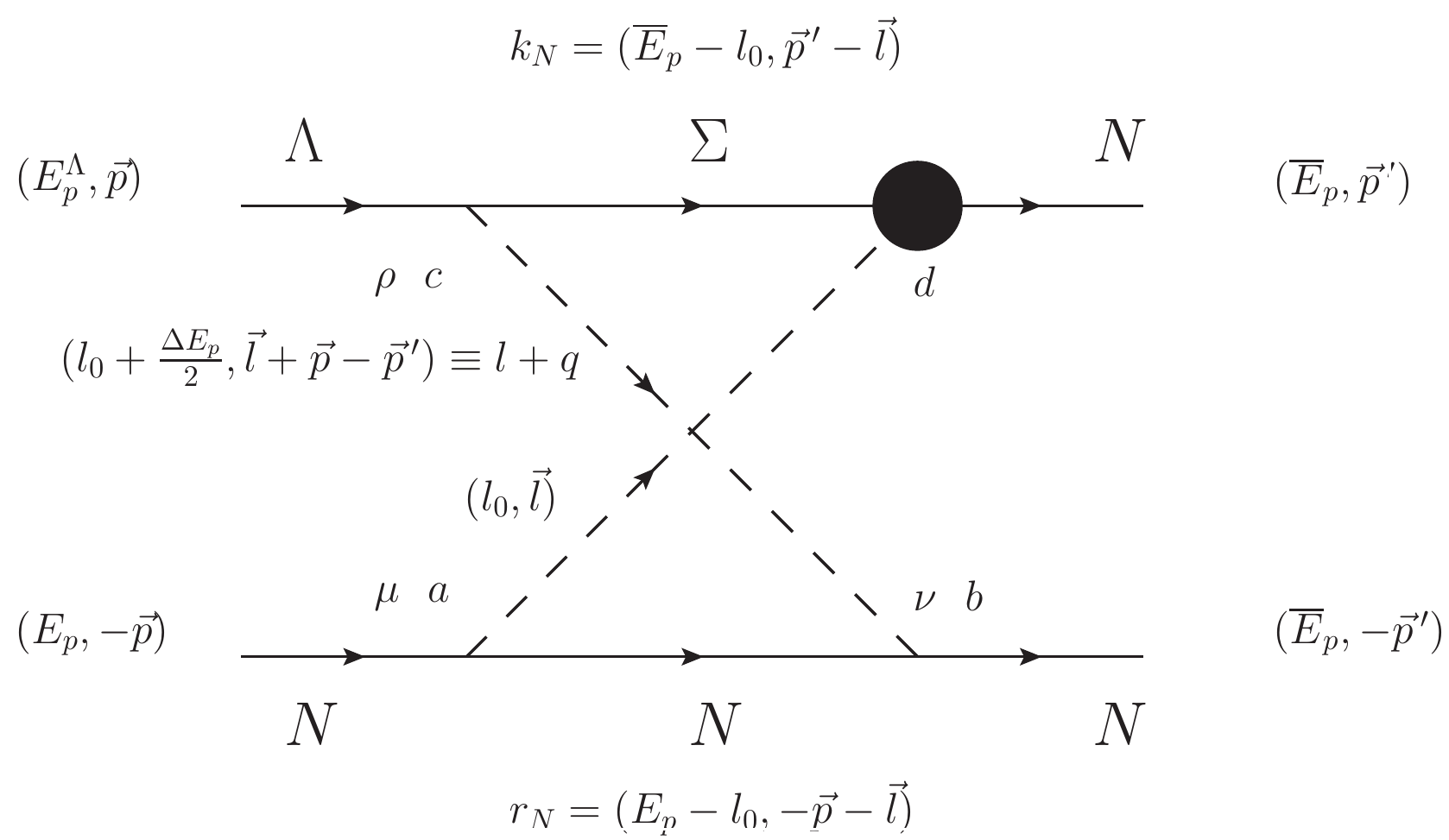}
\caption{Second crossed-box-type Feynman diagram
\label{xbox2}}
\end{figure}
The amplitude for the crossed-box diagram with a $\Sigma$ propagator
is
\begin{align}
V_i=&
-i\frac{G_Fm_\pi^2g_A^2D_s}{16\sqrt{3}M_N^2f_\pi^3}
\intlq
\frac{1}{(l+q)^2-m_\pi^2+i\epsilon}
\nonumber\\\times&\,
\frac{1}{l^2-m_\pi^2+i\epsilon}\,
\frac{1}{r_N^2-M_N^2+i\epsilon}
\\\times&\nonumber\,
\frac{(l^\rho+\vq^\rho)(l^\nu+q^\nu)(l^\mu)}{k_N^2-M_\Sigma^2+i\epsilon}
\\\times&\nonumber\,
\ou_1(\oee,\vpp)(A_\Sigma+B_\Sigma\gamma_5)(\kn+M_\Sigma)\gamma_\rho\gamma_5
 u_1(E_p^\Lambda,\vp)
\\\times&\nonumber\,
\ou_2(\oee_p,-\vpp)\gamma_\nu\gamma_5(\rn+M_N)\gamma_\mu\gamma_5
 u_2(E_p,-\vp)\,.
\end{align}
Using heavy baryon expansion and the master integrals of
Sec.~\ref{sec:mi}, and redefining $\vq\equiv\vpp-\vp$,
\begin{align*}
V_i=&
\frac{G_Fm_\pi^2g_A^2D_s}{16\sqrt{3}M_N f_\pi^3}
\Big[
2B_\Sigma J_{22}\vq^2 \vsu\cdot \vsd
\\-&\nonumber
 2iA_\Sigma J_{22} M_N
\left(\vsu\times \vsd\text{)$\cdot$}\vq\right.
\\-&\nonumber
 A_\Sigma M_N \left(\vq^2 J_{11}+2 \vq^2 J_{23}+5
    J_{34}+\vq^2 J_{35}+4J_{22}\right)\vsu\cdot\vq
\\-&\nonumber
2B_\Sigma J_{22} (\vsu\cdot \vq)(\vsd\cdot \vq) 
\\+&\nonumber
2B_\Sigma \left(
(\vq^2-(3-\eta)q_0(q_0+\Delta M_\Sigma) J_{22}
\right.\\+&\left.\nonumber
(\vq^4-\vq^2q_0^2-\vq^2q_0\Delta M_\Sigma)J_{23}
-\vq^2 J_{31}
\right.\\-&\left.\nonumber
(3-\eta)(2q_0+\Delta M_\Sigma)J_{32}
-(2 \vq^2 q_0 +\vq^2\Delta M_\Sigma)J_{33}
\right.\\+&\left.\nonumber
2(5-\eta)
  \vq^2 J_{34}
+2 \vq^4 J_{35}-(3-\eta) J_{42}-\vq^2 J_{43}
\right.\\+&\left.\nonumber
(15-8\eta)J_{46}+
2(5-\eta) \vq^2 J_{47}+\vq^4 J_{48}
\right.\\-&\left.\nonumber
  \vq^2 q_0 J_{11}
  \left(q_0+\text{$\Delta $M}_{\Sigma }\right)-\vq^2 J_{21}
  \left(2 q_0+\text{$\Delta $M}_{\Sigma }\right)\right)
\Big]\,.
\end{align*}
To take into account the isospin we must replace every $A_\Sigma$ and
$B_\Sigma$ by
\begin{align*}
A_\Sigma\to&
-\sqrt3A_{\Sigma\frac12}+2A_{\Sigma\frac32}
-\frac23(\sqrt3A_{\Sigma\frac12}+2A_{\Sigma\frac32})\vtu\cdot\vtd
\\
B_\Sigma\to&
-\sqrt3B_{\Sigma\frac12}+2B_{\Sigma\frac32}
-\frac23(\sqrt3B_{\Sigma\frac12}+2B_{\Sigma\frac32})\vtu\cdot\vtd\,.
\end{align*}
We have used the master integrals with
$q_0=\frac{M_\Lambda-M_N}{2}$,
$q_0'=M_\Sigma-M_\Lambda+\frac{M_\Lambda-M_N}{2}$,  and 
$\vq=\vp'-\vp$.

\clearpage
\section{Master integrals}
\label{sec:mi}
\subsection{Definitions}
We need the following integrals in order to calculate the 
Feynman diagrams. The $B$'s, $I$'s, $J$'s and $K$'s appear, 
respectively, in the ball, triangle, box and crossed box 
diagrams: 
\[
B_{;\mu;\mu\nu}
\equiv\,
\frac{1}{i}\intl \frac{1 }{l^2-m^2+i\epsilon}\,
\frac{(1;l_\mu;l_\mu l_\nu)}{(l+q)^2-m^2+i\epsilon} \,,
\]
\begin{align*}
I_{;\mu;\mu\nu;\mu\nu\rho}
\equiv\,&
\frac{1}{i}
\intl \frac{1}{l^2-m^2+i\epsilon}\,
\frac{1}{(l+q)^2-m^2+i\epsilon}\,
\nonumber\\&\nonumber
\times\frac{1}{-l_0-q_0'+i\epsilon}
(1;\lm;\lm\lnn;\lm\lnn\lr) \,,
\end{align*}
\begin{align*}
J_{;\mu;\mu\nu;\mu\nu\rho}
\equiv\,&
\frac1i\intl \frac{1}{l^2-m^2+i\epsilon}\,
\frac{1}{(l+q)^2-m^2+i\epsilon}\,
\\&\times
\frac{1}{-l_0-q_0'+i\epsilon}
\,\frac{(1;\lm;\lm\lnn;\lm\lnn\lr)}{-l_0+i\epsilon} \,,
\end{align*}
\begin{align*}
K_{;\mu;\mu\nu;\mu\nu\rho}
\equiv\,&
\frac1i\intl \frac{1}{l^2-m^2+i\epsilon}\,
\frac{1}{(l+q)^2-m^2+i\epsilon}\,
\\&\times
\frac{1}{-l_0-q_0'+i\epsilon}
\frac{(1;\lm;\lm\lnn;\lm\lnn\lr)}{l_0+i\epsilon} \,.
\end{align*}
The strategy is to calculate explicitly the integrals with 
no subindex (no integrated momenta in the numerators), and 
then relate the others to simpler integrals. To do so we 
also need to explicitly calculate the following integrals:
\[
A(m)\equiv\,
\frac{1}{i}
\intl\frac{1}{l^2-m^2+i\epsilon} \,,
\]
\begin{align*}
A_{;\mu;\mu\nu}(q,q')\equiv\,&
\frac{1}{i}\intl\frac{1}{(l+q)^2-m^2+i\epsilon}
\\&\times
\frac{1}{-l_0-q_0'+i\epsilon}(1;l_\mu;l_\mu l_\nu) \,,
\end{align*}
\begin{align*}
C_{;\mu;\mu\nu;\mu\nu\rho}(q_0,q_0')\equiv\,&
\frac{1}{i}\intl\,
\frac{1}{(l+q)^2-m^2+i\epsilon}\,
\\&\times
\frac{1}{-l_0-q_0'+i\epsilon}
\frac{(1;l_\mu;l_\mu l_\nu;l_\mu l_\nu l_\rho)}{-l_0+i\epsilon} \,,
\end{align*}
\begin{align*}
D_{;\mu;\mu\nu;\mu\nu\rho}(q_0,q_0')\equiv\,&
\frac{1}{i}\intl\,
\frac{1}{(l+q)^2-m^2+i\epsilon}
\\&\times
\frac{1}{-l_0-q_0'+i\epsilon}
\frac{(1;l_\mu;l_\mu l_\nu;l_\mu l_\nu l_\rho)}{l_0+i\epsilon} \,.
\end{align*}
The integrals can be divided depending on their subindexes 
being temporal or spatial. We show explicitly all the cases 
for the integrals $J$. The same definitions are used for all 
the other integrals. Therefore, to know any other integral 
one needs to replace in Eq.~(\ref{eq:many}) $J$ by $A$, $B$, $I$, etc.
\begin{align}
J_\mu\equiv\,&
\delta_{\mu0}J_{10}+\delta_{\mu i}J_{11}\vq_i \label{eq:many}\\\nonumber\\
J_{\mu\nu}\equiv\,&
\delta_{\mu0}\delta_{\nu0}J_{20}
+(\delta_{\mu0}\delta_{\nu i}
+\delta_{\mu i}\delta_{\nu 0})J_{21}\vq_i
\nonumber\\&
+\delta_{\mu i}\delta_{\nu j}(J_{22}\delta_{ij}
+J_{23}\vq_i\vq_j)
\nonumber\\\nonumber\\
J_{\mu\nu\rho}\equiv\,&
\delta_{\mu0}\delta_{\nu0}\delta_{\rho0}J_{30}
+\delta\delta\delta_{\{\mu\nu\rho 00i\}}\vq_iJ_{31}
\nonumber\\&
+\delta\delta\delta_{\{\mu\nu\rho 0ij\}}
(\delta_{ij}J_{32}+\vq_i\vq_jJ_{33})
\nonumber\\&
+\delta_{\mu i}\delta_{\nu j}\delta_{\rho k}
(\delta\vq_{\{ijk\}}J_{34}+\vq_i\vq_j\vq_kJ_{35})
\nonumber\\\nonumber\\
J_{\mu\nu\rho\sigma}\equiv\,&
\delta_{\mu0}\delta_{\nu0}\delta_{\rho0}\delta_{\sigma0}J_{40}
+\delta\delta\delta\delta_{\{\mu\nu\rho\sigma000i\}}\vq_iJ_{41}
\nonumber\\&
+\delta\delta\delta\delta_{\{\mu\nu\rho\sigma00ij\}}
(\delta_{ij}J_{42}+\vq_i\vq_jJ_{43})
\nonumber\\&
+\delta\delta\delta\delta_{\{\mu\nu\rho\sigma0ijk\}}
(\delta\vq_{\{ijk\}}J_{44}+\vq_i\vq_j\vq_kJ_{45})
\nonumber\\&
+\delta_{\mu i}\delta_{\nu j}\delta_{\rho k}\delta_{\sigma l}
(\delta\delta_{\{ijkl\}}J_{46}
+\delta\vq\vq_{\{ijkl\}}J_{47}
\nonumber\\&
+\vq_i\vq_j\vq_k\vq_lJ_{48})\,. \nonumber
\end{align}
All coefficients $J_{10}$, $J_{11}$, etc. have been written 
explicitly as functions of $I$, $J$, $K$, which can be 
integrated numerically, and the other simpler functions.
The following definitions have been employed:
\begin{align*}
\delta\vq_{\{ijk\}}=&\,
\delta_{ij}\vq_k
+\delta_{ik}\vq_j
+\delta_{jk}\vq_i\,,
\\
\delta\vq\vq_{\{ijkl\}}=&\,
\delta_{ij}\vq_k\vq_l
+\delta_{ik}\vq_j\vq_l
+\delta_{il}\vq_j\vq_k \nonumber\\
&
+\delta_{jk}\vq_i\vq_l
+\delta_{jl}\vq_i\vq_k
+\delta_{kl}\vq_i\vq_j\,,
\\
\delta\delta_{\{ijkl\}}=&\,
\delta_{ij}\delta_{kl}
+\delta_{ik}\delta_{jl}
+\delta_{il}\delta_{jk}\,.
\end{align*}
The other quantities, $\delta\delta\delta_{\{\mu\nu\rho00i\}}$,
$\delta\delta\delta_{\{\mu\nu\rho0ij\}}$, etc, are not meant to be
contracted with the indexes $i$, $j$, and $k$ appearing in the rest of
the expressions. They only indicate how many of the indexes $\mu$,
$\nu$, $\rho$, and $\sigma$ must be temporal and how many spatial.
It does not matter the order in which $0$, $i$, $j$, and $k$ are
assigned to $\mu$, $\nu$, $\rho$, and $\sigma$, since all the
integrals $J_{\mu\nu}$, $J_{\mu\nu\rho}$, etc, are symmetric with 
respect to these indexes. For example
\[
J_{00i}=J_{0i0}=J_{i00}=\vq_i J_{31} \,.
\]

\subsection{Results for the master integrals}

We have regularized the master integrals via dimensional
regularization, where the integrals depend on the momentum 
dimension $D_\eta$, or more specifically, on the parameter 
$\eta$, defined through $D_\eta=4-\eta$, and on the renormalization
scale $\mu$, for which we have taken $\mu=m_\pi$.
In the following 
we use,
\begin{align*}
R=&-\frac{2}{\eta}-1+\gamma-\log(4\pi) \,,
\\
q_0''=&q_0'-q_0 \,.
\end{align*}
The integrals $A(m)$, $A(q_0,q_0')$ and $B(q_0,|\vq|)$ 
appear, for example, in~\cite{scherer02}. We have checked 
that both results coincide. It is important to maintain 
the $-i\epsilon$ prescription, otherwise the integrals 
may give a wrong result. We take it into account by 
replacing $q_0'\to q_0'-i\epsilon$ when evaluating the 
integrals.

\subsubsection{$A(m), A(q_0,q'_0)$ and $B(q_0,\vq)$}
We have, 
\begin{equation}
A(m)=
-\frac{1}{8\pi^2}m^2\left(\frac12R+\log\left(\frac{m}{\mu}\right)\right) \,.
\end{equation}
\begin{align}
A(q_0,q_0')\equiv
-\frac{q_0''}{8\pi^2}
\left[
\pi\frac{\sqrt{m^2-q_0''^2}}{q_0''}
+1-R-2\log\left(\frac{m}{\mu}\right)
\right. & \nonumber\\\left.
-\frac{2 \sqrt{{q_0''}^2(m^2-{q_0''}^2)} \tan
   ^{-1}\left(\frac{\sqrt{{q_0''}^2}}{\sqrt{m^2-{q_0''}^2
   }}\right)}{{q_0''}^2} 
\right] &
\end{align}
\begin{align}
B(q_0,\vq)&=
-\frac{1}{16\pi^2}\left[-1+R+2\log\left(\frac{m}{\mu}\right)+2L(|q|)\right]
\end{align}
with
\begin{align*}
L(|q|)\equiv &\frac{w}{|q|}\log\left(\frac{w+|q|}{2m}\right) \,,
\end{align*} 
$w\equiv \sqrt{4m^2+|q|^2}$, $|q|\equiv \sqrt{\vq^2-q_0^2}$, and 
$q^2\equiv q_0^2-\vq^2\le0$. 

\subsubsection{$C(q_0,q_0')$ and $D(q_0,q_0')$}
\vspace{-20pt}
\begin{align*}
C(q_0,q_0')\equiv&
-\frac{1}{16\pi^2}\intx\inty
\Bigg[
3y^{-\frac12}(1-y)
\\&
\left[
-\frac43
-\frac12(R-1+\log(4))
-\frac12\log\left(\frac{s_{xy}}{4\mu^2}\right)
\right]
\\&+y^{-\frac12}(1-y)(m^2+q_0''r_0')
s_{xy}^{-1}
\\&
-\pi(q_0''+r_0')s_x^{-\frac12}
\Bigg] \,,
\end{align*}
with
$s_x=m^2-q_0^2+x(q_0^2-q_0''^2)$, 
$s_{xy}=m^2+(1-y)(-q_0^2+x(q_0^2-q_0''^2))$.

\begin{align*}
D(q_0,q_0')=&
-C(q_0,q_0')+\frac{1}{q_0'}\frac{1}{4\pi}\sqrt{m^2-q_0^2} \,.
\end{align*}

\subsubsection{$I(q_0,|\vq|,q_0')$}
\vspace{-20pt}
\begin{align*}
I(q_0,q,q_0')&=
-\frac{1}{8\pi^2}\int_0^1dx\int_0^1dy
\left[
\frac{\pi}{2}\frac{1}{\sqrt{s_x}}
\right.\\&\left.
-\frac34y^{-\frac12}(1-y)\cqp\frac{1}{s_{xy}}
+\frac12y^{\frac12}(1-y)\cqp^3\frac{1}{s_{xy}^2}
\right]
\end{align*}
with
$C_q'=-q_0(1-x)+q_0'$,  
$s_x\equiv -q^2x(1-x)-\left(q_0'-q_0+q_0x\right)^2+m_\pi^2$, and 
$s_{xy}\equiv -q^2x(1-x)-\left(q_0'-q_0+q_0x\right)^2(1-y)+m_\pi^2$.

\subsubsection{$J(q_0,|\vq|,q_0')$ and $K(q_0,|\vq|,q_0')$}
\vspace{-20pt}
\begin{align*}
J\ignore{(q_0,q,q_0',r_0')}=&
-\frac{1}{8\pi^2}\int_0^1dx\inty\,y(1-y)
\Bigg\{
\left(-\cqp^3-\cqp^2\cq
\right.\\&\left.
-\cqp\cq^2-\cq^3
+2 s_x (\cqp+\cq)\right)\frac{3\pi}{8s_{xy}^{\frac52}}
\\&
+(\cqp+\cq)\frac{\pi}{8s_{xy}^{\frac32}}
+
\frac{105}{16}\intz\,z^3\sqrt{1-z}
\Big[
-\frac{3}{s_{xyz}}
\\&
+\left(\cqp^2-5 \cqp\cq+\cq^2-9s_x\right)\frac{2}{7s_{xyz}^2}
\\&+
\left(-9s_x^2+2s_x\left(\cqp^2-5 \cqp\cq+\cq^2\right)+
\right.\\&\left.
3 \cqp^3\cq+\cqp^2\cq^2+3 \cqp \cq^3\right)\frac{8}{35s_{xyz}^3}
\\&+\left(-3s_x^3+s_x^2\left(\cqp^2-5 \cqp\cq+\cq^2\right)
\right.\\&\left.
+s_x\left(3 \cqp^3\cq+\cqp^2\cq^2+3 \cqp \cq^3\right)
\right.\\&\left.
-\cqp^3 \cq^3\right)\frac{16}{35s_{xyz}^4}
\Big]
\Bigg\}
\end{align*}
with
$C_{q}\equiv -q_0(1-x)$, 
$C_q'\equiv -q_0(1-x)+q_0'$, 
$s_x\equiv -q^2x(1-x)+m_\pi^2$, 
$s_{xy}\equiv s_x-\cq^2+y(\cq^2-\cqp^2)$, and 
$
s_{xyz}\equiv s_x+z\cdot y(\cq^2-\cqp^2)-z\cq^2$.
\begin{align*}
K&=
-J
+\frac{1}{8\pi q_0'}\int_0^1 dx
\frac{1}{\sqrt{m^2+(1-x)(\vq^2x-q_0^2)}} \,.
\end{align*}

\subsection{Results for the master integrals with $q_0=q_0'=0$}

\begin{align*}
A(m)&=
-\frac{1}{8\pi^2}m^2\left(\frac12R+\log\left(\frac{m}{\mu}\right)\right)
\\
A(0,0)&=-\frac{m}{8\pi}
\\
B(0,\vq)&=
-\frac{1}{16\pi^2}\left[-1+R+2\log\left(\frac{m}{\mu}\right)+2L(q)\right]
\\
C(0,0)&=-\frac{1}{4\pi^2}
\left(
-\frac{R}{2}-\frac12
-\log(\frac{m}{\mu})\right)
\\
I(0,\vq,0)&=
-\frac{1}{4\pi}At(q)
\\
J(0,\vq,0)&=\frac{1}{2\pi^2\vq^2}L(q),
\end{align*}
where $L(q)$ and $At(q)$ are defined with
\begin{align*}
At(q)\equiv&\frac{1}{2q}\arctan\left(\frac{q}{2m_\pi}\right)
\\
L(q)\equiv&\frac{\sqrt{4m_\pi^2+q^2}}{q}\log\left(\frac{\sqrt{4m_\pi^2+q^2}+q}{2m_\pi}\right)\,.
\end{align*}

\subsection{Relations between master integrals}

\subsubsection{$A_\mu(q_0,q_0')$}

\begin{align*}
A_{10}&=-A(m)-q_0'A
\\
A_{11}&=-A
\end{align*}

\subsubsection{$A_{\mu\nu}(q,q')$}

\begin{align*}
A_{20}&=
\left[
(q_0+q_0')A(m)+
{q_0'}^2A\right]
\\A_{21}&=
A(m)+q_0'A
\\A_{22}&=
\frac{1}{D_\eta-1}\left[
q_0''A(m)+({q_0''}^2-m^2)A\right]
\\A_{23}&=A
\end{align*}

\subsubsection{$B_\mu(q)$}

\begin{align*}
B_{10}&=-\frac{q_0}{2}B
\\
B_{11}&=-\frac12B
\end{align*}

\subsubsection{$B_{\mu\nu}(q)$}

\begin{align*}
B_{20}&=
\frac{1}{2(D_\eta-1)q^2}\Bigg[
(q^2+q_0^2(D_\eta-2))A(m)
\\&
-\left(2\vq^2m^2+\frac12q^2(q^2-D_\eta q_0^2)\right)B
\Bigg]
\\\\
B_{21}&=
\frac{q_0}{2(D_\eta -1)q^2}\left[
(D_\eta -2)A(m)+\left(\frac{ D_\eta}{2}q^2-2m^2\right)B\right]
\\\\
B_{22}&=
-\frac{1}{2(D_\eta -1)}\left[A(m)+\left(2m^2-\frac{q^2}{2}\right)B\right]
\\\\
B_{23}&=
\frac{1}{2(D_\eta -1)q^2}\left[
(D_\eta -2)A(m)+\left(\frac{D_\eta}{2}q^2-2m^2\right)B
\right]
\end{align*}

\subsubsection{$C_\mu(q_0,q_0')$}

\begin{align*}
C_{10}&=-A
\\
C_{11}&=-C
\end{align*}

\subsubsection{$C_{\mu\nu}(q_0,q_0')$}

\begin{align*}
C_{20}&=-A_{10}
\\
C_{21}&\equiv
-A_{11}
\\
C_{22}&=\frac{1}{D_\eta-1}(C_{20}+2q_0C_{10}+(q_0^2-m^2)C)
\\
C_{23}&=C
\end{align*}

\subsubsection{$C_{\mu\nu\rho}(q_0,q_0')$}

\begin{align*}
C_{30}&=-A_{20}
\\
C_{31}&=-A_{21}
\\
C_{32}&=-A_{22}
\\
C_{33}&=-A_{23}
\\
C_{34}&\equiv-C_{22}
\\
C_{35}&=-6C_{11}-3C_{23}-4C
\end{align*}

\subsubsection{$D_\mu(q_0,q_0')$}

\begin{align*}
D_{10}&=A
\\
D_{11}&=-D
\end{align*}

\subsubsection{$D_{\mu\nu}(q_0,q_0')$}

\begin{align*}
D_{20}&\equiv A_{10}
\\
D_{21}&\equiv A_{11}
\\
D_{22}&=\frac{1}{D_\eta-1}(D_{20}+2q_0D_{10}+(q_0^2-m^2)D)
\\
D_{23}&=D
\end{align*}

\subsubsection{$D_{\mu\nu\rho}(q_0,q_0')$}

\begin{align*}
D_{30}&\equiv A_{20}
\\
D_{31}&\equiv A_{21}
\\
D_{32}&\equiv A_{20}
\\
D_{33}&\equiv A_{21}
\\
D_{34}&\equiv -D_{22}
\\
D_{35}&\equiv-6D_{11}-3D_{23}-4D
\end{align*}

\subsubsection{$I_\mu$}

\begin{align*}
I_{10}&=
-B-q_0'I
\\
I_{11}&=
\frac{1}{2\vq^2}
\left[-A(0,q_0',r_0)+A
-2q_0B+(q_0^2-\vq^2-2q_0q_0')I
\right]
\end{align*}

\subsubsection{$I_{\mu\nu}$}

\begin{align*}
I_{20}&=-B_{10}-q_0'I_{10}
\\
I_{21}&=-B_{11}-q_0'I_{11}
\\
I_{22}&=\frac{1}{(D_\eta -2)\vq^{\,2}}
\left[-I_{(\vl\cdot\vq)^2}+\vq^2I_{(\vl^2)}\right]
\\
I_{23}&=\frac{1}{(D_\eta -2)\vq^{\,4}}
\left[(D_\eta -1)I_{(\vl\cdot\vq)^2}-\vq^2I_{(\vl^2)}\right]
\end{align*}
\begin{align*}
I_{(\vl^2)}&=
-A(q,q')-m^2I_0-B_{10}-q_0'I_{10}
\\
I_{(\vl\cdot\vq)^2}&=
\frac12\vq^2
\left[
A_{11}(q,q')
-2q_0B_{11}+(q^2-2q_0q_0')I_{11}
\right]
\end{align*}

\subsubsection{$I_{\mu\nu\rho}$}

\begin{align*}
I_{30}&=-B_{20}-q_0'I_{20}
\\\\
I_{31}&=-B_{21}-q_0'I_{21}
\\\\
I_{32}&=
-B_{22}-q_0'I_{22}
\\\\
I_{33}&=
-B_{23}-q_0'I_{23}
\\\\
I_{34}&=
\frac{-I_{(\vl\cdot\vq)^3}+\vq^2I_{(\vl\cdot\vq)\vl^2}}
{\vq^4(D_\eta -2)}
\\\\
I_{35}&=
\frac{(D_\eta +1)I_{(\vl\cdot\vq)^3}-3\vq^2I_{(\vl\cdot\vq)\vl^2}}
{\vq^6(D_\eta -2)}
\end{align*}

\begin{align*}
I_{(\vl\cdot\vq)^3}&=
\frac12\vq^2
\left[-A_{22}(0,q_0')-\vq^2A_{23}(0,q_0')-\vq^2A(0,q_0')
\right.\\&-2\vq^2  A_{11}(0,q_0')
A_{22}+\vq^2A_{23}+q^2I_{22}+q^2\vq^2I_{23}
\\&\left.
-2q_0B_{22}-2q_0\vq^2B_{23}
-2q_0q_0'I_{22}
-2q_0q_0'\vq^2I_{23}
\right]
\end{align*}

\begin{align*}
I_{(\vl\cdot\vq)\vl^2}&=
\vq^2\left(
-A_{11}-m^2I_{11}-B_{21}-q_0'I_{21}
\right)
\end{align*}

\subsubsection{$J_\mu$}
\begin{align*}
J_{10}&\equiv-I
\\
J_{11}&\equiv \frac{1}{2\vq^2}
\left[
-C(0,q_0')+C
-2q_0I+q^2J\right]
\end{align*}

\subsubsection{$J_{\mu\nu}$}

\begin{align*}
J_{20}&\equiv -I_{10}
\\
J_{21}&\equiv -I_{11}
\\
J_{22}&\equiv\frac{1}{(D_\eta -2)\vq^{\,2}}
\left[-J_{(\vl\cdot\vq)^2}+\vq^2J_{(\vl^2)}\right]
\\
J_{23}&\equiv\frac{1}{(D_\eta -2)\vq^{\,4}}
\left[(D_\eta -1)J_{(\vl\cdot\vq)^2}-\vq^2J_{(\vl^2)}\right]
\end{align*}

\begin{align*}
J_{(\vl^2)}&=
-C-m^2J-I_{10}
\\\\
J_{(\vl\cdot\vq)^2}&=
\frac12
\left[
C_{11}+q^2J_{11}
-2q_0I_{11}
\right]\vq^2
\end{align*}

\subsubsection{$J_{\mu\nu\rho}$}

\begin{align*}
J_{30}&\equiv-I_{20}
\\
J_{31}&\equiv-I_{21}
\\
J_{32}&\equiv-I_{22}
\\
J_{33}&\equiv-I_{23}
\\
J_{34}&\equiv
\frac{-J_{(\vl\cdot\vq)^3}+\vq^2J_{(\vl\cdot\vq)\vl^2}}
{\vq^4(D_\eta -2)}
\\
J_{35}&\equiv
\frac{(D_\eta +1)J_{(\vl\cdot\vq)^3}-3\vq^2J_{(\vl\cdot\vq)\vl^2}}
{\vq^6(D_\eta -2)}
\end{align*}

\begin{align*}
J_{(\vl\cdot\vq)^3}&=
\frac{\vq^2}{2}\left[
-C_{20}(0,q_0')
-\vq^2C_{21}(0,q_0')
-\vq^2C(0,q_0')
\right.\\&\left.
-2\vq^2C_{11}(0,q_0')
+C_{20}+C_{21}\vq^2
\right.\\&\left.
+q^2(J_{22}+J_{23}\vq^2)
-2q_0(I_{22}+I_{23}\vq^2)
\right]
\\
J_{(\vl\cdot\vq)\vl^2}&=
-\vq^2\left[C_{11}
+m^2J_{11}
+I_{21}
\right]
\end{align*}
\subsubsection{$J_{\mu\nu\rho\sigma}$}

\begin{align*}
J_{40}&\equiv -I_{30}
\\
J_{41}&\equiv -I_{31}
\\
J_{42}&\equiv -I_{32}
\\
J_{43}&\equiv-I_{33}
\\
J_{44}&\equiv -I_{34}
\\
J_{45}&\equiv -I_{35}
\\
J_{46}&=
2\,\frac{-J_{\vl^2(\vl\cdot\vq)^2}+\vq^2J_{\vl^4}}
{\vq^2(D-2)(2D+3)}
\\J_{47}&=
\frac{-(2D+3)J_{(\vl\cdot\vq)^4}+2(2+D)\vq^2J_{\vl^2(\vl\cdot\vq)^2}-\vq^4J_{\vl^4}}
{\vq^6(D-2)(2D+3)}
\\J_{48}&=
\frac{(D+4)J_{(\vl\cdot\vq)^4}-6\vq^2J_{\vl^2(\vl\cdot\vq)^2}}
{\vq^8(D-2)}
\end{align*}
\begin{align*}
J_{(\vl\cdot\vq)^4}&=
\frac{\vq^4}{2}\left[
3C_{34}+\vq^2C_{35}
\right.\\&\left.
+q^2(3J_{34}+\vq^2J_{35})
\right.\\&\left.
-2q_0(3I_{34}+\vq^2I_{35})
\right]
\\
J_{\vl^2(\vl\cdot\vq)^2}&=
-\vq^2\Big[C_{22}+\vq^2C_{23}
+m^2(J_{22}+J_{23}\vq^2)
+I_{32}
\\&
+\vq^2I_{33}
\Big]
\\
J_{\vl^4}&=
-(C_{22}(D_\eta -1)+C_{23}\vq^2)
\\&
-m^2(J_{22}(D_\eta -1)+J_{23}\vq^2)
-(I_{32}(D_\eta -1)+I_{33}\vq^2)
\end{align*}

\subsubsection{$K_\mu$}
\begin{align*}
K_{10}&=I
\\
K_{11}
&\equiv
\frac{1}{2\vq^2}
\left[
-D(0,q_0')+D+q^2K
+2q_0I
\right]
\end{align*}

\subsubsection{$K_{\mu\nu}$}

For the first two cases we apply the following tricks,
\begin{align*}
K_{20}&\equiv I_{10}
\\
K_{21}&\equiv I_{11}
\\
K_{22}&\equiv\frac{1}{(D_\eta -2)\vq^{\,2}}
\left[-K_{(\vl\cdot\vq)^2}+\vq^2K_{(\vl^2)}\right]
\\
K_{23}&\equiv \frac{1}{(D_\eta -2)\vq^{\,4}}
\left[(D_\eta -1)K_{(\vl\cdot\vq)^2}-\vq^2K_{(\vl^2)}\right]
\end{align*}

Giving the following results,
\begin{align*}
K_{(\vl^2)}&=-D-m^2K+I_{10}-r_0K_{10}
\\
K_{(\vl\cdot\vq)^2}&=
\frac12
\left[
D_{11}+q^2K_{11}
+2q_0I_{11}
\right]\vq^2
\end{align*}

\subsubsection{$K_{\mu\nu\rho}$}

\begin{align*}
K_{30}&\equiv I_{20}
\\
K_{31}&\equiv I_{21}
\\
K_{32}&\equiv I_{22}
\\
K_{33}&\equiv I_{23}
\\
K_{34}&=
\frac{-K_{(\vl\cdot\vq)^3}+\vq^2K_{(\vl\cdot\vq)\vl^2}}
{\vq^4(D_\eta -2)}
\\
K_{35}&\equiv
\frac{(D_\eta +1)K_{(\vl\cdot\vq)^3}-3\vq^2K_{(\vl\cdot\vq)\vl^2}}
{\vq^6(D_\eta -2)}
\end{align*}
\begin{align*}
K_{(\vl\cdot\vq)^3}&=
\frac{\vq^2}{2}\left[
-D_{22}(0,q_0')
-\vq^2D_{23}(0,q_0')
-\vq^2D(0,q_0')
\right.\\&\left.
-2\vq^2D_{11}(0,q_0')
+D_{22}
+\vq^2D_{23}
\right.\\&\left.
+q^2(K_{22}+K_{23}\vq^2)
+2q_0(I_{22}+I_{23}\vq^2)
\right]
\\
K_{(\vl\cdot\vq)\vl^2}&=
-\vq^2\left[D_{11}
+m^2K_{11}
-I_{21}
+r_0K_{21}\right]
\end{align*}

\subsubsection{$K_{\mu\nu\rho\sigma}$}

\begin{align*}
K_{40}&\equiv I_{30}
\\
K_{41}&\equiv I_{31}
\\
K_{42}&\equiv I_{32}
\\
K_{43}&\equiv I_{33}
\\
K_{44}&\equiv I_{34}
\\
K_{45}&\equiv I_{35}
\\
k_{46}&=
2\,\frac{-K_{\vl^2(\vl\cdot\vq)^2}+\vq^2K_{\vl^4}}
{\vq^2(D-2)(2D+3)}
\\K_{47}&=
\frac{-(2D+3)K_{(\vl\cdot\vq)^4}+2(2+D)\vq^2K_{\vl^2(\vl\cdot\vq)^2}-\vq^4K_{\vl^4}}
{\vq^6(D-2)(2D+3)}
\\K_{48}&=
\frac{(D+4)K_{(\vl\cdot\vq)^4}-6\vq^2K_{\vl^2(\vl\cdot\vq)^2}}
{\vq^8(D-2)}
\end{align*}

\begin{align*}
K_{(\vl\cdot\vq)^4}&=
\frac12\left[
2D_{10}\vq^4+\vq^4D
\right.\\&\left.
+q^2(K_{22}\vq^2+K_{23}\vq^4)
\right.\\&\left.
+2q_0(I_{22}\vq^2+
I_{23}\vq^4)
\right]
\\
K_{\vl^2(\vl\cdot\vq)^2}&=
-\Big[D_{22}+D_{23}\vq^2
+m^2(K_{22}+K_{23}\vq^2)
-I_{32}
\\&
-I_{33}\vq^2
\Big]\vq^2
\\
K_{\vl^4}&=
-(D_{22}(D_\eta-1)+D_{23}\vq^2)
\\&
-m^2(K_{22}(D_\eta-1)+K_{23}\vq^2)
\\&
+(I_{32}(D_\eta-1)+I_{33}\vq^2)
\end{align*}

\end{document}